\documentclass[prd,twocolumn,showpacs,preprintnumbers,nofootinbib,amsmath,eqsecnum]{revtex4}

\usepackage{epsfig}

\newcommand{\req}[1]{(\ref{#1})}

\newcommand{\muR}{\mu_{R}^{2}}
\newcommand{\muF}{\mu_{F}^{2}}
\newcommand{\muO}{\mu_{0}^{2}}
\newcommand{\eps}{\epsilon}

\begin{document}

\preprint{IRB-TH-13/00}

\title{BLM scale for the pion transition form factor}

\author{B. Meli\'{c}}
\thanks{Alexander von Humboldt Fellow.
On leave of absence from the 
Rudjer Bo\v{s}kovi\'{c} Institute,
Zagreb, Croatia.}
\email{melic@thphys.irb.hr}
\affiliation{Institut f\"{u}r Physik,
        Universit\"{a}t Mainz,\\
        D-55099 Mainz, Germany} 
\affiliation{Institut f\"{u}r Theoretische Physik,
        Universit\"{a}t W\"{u}rzburg,\\
        D-97074 W\"{u}rzburg, Germany} 
\author{B. Ni\v{z}i\'{c}}
\email{nizic@thphys.irb.hr}
\author{K. Passek}
\email{passek@thphys.irb.hr}
\affiliation{Theoretical Physics Division, 
        Rudjer Bo\v{s}kovi\'{c} Institute, \\
        P.O. Box 180, HR-10002 Zagreb, Croatia}

\date{\today}

\begin{abstract}
\vspace*{0.5cm}
The NLO Brodsky-Lepage-Mackenzie (BLM) scale 
for the pion transition form factor has been determined.
To achieve that,
a consistent calculation 
up to $n_f$-proportional NNLO contributions to both
the hard-scattering amplitude and the 
perturbatively calculable part of the pion
distribution amplitude has been performed.
By combining and matching the results obtained for 
these two amplitudes,
a proper cancellation of
collinear singularities has been established
and the
$\gamma_5$ ambiguity problem 
(related to the use of the dimensional regularization method)
has been  resolved
by using the naive-$\gamma_5$ as well as the 
't Hooft-Veltman (HV) schemes.
It has been demonstrated that the prediction for the 
pion transition form factor is 
independent of the factorization scale $\muF$ 
at every order in the strong coupling constant,
making it possible to use the simplest choice $\muF=Q^2$
at the intermediate steps of the calculation.
Assuming the pion asymptotic distribution amplitude
and working in the $\overline{MS}$ scheme,
we have found the BLM scale to be $\muR=\mu_{BLM}^2\approx Q^2/9$. 
Based on the same distribution, the complete NLO prediction
for the pion transition form factor
has been calculated in the 
$\alpha_V$ definition of the QCD coupling renormalized at 
$\muR=\mu_V^2 = e^{5/3} \mu_{BLM}^2 \approx Q^2/2$.
It is in good agreement with the presently available
experimental data.
\end{abstract}


\maketitle

\section{Introduction}

The pion transition form factor plays a 
crucial role in testing QCD
predictions for exclusive processes. 
It appears in the
amplitude that relates two, in general virtual,
photons with the lightest hadron, the pion,
$\gamma^* \gamma^* \rightarrow \pi^0$. 
Historically, this process attracted much interest
since the axial anomaly \cite{axan}
fixes the value of the form factor when both
virtualities of the photons are zero
($\gamma \gamma \rightarrow \pi^0$). 
For large virtualities of the photons 
(or at least for one of them) perturbative QCD (pQCD) is
applicable \cite{LeBr80}.
A specific feature of this process
is that the leading-order (LO) prediction 
is zeroth order in the QCD coupling
constant, 
and one expects that pQCD for this
process may work at accessible values of
spacelike photon virtualities \cite{RaR96}.
Experimentally, the most favorable situation 
is when one of the photons is real
($\gamma^* \gamma \rightarrow \pi^0$). 
%and we have restricted our investigation to
%this case.

The framework for analyzing exclusive processes 
at large-momentum transfer within the context
of pQCD was initiated and developed in the late seventies
\cite{excfw,LeBr80}.
It was demonstrated to all orders in perturbation theory
that exclusive amplitudes at
large-momentum transfer factorize into a convolution
of 
a process-dependent and perturbatively calculable
hard-scattering amplitude,
with
a process-independent 
distribution amplitude (DA),
one for each hadron involved in the amplitude.
Whereas the DA is intrinsically nonperturbative
and its form is determined by some nonperturbative methods,
the DA evolution is subject to a perturbative treatment.
In the standard hard-scattering approach (sHSA), a hadron is regarded
as consisting only of valence Fock states, transverse quark momenta
are neglected (collinear approximation) as well as quark masses.

Although the LO predictions 
in the sHSA (as well as in the modified hard-scattering approach
(mHSA) in which the collinear approximation is abandoned \cite{mHSA}) 
have been obtained for many exclusive processes,
only a few processes have been analyzed at the 
next-to-leading order (NLO):
the pion electromagnetic form factor
\cite{pff,KaM86,MNP99},
the pion transition form factor \cite{AgC81,Bra83,KaM86}
(and \cite{MuR97} in the mHSA),
and the process
$\gamma \gamma \rightarrow M \overline{M}$ ($M=\pi$, $K$) \cite{Ni87}.

It is well known that, unlike in QED, one cannot rely upon
the LO predictions in pQCD (the expansion parameter, i.e.,
the running coupling constant is rather large at current
energies), and that higher-order corrections are important.
The size of the NLO correction as well as the size of
the expansion parameter, i.e. the QCD running coupling
constant, can serve as sensible indicators of the convergence
of the expansion. 
However, as the truncation of the perturbative expansion 
at any finite order causes the residual dependence 
of the prediction on the choice of the 
renormalization scale and scheme, 
these choices introduce
an ambiguity in the interpretation of the finite-order 
perturbative prediction.
In general, including higher-order corrections
has a stabilizing effect 
(see \cite{MNP99}, for illustration)
reducing the dependence of the
predictions on the schemes and scales 
(since the all-order prediction is independent of the scheme
and scale choice).
However, to asses the convergence of the perturbative expansion,
it is necessary not only to extend the calculation beyond the LO
(which is a very demanding task in many cases),
but also  to optimize the choices of the scale and scheme
according to some sensible criteria.

In the Brodsky-Lepage-Mackenzie (BLM) 
procedure \cite{BLM83,BrLu95},
all vacuum-polarization effects from the
QCD $\beta$-function are resummed 
into the running coupling constant. 
Since the coefficients $\beta_0, \beta_1, \cdots $ are functions
of $n_f$ (number of flavors),
according to BLM procedure,
the renormalization scale best suited
to a particular process in a given order can be, in practice,  
determined by computing vacuum-polarization insertions 
in the diagrams of that order, and by setting the scale
demanding that $n_f$-proportional terms should vanish.
The renormalization scales in the BLM method are physical
in the sense that they reflect the mean virtuality
of the gluon propagators 
and the important advantage of this method is
``pre-summing'' the large ($\beta_0 \alpha_S$)$^n$ terms,
i.e., the infrared renormalons associated with the coupling constant
renormalization (\cite{BrJ98} and references therein). 

The optimization of the renormalization scale and scheme
for exclusive processes by employing the BLM scale fixing
was elaborated  in Ref. \cite{BrJ98}.
It was stated that exclusive processes are especially sensitive 
to the choice of the renormalization scale for the underlying 
hard-scattering amplitude and
since each external momentum entering an exclusive reaction
is partitioned among many propagators of the
underlying hard-scattering amplitude, the physical scales
that control these processes are inevitably much softer
than the overall momentum transfer.
The BLM method was applied to the pion electromagnetic form factor
and the
$\gamma \gamma \rightarrow \pi^+ \pi^-$ process.
For the pion transition form factor, the size of the BLM scale
was only assumed (taken the same as for the pion 
electromagnetic form factor).
Since the LO prediction for the pion transition form factor 
is zeroth order in the QCD coupling constant, 
the NLO corrections \cite{AgC81,Bra83,KaM86} 
represent leading QCD corrections 
and the vacuum polarization contributions
appearing at the next-to-next-to-leading order
(NNLO) are necessary for determining 
the BLM scale for this process. 
 
The purpose of this work is to determine the
BLM scale for the pion transition form factor, 
i.e., for the $\gamma^* \gamma \rightarrow \pi$ process.
Although, the structure of the process 
is simple, the calculation of higher-order corrections
to the hard-scattering amplitude
is complicated by the $\gamma_5$ ambiguity, which appears
when using dimensional regularization.
In our calculation we use dimensional regularization
in $D=4 - 2 \eps$ dimensions to regularize both
ultraviolet (UV) and collinear singularities.
We have obtained the LO, NLO, and 
$n_f$-proportional NNLO terms
for the hard-scattering amplitude 
using the Feynman gauge and
modified minimal-subtraction scheme ($\overline{\text{MS}}$)
for which a suitable compact form has been adopted.
In order to correctly subtract the collinear
singularities and also to verify the right choice of
the $\gamma_5$ prescription, we have also determined the
LO, NLO, and $n_f$-proportional NNLO terms of the 
perturbatively calculable (evolutional) part
of the distribution amplitude.
The $\gamma_5$ ambiguity present in the
calculation of the hard-scattering part has been resolved
by combining and matching the results for the hard-scattering amplitude
with the results for the distribution amplitude
part.
The proper cancellation of singularities has been established
and the $\gamma_5$ problem has been resolved
using both the so-called naive-$\gamma_5$ \cite{ChF79} 
and the 
't Hooft-Veltman (HV) schemes \cite{tHV72,BM77}.
It has been demonstrated that the prediction for the pion transition
form factor is independent
of the factorization scale $\muF$ 
at every order in strong coupling constant.
Finally,
we have been able 
to justify the natural choice 
$\muF=Q^2$ for  the factorization scale
and to determine the renormalization
scale using the BLM scale setting method.

The plan of the paper is as follows.
Section II is devoted to some preliminary considerations.
In Sec. III the calculational procedure is briefly
outlined.
In Sec. IV the LO, NLO, and the $n_f$-proportional NNLO
unrenormalized contribution to both the hard-scattering amplitude
and the perturbatively calculable part of the pion distribution
amplitude are obtained.
Renormalization of the UV divergences and factorization of the
collinear divergences present in the hard-scattering and
the pion distribution amplitude are performed in Sec. V.
The results for both amplitudes are obtained in the naive-$\gamma_5$
as well as the HV scheme.
The complete leading-twist analytical expression 
for the pion transition form factor
up to $n_f$-proportional NNLO terms is obtained in Sec. VI.
Section VII is devoted to determining the BLM scale for the pion
transition form factor based on which the complete NLO
numerical predictions are then obtained
in the $\overline{MS}$ and $\alpha_V$ renormalization schemes.
The concluding remarks are given in Sec. VII.
The $\gamma_5$ problem is addressed in detail in Appendix A.
In Appendix B the Feynman rules for the perturbatively calculable
part of the distribution amplitude are derived.
Finally, in Appendix C, we clarify some often obscured points on the
coupling constant renormalization and justify our renormalization
convention.

\section{Preliminaries}
\label{s:formalism}

The pion transition form factor
$F_{\gamma* \gamma \pi}(Q^2)$ 
for a pseudoscalar meson $\pi^0$
is defined in terms
of the amplitude $\Gamma^{\mu \nu}$ for
$\gamma^*(q,\mu) + \gamma(k,\nu) \rightarrow \pi(P)$, as
\begin{equation}
    \Gamma^{\mu \nu}=i \, e^2 \; F_{\gamma^* \gamma \pi}(Q^{2}) 
         \; \eps^{\mu \nu \alpha \beta} \; P_{\alpha}  q_{\beta}
               \, .
\label{eq:Gmunu}
\end{equation}
For large-momentum transfer $Q^2=-q^2$, 
the form factor 
can be represented \cite{excfw,LeBr80} 
as a convolution 
\begin{equation}
    F_{\gamma^* \gamma \pi}(Q^{2})= 
       \Phi^{*}(x,\muF) \, \otimes \, T_{H}(x,Q^{2},\muF) 
                    \,, 
\label{eq:tffcf}
\end{equation}
where $\otimes$ stands for the usual convolution symbol
defined by 
\begin{equation}
A(z) \otimes B(z) = \int_0^1 dz A(z) B(z) \, .
\label{eq:defconv}
\end{equation}
In Eq. \req{eq:tffcf}, the function
$T_{H}(x,Q^{2},\muF)$ is 
the hard-scattering amplitude
for producing a collinear $q \overline{q}$ pair
from the initial photon pair;
$\Phi^{*}(x,\muF)$ is the pion distribution amplitude
representing the amplitude for the final state
$q \overline{q}$ to fuse into a pion,
i.e., the probability amplitude for finding the valence
$q \overline{q}$ Fock state in the final pion with the 
constituents carrying fractions $x$ and $(1-x)$
of the meson's total momentum $P$; 
$\muF$ is
the factorization (or separation) scale at which
soft and hard physics factorize.

\begin{figure}
  \centerline{\epsfig{file=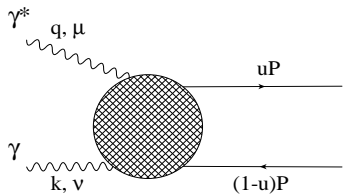,height=85pt,width=130pt,silent=}}
 \caption{Feynman diagram describing the
$\gamma^* \gamma \to q \overline{q}$ amplitude in terms of which the
hard-scattering amplitude for the
$\gamma^* \gamma \to \pi$ transition is obtained.}
 \label{f:T}
\end{figure}

The hard-scattering amplitude $T_{H}$ is obtained by
evaluating the  $\gamma^* \gamma \to q \overline{q}$
amplitude, which is described by the Feynman diagrams in
Fig. \ref{f:T}, with massless on-shell quarks collinear with
outgoing meson.
By definition, $T_H$ is free of collinear singularities and has a
well--defined expansion in $\alpha_S(\muR)$,
with $\muR$ being the renormalization (or coupling constant) scale
of the hard-scattering amplitude.
Thus, one can write
\begin{eqnarray}
  \lefteqn{T_{H}(x,Q^2,\muF)} \nonumber \\ 
       &\qquad =&
         T_{H}^{(0)}(x,Q^2)
         + \frac{\alpha_{S}(\muR)}{4 \pi} \, 
             T_{H}^{(1)}(x, Q^2,\muF) 
             \nonumber \\ & &
         + \frac{\alpha_{S}^2(\muR)}{(4 \pi)^2} \, 
             T_{H}^{(2)}(x, Q^2,\muF,\muR) 
                + \cdots  \, . 
\label{eq:TH}
\end{eqnarray}

Although the function $\Phi(x,\muF)$ is intrinsically 
nonperturbative
(containing the effects of confinement, 
nonperturbative interactions,
and meson bound--state dynamics), 
it  satisfies an evolution equation of the
form
\begin{equation}
  \muF \frac{\partial}{\partial \muF} \Phi(x,\muF)   =
   V(x,u,\alpha_S(\muF)) \, \otimes \, \Phi(u,\muF)
         \, ,
\label{eq:eveq}
\end{equation}
where $V(x,u,\alpha_S(\muF))$ is the 
perturbatively calculable evolution kernel
\begin{eqnarray}
  \lefteqn{V(x,u,\alpha_S(\muF))}
           \nonumber \\ & & \quad
        =  \frac{\alpha_S(\muF)}{4 \pi} \, V_1(x,u) +
                \frac{\alpha_S^2(\muF)}{(4 \pi)^2}  V_2(x,u) 
             + \cdots \, . \qquad
\label{eq:kernel}
\end{eqnarray}
If the distribution amplitude $\Phi(x,\muO)$ 
is determined at an initial momentum scale $\muO$ 
(using some nonperturbative methods),
then the differential-integral evolution equation \req{eq:eveq} 
can be integrated using the moment method to give $\Phi(x,\muF)$
at any momentum scale $\muF > \muO$. 
The one-\cite{LeBr80} and 
two-loop \cite{DiR84etc,Ka85etc,MiR85} corrections to the
evolution kernel are known,
but because of the complicated structure of the two-loop corrections, 
it was possible to obtain numerically only the first few moments 
of the evolution kernel \cite{MiR86}.
However,
based on the conformal spin expansion,
the conformal Ward identities, 
and the conformal consistency relation,
the complete analytical form of the NLO solution 
of the evolution equation \req{eq:eveq}
has been obtained in Ref. \cite{Mu94etc}.

%The form  of the pion distribution amplitude ${\Phi}(x,\muF)$
%is not yet accurately known.
Instead of using ${\Phi}(x, \muF)$, 
one often introduces the distribution amplitude
$\phi(x, \muF)$ normalized to unity 
\begin{equation}
\int_0^1 dx \, {\phi}(x, \muF) =1 \, ,
\label{eq:intphi}
\end{equation}
and related to ${\Phi}(x, \muF)$ by
\begin{equation}
     {\Phi}(x, \muF) = N_{\Phi} \; \phi(x,\muF) 
      \, , 
\label{eq:Phiphi}
\end{equation}
where 
\begin{equation} 
  N_{\Phi}= \frac{f_{\pi}}{2 \sqrt{2 N_c}}
\label{eq:NPhi}
\end{equation}
is the normalization constant imposed by the leptonic
$\pi^+ \rightarrow \mu^+ \nu_{\mu}$ decay, 
$f_{\pi}=0.131$ GeV is the pion decay constant,
and $N_c$($=3$) is the number of QCD colors.

It is convenient to expand
the distribution amplitude $\phi(x,\muO)$ 
(determined at the initial scale $\muO$)
in terms of the
Gegenbauer polynomials $C_n^{3/2}(2 x -1)$,
representing the
eigenfunctions of the LO evolution kernel $V_1$ :
\begin{equation}
     \phi(x,\mu_0^2)= 6 x (1-x) \,
	  \sum_{n=0}^{\infty} {}' B_n \:  C_n^{3/2}(2 x - 1) 
	   \, .
\label{eq:phiexp}
\end{equation}
The nonperturbative input is now contained in
the $B_n$ coefficients
and
$\sum {}'$ denotes the sum over even indices.
The DA \req{eq:phiexp}, when evoluted 
to the scale $\muF$, 
is represented by the perturbative expansion 
\begin{equation}
   \phi(x,\muF)=\phi^{LO}(x,\muF) + \frac{\alpha_S(\muF)}{4 \pi} \,
               \phi^{NLO}(x,\muF) + \cdots \, , 
\label{eq:evDA}
\end{equation}
where
\begin{eqnarray}
     \phi^{LO}(x,\muF)&=& 6 x (1-x) \,
	  \sum_{n=0}^{\infty} {}' B_n^{LO}(\muF) \:  C_n^{3/2}(2 x - 1) 
	  \, ,
        \nonumber \\ 
\label{eq:phiLOexp} \\
     \phi^{NLO}(x,\muF) &=&  6 x (1-x) \, 
	  \sum_{k=2}^{\infty} {}' B_k^{NLO}(\muF) \; C_k^{3/2}(2 x - 1) 
	  \, .
        \nonumber \\ 
\label{eq:phiNLOexp}
\end{eqnarray}
The coefficients $B_n^{LO}(\muF)$ and $B_k^{NLO}(\muF)$ 
depend on the nonperturbative
input $B_n$, as well as on the scales $\muO$ and $\muF$.
Their exact form
can be read from the results 
obtained from \cite{Mu94etc} 
and listed in \cite{MNP99}%
\footnote{It should, however, be pointed out 
that, in contrast to 
the expansion parameter $\alpha_S(\muF)/(4 \pi)$ 
from \protect\req{eq:evDA},  
the expansion parameter
$\alpha_S(\muF)/\pi$ was chosen
in \protect\cite{MNP99}. Hence, the expressions for
$B_n^{LO (NLO)}$ from \protect\cite{MNP99} should be modified
accordingly.}.

\section{Calculational procedure}
\label{s:cprocedure}

Before proceeding with the calculation, 
we would like to point out
some subtleties connected with 
the calculational procedure that (we think) deserve more
explanation. 

The hard-scattering amplitude $T_H$ is
obtained by evaluating 
the $\gamma^* + \gamma \rightarrow q \overline{q}$
amplitude, which 
contains collinear singularities,
owing to the fact that final state quarks are taken to be
massless and on-shell. 
Since, by definition, $T_H$ is a finite quantity free of collinear
singularities, these singularities should be subtracted.
The 
$\gamma^* + \gamma \rightarrow q \overline{q}$
amplitude 
with the Lorentz structure factored out as in Eq. \req{eq:Gmunu}
and denoted by $T$ factorizes as
\begin{equation}
    T(u,Q^2) = T_H(x, Q^2, \muF) \, \otimes \, Z_{T,col}(x, u; \muF)
        \, ,
\label{eq:TTHZ}
\end{equation}
where, 
as usual, $u$ and $1-u$ denote the quark/antiquark longitudinal
momentum fractions,
$\muF$ is a factorization scale at which the separation
of collinear singularities takes place and
all collinear singularities are factorized in $Z_{T,col}$.

On the other hand, 
a process-independent distribution amplitude
for a pion 
in a frame where
$P^+=P^0+P^3=1$, $P^-=P^0-P^3=0$, and $P_{\perp}=0$
is defined \cite{LeBr80,Ka85etc,BrD86} as
\begin{eqnarray}
 \Phi (u)
 & = &  \int \frac{dz^-}{2 \pi} e^{i(u-(1-u))z^- /2}
      \nonumber \\ & & \times
    \left< 0 \left| 
  \bar{\Psi}(-z) \, \frac{\gamma^+ \gamma_5}{2\sqrt{2}}
           \, \Omega \, \Psi(z) 
    \right| \pi \right> _{(z^+=z_{\perp}=0)}
                \, , \qquad 
\label{eq:PhiOPi}
\end{eqnarray}
where 
\begin{eqnarray}
 \Omega & = & 
%P \text{exp} \left\{ i g \int_{-1}^{1} ds A(z s)z \right\}
%               \nonumber \\
%        & = &  
  \text{exp} \left\{ i g \int_{-1}^{1} ds A^+(z s)z^-/2 \right\}
\label{eq:Omega}
\end{eqnarray}
is a path-ordered factor
making $\Phi$ gauge invariant.
The matrix element in \req{eq:PhiOPi} contains an ultraviolet
divergence coming from the light-cone singularity
at $z^2=0$ \cite{LeBr80,BrD86}. This divergence should be regulated, 
and after renormalization, which introduces
a renormalization scale $\tilde{\mu}_R^2$, 
$z^2$ is effectively smeared over a region of order
$z^2=-z_{\perp}^2\sim 1/\tilde{\mu}_R^2$.
As a result,
the pion distribution amplitude $\Phi(v,\tilde{\mu}_R^2)$ 
is obtained corresponding to 
the pion wave function integrated
over the pion intrinsic transverse momentum up to the scale 
$\tilde{\mu}_R^2$.
The distribution amplitude $\Phi(v,\tilde{\mu}_R^2)$ 
is a finite quantity and enters the convolution expression
\req{eq:tffcf}.

The unrenormalized
pion distribution amplitude $\Phi(u)$ 
given in \req{eq:PhiOPi} and the distribution amplitude
$\Phi(v,\tilde{\mu}_R^2)$ renormalized at the scale 
$\tilde{\mu}_R^2$ are 
(owning to the multiplicative renormalizability of the
composite operator $\overline{\Psi} \gamma^+ \gamma_5 \Omega \Psi$) 
related by a multiplicative renormalizability equation
\begin{eqnarray}
   \Phi(u) & = &
            Z_{\phi,ren}(u,v; \tilde{\mu}_R^2) \otimes
              \Phi(v, \tilde{\mu}_R^2)
             \, .
\label{eq:PhiZPhi}
\end{eqnarray}
By differentiating this equation 
with respect to $\tilde{\mu}_R^2$ one obtains
the evolution equation \req{eq:eveq},
with the evolution potential $V$ given by 
\begin{equation}
   V = -Z_{\phi, ren}^{-1} \left( \tilde{\mu}_R^2
       \frac{\partial}{\partial \tilde{\mu}_R^2} Z_{\phi, ren}
              \right)
          \, .
\label{eq:VZ}
\end{equation}
For notational simplicity, here and where appropriate, 
we use the notation 
in which the convolution ($\otimes$)
is replaced by the matrix multiplication in x-y space 
(unit matrix  is given by $\openone \equiv \delta(x-y)$).

The pion distribution amplitude as given in
\req{eq:PhiOPi}, with $\left| \pi \right>$ being the physical
pion state, of course, cannot be determined using perturbation theory. 
If the meson state $\left| \pi \right>$ is replaced
by a $\left| q \overline{q}; t \right>$ state composed of a 
free (collinear, massless, and on-shell) quark and antiquark 
(carrying momenta $t P$ and $(1-t) P$),
the amplitude \req{eq:PhiOPi} becomes
\begin{eqnarray}
  \tilde{\phi} (u, t)
& = &  \int \frac{dz^-}{2 \pi} e^{i(u-(1-u))z^- /2}
   \nonumber \\ & & \times
    \left< 0 \left| 
  \overline{\Psi}(-z) \, \frac{\gamma^+ \gamma_5}{2\sqrt{2}}
           \, \Omega \, \Psi(z) \right| q \overline{q}; t \right>
          \, \frac{1}{\sqrt{N_c}}
              \, . \qquad
\label{eq:phiOqq}
\end{eqnarray}
Taking \req{eq:phiOqq} into account,
Eq. \req{eq:PhiOPi} can be written in the form
\begin{equation}
   \Phi(u)  = 
       \tilde{\phi}(u,t) \otimes
            \, \left< q\bar{q}; t | \pi \right> \, \sqrt{N_c}
               \, .
\label{eq:Phitphirest}
\end{equation}
The amplitude \req{eq:phiOqq} can be treated perturbatively,
making it possible to investigate
the high-energy tail of
the pion DA, to obtain $Z_{\phi,ren}$ 
and to determine the DA evolution.

The distribution $\tilde{\phi}(u,t)$ is multiplicatively
renormalizable and the UV singularities 
that are not removed by the renormalization of the fields
and by the coupling constant renormalization, 
factorize in the renormalization constant 
$Z_{\phi,ren}$ at the renormalization scale
$\tilde{\mu}_R^2$.
Apart from UV singularities, the matrix element in 
\req{eq:phiOqq} contains collinear singularities 
(since the initial state quarks are, as before, taken to be
massless and on-shell),
which are absorbed
in $Z_{\phi,col}$ at the factorization scale $\mu_0^2$.
Hence, one obtains
\begin{eqnarray}
   \lefteqn{\tilde{\phi}(u,t)} \nonumber \\
& = & Z_{\phi,ren}(u,v; \tilde{\mu}_R^2) \otimes 
              \phi_V(v,s; \tilde{\mu}_R^2, \mu_0^2) \otimes
               Z_{\phi,col}(s,t; \mu_0^2)
                 \, .
           \nonumber \\ 
\label{eq:ZfVZ}
\end{eqnarray}
Upon combining Eqs. \req{eq:Phitphirest} and \req{eq:ZfVZ},
the distribution $\Phi(u)$ can be written in the form
\begin{equation}
   \Phi(u)=
         Z_{\phi,ren}(u,v; \tilde{\mu}_R^2) \otimes
         \phi_V(v,s; \tilde{\mu}_R^2, \mu_0^2) \otimes
         \Phi(s, \mu_0^2)
           \, .
\label{eq:PhiZfVPhi}
\end{equation}
Here,
\begin{equation}
\Phi(s, \mu_0^2) = Z_{\phi,col}(s,t; \mu_0^2) \otimes 
 \, \left< q\bar{q}; t | \pi \right> \, \sqrt{N_c}
               \, 
\label{eq:PhiNP}
\end{equation}
represents the nonperturbative input (containing all effects 
of collinear singularities, confinement, and pion bound-state 
dynamics) determined at the scale 
$\mu_0^2$, while 
$\phi_V(v,s; \tilde{\mu}_R^2, \mu_0^2)$
governs the evolution of distribution amplitude
to the scale $\tilde{\mu}_R^2$:
\begin{equation}
   \Phi(v,\tilde{\mu}_R^2)=
         \phi_V(v,s; \tilde{\mu}_R^2, \mu_0^2) \otimes
         \Phi(s, \mu_0^2)
           \, ,
\label{eq:PhifVPhi}
\end{equation}
and satisfies the evolution equation
\begin{eqnarray}
       \lefteqn{\tilde{\mu}_R^2 
       \frac{\partial}{\partial \tilde{\mu}_R^2} 
                        \phi_V (v,s,\tilde{\mu}_R^2,\muO)}
           \nonumber \\  
        \qquad & \qquad =& V(v,s',\tilde{\mu}_R^2) \, \otimes \, 
                           \phi_V(s',s,\tilde{\mu}_R^2,\muO)
             \, . \qquad 
\label{eq:VfV}
\end{eqnarray}

\begin{figure}
  \centerline{\epsfig{file=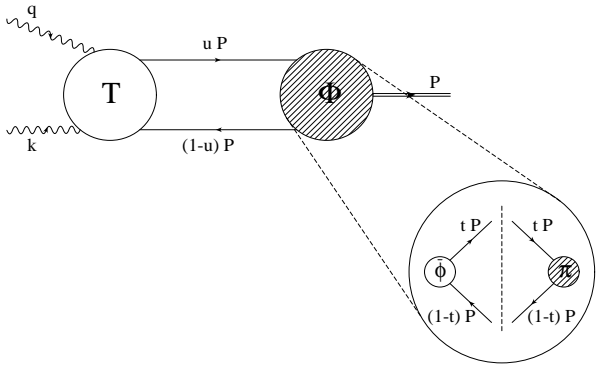,height=150pt,width=240pt,silent=}}
 \caption{Pictorial representation of the 
    of the pion transition form factor 
    calculational ingredients:
    $T$ represents the perturbatively calculable hard-scattering
    amplitude , while $\Phi$ is the pion distribution
    amplitude given by \protect\req{eq:PhiOPi} which can be
    expressed, as in \protect\req{eq:Phitphirest},
    in terms of the perturbatively calculable part
    $\tilde{\phi}$
    \protect\req{eq:phiOqq}
    and the perturbatively uncalculable part.}  
 \label{f:FpiDA}
\end{figure}
By convoluting
the ``unrenormalized'' (in the sense of collinear singularities)
hard-scattering amplitude $T(u,Q^2)$   
with the unrenormalized pion distribution amplitude
$\Phi(u)$,
given by Eqs. \req{eq:TTHZ} and \req{eq:PhiZPhi}, respectively,
one obtains (in a way 
analogous to \cite{CuF80,LeBr80}) 
the following expression for 
the pion transition form factor $F_{\gamma^* \gamma \pi}(Q^2)$:
\begin{equation}
  F_{\gamma^* \gamma \pi}(Q^2) =
       \Phi^{\dagger}(u) \, \otimes \,  T(u, Q^2)
          \, .
\label{eq:Fpiur1}
\end{equation}
The divergences of $T(u,Q^2)$ and $\Phi(u)$ cancel
for $\tilde{\mu}_R^2=\muF$
\begin{equation}
      Z_{T,col}(x,u; \muF) \otimes Z_{\phi, ren}(u,v; \muF)
           = \delta(x-v)
              \, ,
\label{eq:ZTZf}
\end{equation}
and the usual expression \req{eq:tffcf} emerges,
where the pion transition form factor is expressed in
a form of the convolution of two finite amplitudes:
\begin{displaymath}
    F_{\gamma \pi}(Q^{2})=
      T_{H}(x,Q^{2},\muF)  \, \otimes \,  \Phi^{*}(x,\muF)
                    \,.
%\label{eq:tffcf1}
\end{displaymath}
It is worth pointing out that
the scale $\muF$ representing the boundary between the low- and 
high-energy parts in
\req{eq:tffcf} 
is, at same time, the separation scale for
collinear singularities
in $T(u,Q^2)$, on the one hand,
and the renormalization scale for
UV singularities appearing in the perturbatively calculable part
of the distribution amplitude $\Phi(u)$, on the other hand.
The calculational procedure explained above is 
illustrated in Fig. \ref{f:FpiDA}.

Our main goal in this work is to determine the
BLM scale for the pion transition form factor.
To achieve that, 
we make use of the calculational procedure
outlined above and in the following sections
calculate
the LO, NLO, and $n_f$-proportional NNLO contributions to the 
perturbative expansions of both the hard-scattering
amplitude and the distribution amplitude. 

This is the first calculation of the hard-scattering amplitude
$T(u,Q^2)$ of an exclusive process with the NNLO terms taken into account.
The subtraction (separation) of collinear divergences 
at the NNLO is significantly more demanding than that at the NLO.
Owing to the fact that the process under consideration
contains one pseudoscalar meson, the calculation is further
complicated by the $\gamma_5$ ambiguity related to the use
of the dimensional regularization method to treat divergences.

In order to correctly subtract the
collinear divergences and determine the right treatment of the
$\gamma_5$ matrix, we  determine the LO, NLO,
and $n_f$-proportional NNLO contributions to the
distribution $\tilde{\phi}(u,t)$ given in \req{eq:phiOqq},
and by that, following 
(\ref{eq:Phitphirest}-\ref{eq:PhiZfVPhi}),
the renormalization constant $Z_{\phi,ren}$
and the evolutional part $\phi_V$ of the distribution amplitude $\Phi(u)$
\req{eq:PhiOPi}.  
Since, there is no $\gamma_5$ ambiguity in the DA calculation,
the $\gamma_5$ ambiguity present in the hard-scattering
calculation is resolved using \req{eq:ZTZf}.
As an additional check, we employ two
$\gamma_5$ schemes in our calculation.
Finally, we obtain 
the ($\gamma_5$ scheme independent) 
prediction for the pion transition form factor 
$F_{\gamma^* \gamma \pi}$ up to the 
$n_f$-proportional NNLO terms, 
expressed in terms of the finite quantities
$T_H(x,Q^2,\muF)$ and $\Phi(x,\muF)$.

\section{LO, NLO, and 
\lowercase{$n_f$}-proportional NNLO unrenormalized contributions
to the hard-scattering and the distribution amplitudes}
\label{s:T}

In this section we present the calculation
of the LO, NLO, and $n_f$-proportional NNLO contributions
to the hard-scattering amplitude and the perturbatively
calculable part of the distribution amplitude. 

\subsection{Contributions to the hard-scattering amplitude}
\label{ss:Tfd}

The hard-scattering amplitude $T_H$ for the pion transition
form factor is obtained by evaluating the
$\gamma^* \gamma \rightarrow q\overline{q}$ 
amplitude for the parton subprocess, 
which is described by the Feynman diagrams
of Fig. \ref{f:T}.

The $q \overline{q}$ pair has to be projected into
a negative-parity and spin 0 (pseudoscalar) state. 
This is achieved by introducing 
the projection operator $\gamma_5 \not \! P/\sqrt{2}$ 
and taking the trace over a fermion loop.
On the other hand, the color-singlet nature of 
the $q \overline{q}$ state is taken into account
by introducing the factor 
$\sum_{\alpha=1}^{3} \delta_{\alpha \beta}/\sqrt{N_c}$, 
and taking the trace over the color indices.
Also, the flavor function $(u \overline{u}-d\overline{d})/\sqrt{2}$
should be included.

The hard-scattering amplitude $T(u,Q^2)$
can generally be expressed as an expansion
in $\alpha_S$
\begin{eqnarray}
   T(u, Q^2)& = & \frac{N_T}{Q^2} \left\{
        T^{(0)}(u) + \frac{\alpha_S}{4 \pi}\,  \, T^{(1)}(u)
             \right. \nonumber \\ & & 
        + \frac{\alpha_S^2}{(4 \pi)^2} 
          \left[  \left( -\frac{2}{3} n_f \right) 
              \,  T^{(2,n_f)}(u)
              + \cdots \right] 
              \nonumber \\ & & \left.
              + \cdots \right\}
              \, .
\label{eq:T}
\end{eqnarray}

\subsubsection{LO contributions}

\begin{figure}
\centerline{\epsfig{file=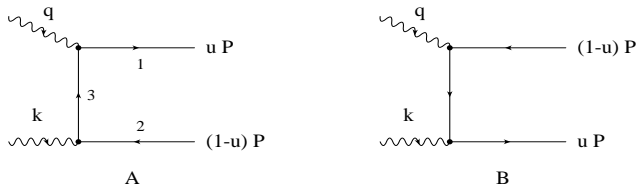,height=2.8cm,width=9cm,silent=}}
 \caption{Lowest-order Feynman diagrams contributing to the
  $\gamma^* \gamma \rightarrow q \overline{q}$ amplitude.}
 \label{f:Tlo}
\end{figure}

In the LO approximation there are only two Feynman diagrams
that contribute to 
the $\gamma^* \gamma \rightarrow q\overline{q}$
amplitude. They are shown in Fig.\ref{f:Tlo}.
The contribution of diagram $A$ 
(after $i e^2 \eps^{\mu \nu \alpha \beta} P_{\alpha}q_{\beta}$
is factored out) is given by
\begin{equation}  
      T_A= \frac{N_T}{Q^2} \; \frac{1}{1-u}
         \, ,
\label{eq:TA}
\end{equation}
where
\begin{equation}
    N_T=2 \sqrt{2 N_c} \, C_{\pi}
     \, ,
\label{eq:NT}
\end{equation}
and
\begin{equation}
  C_{\pi}=\frac{e_u^2 -e_d^2}{\sqrt{2}}=\frac{\sqrt{2}}{6} 
\label{eq:Cpi}
\end{equation}
is the factor
taking into account the flavor content
of the $q \overline{q}$ pair.
The contribution of diagram $B$ is obtained 
by making the replacement $u \rightarrow (1-u)$ in \req{eq:TA}.
Therefore, the lowest-order (QED)
contribution to the
$\gamma^* \gamma \rightarrow q\overline{q}$
amplitude , 
i.e., to $T(u,Q^2)$ given in \req{eq:T}, 
is
\begin{equation}
   T^{(0)}(u)=
         \frac{1}{1-u} + (u \rightarrow 1-u)
       \, .
\label{eq:T0}
\end{equation}

\subsubsection{NLO contributions}

\begin{figure}
  \centerline{\epsfig{file=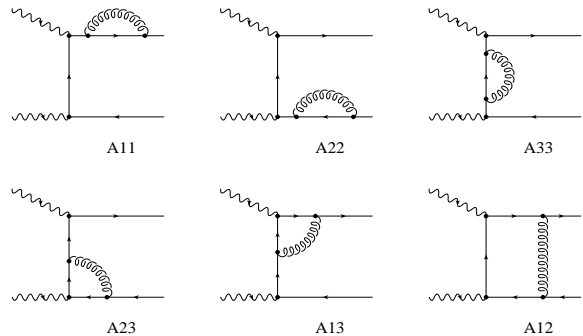,height=5cm,width=8cm,silent=}}
 \caption{Distinct one-loop Feynman diagrams contributing to the
  $\gamma^* \gamma \rightarrow q \overline{q}$ amplitude,
  generated from diagram $A$ of Fig. \protect\ref{f:Tlo}
  by inserting a gluon line.}
 \label{f:Tnlo}
\end{figure}

At NLO there are 12 one-loop Feynman diagrams
contributing to 
the $\gamma^* \gamma \rightarrow q\overline{q}$ 
amplitude. 
They can be generated by inserting an internal gluon line into 
the lowest-order diagrams of Fig. \ref{f:Tlo}.
We use the notation where $Aij$ is the diagram 
obtained from diagram $A$
by inserting the gluon line connecting the lines $i$ and $j$,
where $i,j=1,2,3$.
Since NLO, and all higher-order, diagrams generated 
from diagram $B$ can be obtained 
from the corresponding diagrams generated from
diagram $A$ by using the substitution 
$u \rightarrow (1-u)$,
the total number of NLO diagrams to be evaluated is 6.
They are shown in Fig. \ref{f:Tnlo}. 
%
%%%%%
\renewcommand{\arraystretch}{1.7}
\begin{table*}  
\caption{Contributions $\widehat{T}_{Aij}^{UV}$
and $\widehat{T}_{Aij}^{IR}$ (defined in \protect\req{eq:Aij}) 
of Feynman diagrams  shown in
Fig. \protect\ref{f:Tnlo}. 
Apart from the  result denoted by $12_{\delta=1}$,
the listed results correspond
to the results obtained
using the HV scheme, or equivalently,
to the results obtained 
in the naive-$\gamma_5$ scheme 
with $\gamma_5$ being outside the contracting $\gamma$ matrices.
The result denoted by $12_{\delta=1}$
has been obtained using the naive-$\gamma_5$ scheme
with $\gamma_5$ placed between the contracting $\gamma$ matrices
and between two Dirac slashed loop momenta.}
%\small
 
          \begin{ruledtabular} \begin{tabular}      {cll}
$ij$ & \multicolumn{1}{c}{$\widehat{T}_{Aij}^{UV}$} & 
\multicolumn{1}{c}{$\widehat{T}_{Aij}^{IR}$}
 \\ \hline 
$11(22)$ &
 $\displaystyle - \frac{1}{2} (1- \eps) \left(1 - \frac{\eps}{2}\right)
           \, (1-u)^{-\eps}$ &
 $\displaystyle - \frac{1}{2} (1- \eps) \left(1 - \frac{\eps}{2}\right)
           \, (1-u)^{-\eps}$ 
                 \\[0.3cm]
$33$ &
 $\displaystyle - (1- \eps) \, (1-u)^{-\eps}$ & $\qquad 0 $
                  \\[0.3cm]
$23$ &
  $\displaystyle (1- \eps)  \, (1-u)^{-\eps}$ &
   $\displaystyle (2 + \eps)  \, (1-u)^{-\eps}$ 
                  \\[0.3cm]
$13$ &
  $\displaystyle (1- \eps)  \, \left[ (1-u)^{-\eps} +
        \frac{1}{u}(1-(1-u)^{-\eps}) \right]$ &
    $\displaystyle \left[
     (2 + \eps) (1-u)^{-\eps} + \left( \frac{2}{\eps} + \eps \right)
        \frac{1}{u}(1-(1-u)^{-\eps}) \right]$ 
                  \\[0.3cm]
$12_{\delta=0}$ & $\qquad 0 $ &
 $\displaystyle  -2 (1 + \eps) \,  \left( \frac{1}{\eps} + 2 \right)
        \, \frac{1-u}{u}(1-(1-u)^{-\eps})$ 
                  \\[0.3cm]
$12_{\delta=1}$ & $ \qquad 0 $ &
   $\displaystyle -2 (1 - \eps)  \, \frac{1}{\eps}
        \, \frac{1-u}{u}(1-(1-u)^{-\eps})$  \\[0.3cm]
\end{tabular} \end{ruledtabular}                
 
%\normalsize
\label{t:tTAij}
\end{table*} 
\renewcommand{\arraystretch}{1}

These diagrams contain ultraviolet (UV) singularities, and
owing to the fact that the final state quarks are
massless and on shell they also contain collinear
singularities.
To regularize these singularities, we
use dimensional regularization in
$D=4-2\epsilon$ space-time dimensions.

As is well known, dimensional regularization leads to an ambiguity
when dealing with the pseudoscalar matrix
$\gamma_5$.
The reason for this lies in the fact that the matrix $\gamma_5$
cannot be unambiguously defined in $D \ne 4$ dimensions.
In practice, the difficulty arises in evaluating a trace containing
a single $\gamma_5$.
We address this problem in detail in Appendix
\ref{app:gamma5}.

In order to make sure that our results for the
pion transition form factor are
$\gamma_5$ scheme independent,
we have evaluated all  the contributions using two schemes:
the naive-$\gamma_5$ scheme \cite{ChF79}
and the 't Hooft-Veltman (HV) 
scheme \cite{tHV72,BM77}, defined by
\req{eq:ng5sh} and \req{eq:BMsh1}, respectively. 

A few remarks concerning the diagrams with quark self-energy 
corrections where the quark momentum $p$ is on-shell
($A11$, $A22$) are in order.
Since these corrections modify external legs, 
each of these diagrams is accompanied with a factor
of $1/2$ coming from the expansion of the
quark field renormalization constant $\sqrt{Z_2}$.
In dimensional regularization, the contributions of each of these
diagrams turn out to be proportional to
$\left( p^2 \right)^{-\eps}$ and, therefore,
vanish when $p^2=0$.
On closer inspection, however,
one finds that this vanishing is a result of the cancellation
of a UV pole with collinear pole.
The UV pole contributes to the 
renormalization of the quark fields 
(already taken into account by the factor $\sqrt{Z_2}$) 
and eventually leads to a correct running of the
coupling constant.

The contribution of any of the diagrams $Aij$ 
shown in Fig. \ref{f:Tnlo} can be generally
expressed as
\begin{subequations}
\begin{equation}
   T_{Aij} =
      \frac{N_T}{Q^2} \; \frac{1}{1-u}
      \;\frac{\alpha_S}{4 \pi} \; C_F
                  \; \widetilde{T}_{Aij}
       \, ,
\label{eq:TAij}
\end{equation}
where $C_F=4/3$ is the color factor
(the same for all diagrams), while 
$\widetilde{T}_{Aij}$ is defined by
\begin{eqnarray}
    \widetilde{T}_{Aij} &= &
      \left( \Gamma_{UV}^{(0)}(\eps) \;
      \frac{1}{1-2 \eps}\;
      \widehat{T}_{Aij}^{UV} 
     \right. \nonumber \\ & & \left.
        +
      \Gamma_{IR}^{(0)}(\eps) \;
      \frac{1}{1-2 \eps} \;
      \widehat{T}_{Aij}^{IR} \right)
      \left( \frac{\mu^2}{Q^2} \right)^{\eps}
         \, ,
\label{eq:tTAij}
\end{eqnarray}
\label{eq:Aij}
\end{subequations}%
with the following abbreviations
\begin{subequations}
\begin{eqnarray}
     \Gamma_{UV}^{(0)}(\eps)&=&\Gamma(\eps)
\frac{\Gamma(1-\eps) \Gamma(1-\eps)}{\Gamma(1- 2 \eps)}
          (4 \pi )^{\eps} \, ,
\label{eq:Gamma0UV} \\
     \Gamma_{IR}^{(0)}(\eps)&=&\Gamma(1+\eps)
\frac{\Gamma(-\eps) \Gamma(1-\eps)}{\Gamma(1- 2 \eps)}
          (4 \pi )^{\eps}
        \, .
\label{eq:Gamma0IR}
\end{eqnarray}
\label{eq:Gamma0}
\end{subequations}
The first $\Gamma$ function on the right-hand side of
Eqs. \req{eq:Gamma0} originates
from the loop momentum integration, while 
the integration over Feynman parameters
%(which we employ for evaluating Feynman integrals) 
produces $\Gamma$s collected in a fraction.
Consequently, the singularity contained in $\Gamma(\eps)$ 
appearing in \req{eq:Gamma0UV} is of UV origin, while the
singularity contained in 
$\Gamma(-\eps)$ appearing in \req{eq:Gamma0IR}
is of infrared (IR) origin. 
It should be pointed out, however, that none of the diagrams of
Fig. \ref{f:Tnlo} contains a soft (genuine IR) singularity, 
so that the, here and in the following,
subscript (and / or the superscript) IR
signifies the collinear singularity. 
If the relation
\begin{equation}
  \Gamma(z) \Gamma(1-z) = \frac{\pi}{\sin \pi z}
\label{eq:GzG1z}
\end{equation}
is taken into account in \req{eq:Gamma0},
one finds that
\begin{equation}
\Gamma_{UV}^{(0)}(\eps)=-\Gamma_{IR}^{(0)}(\eps)
         \, .
\label{eq:G0uvG0ir}
\end{equation}
Nevertheless, we continue to keep track of the
origin of the UV and collinear singularities.

The contributions $\widehat{T}_{Aij}^{UV}$
and $\widehat{T}_{Aij}^{IR}$ of the individual diagrams are
given  in Table \ref{t:tTAij}.
Following the explanations and notation given
in App. \ref{app:gamma5},
we list the contributions
obtained using the HV scheme, which are equivalent
to the results obtained in the naive-$\gamma_5$ scheme 
with $\gamma_5$ being positioned
outside the contracted $\gamma$ matrices.
For diagram $A12$ we also list 
the contribution obtained in the naive-$\gamma_5$ scheme
corresponding to the case where $\gamma$ matrices are
contracted through the string of $\gamma$ matrices
of the form $\not l \gamma_5 \not l$. 
As elaborated in Appendix \ref{app:gamma5},
the $\gamma_5$ ambiguity in 
diagrams $A11$, $A22$, $A33$, $A23$, and $A13$
has been resolved
with the help of QED Ward identities. 
The remaining ambiguity in diagram $A12$
is parameterized by $\delta$, taking
the value $0$ for the first choice for handling $\gamma_5$
in diagram $A12$, and $1$ for the second.
Our results listed in Table \ref{t:tTAij}
are in agreement with \cite{Bra83} (but see the comments 
in App. \ref{app:gamma5}).

The NLO contribution $T^{(1)}(u)$ from Eq. \req{eq:T}
is of the form
\begin{subequations}
\begin{eqnarray}
   T^{(1)}(u) &=&
         \left[ \Gamma_{UV}^{(0)}(\eps) \frac{1}{1-2 \eps} 
                  \overline{T}^{(1)}_{UV}(u)
             \right. \nonumber \\ & & \left. +
               \Gamma_{IR}^{(0)}(\eps) \frac{1}{1-2 \eps}
                  \overline{T}^{(1)}_{IR}(u)
               \right] 
         \left(\frac{\mu^2}{Q^2}\right)^{\eps}
            \, , \qquad 
\label{eq:T1} 
\end{eqnarray}
where
\begin{eqnarray}
  \lefteqn{\overline{T}^{(1)}_{UV}(u)} \nonumber \\ &=&
       C_F \, \frac{1}{1-u}
 (1 - \eps) \left[ \frac{\eps}{2} (1-u)^{-\eps} 
     + \frac{1}{u} \left( 1 - (1-u)^{-\eps}\right) \right]
        \nonumber \\ & & 
      + (u \rightarrow 1-u) \, ,
   \label{eq:bT1UV} 
\end{eqnarray}
and
\begin{eqnarray}
  \lefteqn{\overline{T}^{(1)}_{IR}(u)} \nonumber \\ &=&
       C_F \, \frac{1}{1-u}
   \left[ 
    4 + 2 \eps 
- \left( 1 - \frac{3}{2}\eps + \frac{1}{2} \eps^2 \right)
        (1-u)^{-\eps}  
    \right.   \nonumber \\ & &  \left. 
+ \left( \frac{2}{\eps} - 4 - \eps + (4 \delta -3) \frac{1-u}{u} (2 + \eps)
\right) 
        \right. \nonumber \\ & &  \quad \times 
     \left( 1 - (1-u)^{-\eps} \right) \Big]
      + (u \rightarrow 1-u)
          \, .
   \label{eq:bT1IR} 
\end{eqnarray}
\label{eq:T1all} 
\end{subequations}

\subsubsection{$n_f$-proportional NNLO contributions}
\begin{figure}
  \centerline{\epsfig{file=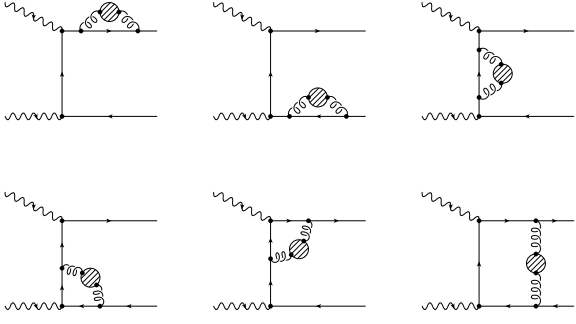,height=4.7cm,width=8cm,silent=}}
 \caption{Distinct vacuum-polarization
  two-loop Feynman diagrams contributing to the
  $\gamma^* \gamma \rightarrow q \overline{q}$ amplitude,
  which have been obtained from the one-loop diagrams
  of Fig. \ref{f:Tnlo} by inserting the vacuum-polarization
  bubbles.}
 \label{f:Tbub}
\end{figure}
%
%%%%%
\renewcommand{\arraystretch}{1.7}
\begin{table*} 
\caption{
Same as Table \ref{t:tTAij} but for 
contributions $\widehat{T}_{(Aij)_{nf}}^{UV}$
and  $\widehat{T}_{(Aij)_{nf}}^{IR}$
(defined in \protect\req{eq:nfAij}), 
corresponding to the
Feynman diagrams shown in
Fig. \protect\ref{f:Tbub}.}.
%\small
 
          \begin{ruledtabular} \begin{tabular}      {cll}
$ij$ & \multicolumn{1}{c}{$\widehat{T}_{(Aij)_{nf}}^{UV}$} & 
\multicolumn{1}{c}{$\widehat{T}_{(Aij)_{nf}}^{IR}$}
 \\[0.1cm] \hline 
$11(22) \quad$ &
 $ \displaystyle  - \frac{1}{2} 
 \frac{(1-\eps)^2 (1-\frac{\eps}{2})}{
                      (1+\frac{\eps}{2})(1-\frac{3}{2}\eps)}
           \, (1-u)^{-2 \eps} \quad$ & 
 $ \displaystyle  - \frac{1}{2} 
 \frac{(1-\eps)^2 (1-\frac{\eps}{2})}{
                      (1+\frac{\eps}{2})(1-\frac{3}{2}\eps)}
           \, (1-u)^{-2 \eps}$  
                 \\[0.5cm]
$33$ &
 $ \displaystyle - \frac{(1-\eps)^2}{(1- \frac{3}{2}\eps)}
           \, (1-u)^{-2 \eps} $ & $\qquad 0 $
                  \\[0.5cm]
$23$ &
 $ \displaystyle  \frac{(1-\eps)^2}{(1-\frac{3}{2}\eps)}
           \, (1-u)^{-2 \eps}$  & 
 $ \displaystyle  \frac{(2 - \eps - 2 \eps^2)}{(1-\frac{3}{2}\eps)}
           \, (1-u)^{-2 \eps}$   
                  \\[0.5cm]
$13$ &
 $ \displaystyle  \frac{(1-\eps)^3}{(1-\frac{3}{2}\eps)}
           \, I_1(u,\eps)$  & 
 $ \displaystyle  \frac{(1-\eps)}{(1-\frac{3}{2}\eps)}
      \left[ 2 \eps (1-\eps) \,I_2(u,\eps) 
       - (2-\eps-2 \eps^2)\, u \, I_3(u,\eps) \right]$
                  \\[0.5cm]
$12_{\delta=0}$ &  $\qquad 0 $ & 
 $ \displaystyle  -2 (1+\eps)
      \left[ - (1-u)^{-2 \eps} 
             - 2 \eps \,  I_2(u,\eps)
             - (1-\eps) \, u \, I_3(u,\eps)
             \right] (1-u)$
%         \right. $
%                 \\[0.1cm]
% & & $\displaystyle
%             \left. \; \;
%             - 2 \eps \,  I_2(u,\eps)
%             \right] (1-u)$
                  \\[0.5cm]
$12_{\delta=1}$ &  $\qquad 0 $ & 
 $ \displaystyle  -2 (1-\eps)
      \left[ - (1-u)^{-2 \eps} 
             - (1-\eps) \, u \, I_3(u,\eps)
             \right] (1-u)$
                  \\[0.1cm]
\end{tabular} \end{ruledtabular}                
 
%\normalsize
\label{t:tTnfAija}
\end{table*} 
\renewcommand{\arraystretch}{1}

By inserting the vacuum polarization bubbles in the NLO diagrams
of Fig. \ref{f:Tnlo}, the NNLO diagrams displayed 
in Fig. \ref{f:Tbub} are obtained.
The vacuum polarization insertion is (in the Feynman gauge)
given by the replacement
\begin{equation}
-i \, g^{\kappa \lambda} \, \frac{\delta_{a b}}{l^2+i \eta}
     \rightarrow
-i\, \left( g^{\kappa \lambda}  - 
   \frac{l^{\kappa} l^{\lambda}}{l^2+i \eta}  \right) \, 
     \frac{\delta_{a b}}{l^2+i \eta}  \, \Pi(l^2)
            \, ,
\label{eq:GBUB}
\end{equation}
where %$a,b$ are color factors and
\begin{eqnarray}
 \Pi(l^2) & = & \frac{g^2}{(4 \pi)^2}
      \left( \frac{\mu^2}{-l^2 - i \eta} \right)^{\eps}
      \left[ ( 5 - 3 \eps ) - \frac{2}{3} ( 1 -  \eps ) n_f \right]
             \nonumber \\ & & \times
      \frac{1}{(1-\frac{2}{3}\eps )(1 - 2 \eps)}
       \Gamma(\eps) 
      \frac{\Gamma(1-\eps) \Gamma(1-\eps)}{\Gamma(1- 2 \eps)}
          (4 \pi )^{\eps} 
             \, ,
             \nonumber \\ & & 
\label{eq:BUB}
\end{eqnarray}
and, due to gauge invariance 
of the complete finite order contribution 
(which we have used as an additional check of our calculation), 
it can effectively be described by 
\begin{equation}
-i \, g^{\kappa \lambda} \, \frac{\delta_{a b}}{l^2+i \eta}
     \rightarrow
-i\,  g^{\kappa \lambda}   
    \, 
     \frac{\delta_{a b}}{l^2+i \eta}  \, \Pi(l^2)
            \, .
\label{eq:GBUBef}
\end{equation}
We are interested only in $n_f$-proportional part
(from the quark loops inserted in the gluon propagator)
\begin{subequations}
\begin{equation}
 \Pi_{n_f}(l^2)  =  \left(- \frac{2}{3} n_f \right) \,
          \frac{1}{(l^2 + i \eta)^{\eps}} \,
         \frac{g^2}{(4 \pi)^2} \,
            f_{n_f}(\eps, \mu^2)
             \, ,
\label{eq:Inf} 
\end{equation}
where $f_{n_f}$ is defined by
\begin{equation}
  f_{n_f}(\eps, \mu^2)  =   
      \left( \frac{\mu^2}{-1 - i \eta'} \right)^{\eps}
      \frac{1-\eps}{(1-\frac{2}{3}\eps )(1 - 2 \eps)}
%       \Gamma(\eps) 
%      \frac{\Gamma(1-\eps) \Gamma(1-\eps)}{\Gamma(1- 2 \eps)}
%          (4 \pi )^{\eps} 
        \Gamma_{UV}^{(0)}(\eps)
         \, .
\label{eq:fnf}
\end{equation}
\label{eq:nf}
\end{subequations}

The contributions of the two-loop Feynman diagrams 
shown in Fig. \ref{f:Tbub} can than be generally written as
\begin{subequations}
\begin{equation}
 T_{(Aij)_{nf}}=
       \frac{N_T}{Q^2} \;  \frac{1}{1-u} \;
       \frac{\alpha_S^2}{(4 \pi)^2}  \; C_F \;  
         \left( - \frac{2}{3} n_f \right) \;
             \widetilde{T}_{(Aij)_{nf}}
         \, ,
\label{eq:TnfAij}
\end{equation}
where
%\begin{equation}
%   \widetilde{T}_{(Aij)_{nf}}=\widetilde{T}_{(Aij)_{nf}}^{(a)}
%                        -\widetilde{T}_{(Aij)_{nf}}^{(b)}
%        \, ,
%\label{eq:nfab}
%\end{equation}
%and
%$\widetilde{T}_{(Aij)_{nf}}^{(a)}$ and
%$\widetilde{T}_{(Aij)_{nf}}^{(b)}$  are obtained
%from the
%$g^{\kappa \lambda} l^2$ and $l^{\kappa} l^{\lambda}$
%proportional parts of \req{eq:Inf}, respectively.
%They are given by
\begin{eqnarray}
%    \lefteqn{\widetilde{T}_{(Aij)_{nf}}^{(a,b)}}
    \lefteqn{\widetilde{T}_{(Aij)_{nf}}}
             \nonumber \\ &=&
      \left( \Gamma_{UV}^{(0)}(\eps) \,
             \Gamma_{UV}^{(1)}(\eps) \,
      \frac{(1-\eps)}{(1-\frac{2}{3} \eps)(1-2 \eps)(1-3 \eps)} \, 
      \widehat{T}_{(Aij)_{nf}}^{UV} 
        \right. \nonumber \\ & & \left.
      + \; \Gamma_{UV}^{(0)}(\eps) \,  
             \Gamma_{IR}^{(1)}(\eps) \,
      \frac{(1-\eps)}{(1-\frac{2}{3} \eps)(1-2 \eps)(1-3 \eps)} \, 
      \widehat{T}_{(Aij)_{nf}}^{IR} \right) 
             \nonumber \\ & & \times \,
      \left( \frac{\mu^2}{Q^2} \right)^{2 \eps}
           \, ,
\label{eq:tTnfAij}
\end{eqnarray}
\label{eq:nfAij}
\end{subequations}
while, similarly to \req{eq:Gamma0}, the abbreviations
\begin{subequations}
\begin{eqnarray}
     \Gamma_{UV}^{(1)}(\eps)&=&
   \frac{\Gamma(2 \eps)}{\Gamma(1+\eps)} \;
\frac{\Gamma(1-2 \eps) \Gamma(1-\eps)}{\Gamma(1- 3 \eps)}
          (4 \pi )^{\eps} \quad  
\label{eq:Gamma1UV} \\
     \Gamma_{IR}^{(1)}(\eps)&=&
  \frac{\Gamma(1+2 \eps)}{\Gamma(1+\eps )} \;
\frac{\Gamma(-2 \eps) \Gamma(1-\eps)}{\Gamma(1- 3 \eps)}
          (4 \pi )^{\eps} \quad
\label{eq:Gamma1IR}
\end{eqnarray}
\label{eq:Gamma1}
\end{subequations}
have been introduced. 

The contributions $\widehat{T}_{(Aij)_{nf}}^{UV,IR}$
%and $\widehat{T}_{(Aij)_{nf}}^{(b)UV,IR}$ 
of the individual diagrams are
listed  in Table \ref{t:tTnfAija}.
The integrals $I_i(u,\eps)$ ($i=1,2,3$) appearing
in this table are defined as
\begin{eqnarray}
   I_1(u,\eps) &\equiv& I(u; \eps, 2 \eps)
             \\
   I_2(u,\eps) &\equiv& I(u; \eps, 1+2 \eps)
             \\
   I_3(u,\eps) &\equiv& I(u; 1+\eps, 1+2 \eps)
             \, ,
\label{eq:I123}
\end{eqnarray}
where
%\newpage
\begin{eqnarray}
I(u; a, c)&=&\int_0^1 dy \frac{y^a}{(1-u y)^c} 
            \nonumber \\
          &=& \frac{1}{1+a}
       \ _2F_1(c, 1+a, 2+a; u)
               \, .
\label{eq:Iac}
\end{eqnarray}
As far as the $\gamma_5$-scheme dependence of the NNLO
diagram contributions is concerned, it is the same as for
the NLO diagrams from Table \ref{t:tTAij}.

%As expected from the gauge invariance of the hard-scattering
%amplitude, the sum of the 
%$\widetilde{T}_{(Aij)_{nf}}^{(b)}$ contributions
%is zero, 
%%$\sum \widetilde{T}_{(Aij)_{nf}}^{(b)} = 0$,
%and therefore
%$\sum \widetilde{T}_{(Aij)_{nf}} =  \sum
%\widetilde{T}_{(Aij)_{nf}}^{(a)}$.

\begin{widetext}
The $n_f$-proportional NNLO contribution
$T^{(2)}_{n_f}(u)$ from Eq. \req{eq:T}
takes the form
\begin{subequations}
\begin{eqnarray}
   T^{(2,n_f)}(u)&=&
         \left[ \Gamma_{UV}^{(0)}(\eps) \Gamma_{UV}^{(1)}(\eps) 
 \frac{(1-\eps)}{(1-\frac{2}{3}\eps)(1-2\eps) 
                                       (1-3 \eps)} 
                  \overline{T}^{(2,n_f)}_{UV}(u)
            \right. \nonumber \\ &  & \left. 
                    +
               \Gamma_{UV}^{(0)}(\eps) \Gamma_{IR}^{(1)}(\eps) 
 \frac{(1-\eps)}{(1-\frac{2}{3}\eps)(1-2\eps)
                                      (1-3 \eps)} 
                  \overline{T}^{(2,n_f)}_{IR}(u)
               \right] 
         \left(\frac{\mu^2}{Q^2}\right)^{2 \eps}
           \, ,
      \label{eq:T2nf}
\end{eqnarray}
where
\begin{eqnarray}
  \overline{T}^{(2,n_f)}_{UV}(u) &=&
       C_F \, \frac{1}{1-u}
      \frac{(1-\eps)^2}{1-\frac{3}{2} \eps}
   \left[ (1-\eps) \, I_1(u,\eps)  
         - \frac{1-\frac{\eps}{2}}{1+\frac{\eps}{2}} 
             \,  (1-u)^{-2 \eps} \right]
      + (u \rightarrow 1-u)
         \, ,
   \label{eq:bT2nfUV} \\[0.3cm]
  \overline{T}^{(2,n_f)}_{IR}(u) &=&
       C_F \, \frac{1}{1-u}
   \left\{ \left[ \frac{1+\frac{5}{2} \eps -\frac{9}{2} \eps^2
             - \frac{1}{2} \eps^3}{(1+\frac{\eps}{2})
                      (1-\frac{3}{2}\eps)}
               + 2  \Big(1 - (2 \delta -1) \eps \Big) (1-u) \right]
         (1-u)^{-2 \eps} 
           \right.  \nonumber \\ & & \left.
       + 2 \eps \left[ \frac{(1-\eps)^2}{1-\frac{3}{2}\eps}
              - (2 \delta -2) \, (1+\eps)\,  (1-u) \right]
          \,  I_2(u,\eps)
           \right.  \nonumber \\ & & \left.
 -  2 \left[ \frac{(1-\frac{\eps}{2}- \eps^2)(1-\eps)}{1-\frac{3}{2}\eps} 
              - \Big(1-(2 \delta-1 )\eps \Big) (1-\eps)\,   (1-u)
               \right] \, u \,  I_3(u,\eps)
   \right\} 
        \nonumber \\ & & 
      + (u \rightarrow 1-u)
          \, .
     \label{eq:bT2nfIR}
\end{eqnarray}
\label{eq:T2nfall}
\end{subequations}
\end{widetext}

Our results are expressed in a compact form in which the
complete functional dependence on the dimensional
parameter $\eps$ is retained.
This is in contrast to the expansion over $\eps$
often encountered in the literature.
In this expansion, nonleading terms in $\eps$ are neglected
before renormalization and factorization of collinear
singularities.
As we show in Sec. \ref{ss:ren},  it is advantageous 
(both for the simplicity and accuracy check of the calculation)
not to  expand the functions $\Gamma^{(0,1)}_{UV,IR}$ over $\eps$.

\subsection{Contributions to the perturbatively
calculable part of the distribution amplitude}
\begin{figure}
  \centerline{\epsfig{file=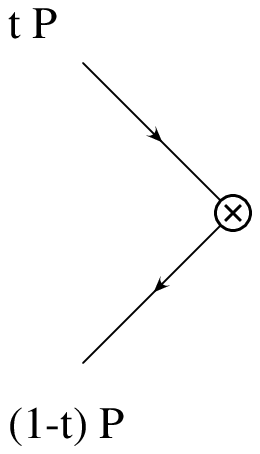,height=102pt,width=60pt,silent=}}
 \caption{The LO diagram contributing to 
          the $\tilde{\phi}(u,t)$ distribution
          \protect\req{eq:phiOqq}, i.e., \protect\req{eq:phiOqqFr}.} 
 \label{f:DAlo}
\end{figure}

In Sec. \ref{s:formalism} we have defined the 
distribution amplitude $\tilde{\phi} (u, t)$
\req{eq:phiOqq} representing the
perturbatively calculable part of the pion distribution
amplitude. 
Following \cite{Ka85etc}, we have rederived the
Feynman rules for this operator in the Feynman gauge. 
They are listed in Appendix \ref{app:DA}.

We now proceed to calculate LO, NLO, and $n_f$-proportional
NNLO contributions to the $\tilde{\phi}(u,t)$ distribution
amplitude defined in \req{eq:phiOqq}, or equivalently in \req{eq:phiOqqFr}.

Contrary to \cite{Ka85etc,BrD86}, we  
use dimensional regularization to regularize
both UV and collinear singularities
\footnote{
The evolutional behavior of the DA can be
extracted from \protect\req{eq:phiOqq} 
even when using dimensional regularization
for both UV and mass singularities.
We introduce the auxiliary scale $\tilde{Q}^2$
and we insist on
discriminating between
UV and collinear singularities. 
Otherwise, the UV and collinear part of 
higher-order corrections would cancel,
leading to $\tilde{\phi}(u,t)=\delta(u-t)$.
}.
This enables us to combine these results
with the hard-scattering results, also 
obtained by employing the dimensional regularization.
Compared to the hard-scattering amplitude calculation, 
calculation of the $\tilde{\phi}$ amplitude is
complicated by the fact that noncovariant
$l^+$, $l^-$ and $\delta$-function
terms (see \req{eq:OvertexS} and \req{eq:Fr0})
appear in the loop-momenta.
To deal with these types of terms and in order to simplify 
the expressions we follow the prescription
given in \cite{Ka85etc}.

The presence of two $\gamma_5$ matrices 
in the traces  has enabled their unambiguous
treatment in the naive-$\gamma_5$ scheme
(see Appendix \ref{app:gamma5}).
Additionally, we have obtained the results 
using the HV scheme, which, however, introduces
the ``spurious'' anomalous terms,
and hence
the additional renormalization is required.
The corresponding renormalization constant
will be determined by comparing the results
obtained in the naive-$\gamma_5$ with those
obtained in the HV scheme.

The perturbatively calculable $\tilde{\phi} (u, t)$ amplitude 
can be represented
as a series in $\alpha_S$ 
\begin{eqnarray}
   \tilde{\phi} (u, t)& = &\tilde{\phi}^{(0)}(u,t) 
     + \frac{\alpha_S}{4 \pi}\,  \tilde{\phi}^{(1)}(u,t)
           \nonumber \\ & &
     + \frac{\alpha_S^2}{(4 \pi)^2} 
  \left[ \, \left( - \frac{2}{3} n_f \right)\, \tilde{\phi}^{(2,n_f)}(u,t)
         + \cdots \right] + \cdots \, . 
            \nonumber \\ & &
\label{eq:phiutraz}
\end{eqnarray}

\begin{figure}
  \centerline{\epsfig{file=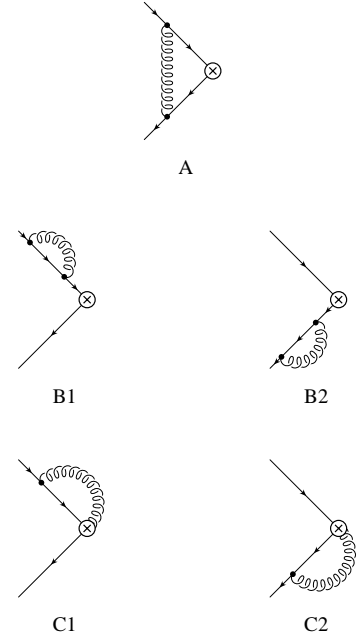,height=255pt,width=140pt,silent=}}
 \caption{The one-loop diagrams contributing to  
          the $\tilde{\phi}(u,t)$ distribution
          \protect\req{eq:phiOqq}, i.e., \protect\req{eq:phiOqqFr} 
          (in the Feynman gauge).}
 \label{f:DAnlo}
\end{figure}

\subsubsection{LO contribution}

The contribution of the LO Feynman diagram 
displayed in Fig. \ref{f:DAlo}
reads 
\begin{equation}
   \tilde{\phi}^{(0)}(u,t)  =  \delta(u - t)
            \, .
%\label{eq:phi0}
\end{equation}

\subsubsection{NLO contributions}

%%%%%
\renewcommand{\arraystretch}{1.7}
\begin{table*} %
\caption{Contributions $F_X$ 
        appearing in Eqs.
        \protect\req{eq:KX} and \protect\req{eq:KXnfa}
        and corresponding to the diagrams ($X$) displayed
        in Fig. \ref{f:DAnlo} and Fig. \ref{f:DAbub}.
        We list also the general results for the integrals 
        appearing in Eqs.
        \protect\req{eq:KX} and \protect\req{eq:KXnfa},
        and parameterized by $\kappa=0$ and $\kappa=\eps$,
        respectively.} 
 
        \begin{ruledtabular} \begin{tabular}     {ccc}
X & \multicolumn{1}{c}{$F_X(u,t;y)$}  
 &\multicolumn{1}{c}{
 \rule[-4mm]{0mm}{10mm} $\displaystyle 
             \left[ \int_0^1 dy \, y^{\kappa}
                           F_X(u,t;y)\right]_{\kappa=0, \, \eps} $} 
 \\ \hline 
$A$ & $(1-\eps) \left[ u_1  \delta(u_1 - y \,t_1) + 
                     u_2  \delta(u_2 - y \,t_2) \right]
                   $ \hspace{0.7cm}
 & 
 $(1-\eps)\displaystyle 
    \left[\frac{u_1}{t_1} \theta(t_1-u_1) 
          \left(\frac{u_1}{t_1}\right)^{\kappa}
                    + 
\renewcommand{\arraystretch}{1}
        \left( \begin{array}{c} u_1 \rightarrow u_2 \\
                                t_1 \rightarrow t_2
               \end{array} \right) \right]$  
\renewcommand{\arraystretch}{2} 
           \\[0.5cm]
$A + B1 + B2$ \hspace{0.25cm}
 & $(1-\eps) \Big\{ u_1  \delta(u_1 - y\, t_1) + 
                     u_2  \delta(u_2 - y\, t_2) \Big\}_+
                   $ 
 &
 $(1-\eps)\displaystyle 
    \left\{\frac{u_1}{t_1} \theta(t_1-u_1) 
          \left(\frac{u_1}{t_1}\right)^{\kappa}
                    + 
\renewcommand{\arraystretch}{1}
        \left( \begin{array}{c} u_1 \rightarrow u_2 \\
                                t_1 \rightarrow t_2
               \end{array} \right) \right\}_+$  
\renewcommand{\arraystretch}{2} 
           \\[0.5cm]
$C1$ & $-\left\{ \displaystyle\frac{u_1}{u_1-t_1} \delta(u_1-y\, t_1) 
              \right\}_+  $
 &
 $\displaystyle 
    \left\{\frac{u_1}{t_1} 
           \frac{1}{t_1-u_1} \theta(t_1-u_1) 
          \left(\frac{u_1}{t_1}\right)^{\kappa}
               \right\}_+ $  
           \\[0.5cm]
\end{tabular} \end{ruledtabular}             
 
\label{t:FXintFX}
\end{table*} %
\renewcommand{\arraystretch}{1}
%%%%%

At NLO there are 5 one-loop Feynman diagrams 
contributing when the Feynman gauge is used.
They are displayed in Fig. \ref{f:DAnlo}.
The general form of these individual
contributions (denoted by $X$) is given by
\begin{eqnarray}
   \lefteqn{\tilde{\phi}_X(u,t)} \nonumber \\ & = &   
     \frac{\alpha_S}{4 \pi} \, 
        K_X (u,t) \; 
      \left\{ \frac{(4 \pi)^2}{i} \,
         \left[ \mu^{2 \eps}
             \int \frac{d^D l}{(2 \pi)^D}
        \frac{1}{(l^2 + i \eta )^2} \right] \right\}
              \, ,
           \nonumber \\
\label{eq:phiX}
\end{eqnarray}

where
\begin{equation}
   K_X (u,t)  = 2\, C_F \; \int_0^1 dy \, F_X(u,t; y)
         \, , 
\label{eq:KX}
\end{equation}
and the $F_X(u,t; y)$ and $K_X (u,t)$ contributions
can be read from Table \ref{t:FXintFX}.
The notation $u_1=u$, $u_2=1-u$ has been introduced ,
as well as the usual definition of % 
%\newpage
the ``${}_+$'' form
\begin{equation}
   \Big\{ F(x,y) \Big\}_+ \equiv F(x,y) 
        - \delta(x-y) \int_0^1 dz \, F(z,y)
              \, ,
%\label{eq:F+}
\end{equation}
the presence of which reflects the chiral symmetry
conservation.
This ``${}_+$'' form is a consequence of the fact
that the axial current is conserved in the chiral limit,
and represents a general all-order property
\cite{MiR85}.

By definition, the $D$ dimensional integral
in Eq. \req{eq:phiX}
gives zero in dimensional regularization,
but only if we do not distinguish
between UV and collinear singularities
\footnote{
The $D$ dimensional integrals appearing in
Eqs. \protect\req{eq:phiX} and \protect\req{eq:phiXnfab}
are of the form
%\begin{subequations}
\begin{displaymath}
I_{[\alpha]} =   \, 
           \mu^{2 \eps}
             \int \frac{d^D l}{(2 \pi)^D}
        \frac{1}{(l^2 + i \eta )^{\alpha}} 
               \, .
%\label{eq:Ialphadef}
\end{displaymath}
By employing 
\begin{displaymath}
I_{[\alpha]} =   
          \left[ \mu^{2 \eps}
             \int \frac{d^D l}{(2 \pi)^D}
        \frac{(l - p)^2}{(l^2 + i \eta )^{\alpha}((l-p)^2 + i \eta )} 
                \right]_{p^2\ne0} 
              \, ,
%\label{eq:Ialphap}
\end{displaymath}
and insisting on distinguishing the $\Gamma$ functions
obtained from the loop-momentum integrations and
$\Gamma$ functions from the Feynman parameter integration,
it can be shown that 
\begin{eqnarray*}
I_{[\alpha]}& = &  
         \frac{i}{(4 \pi)^2}
         \frac{1}{(p^2+i \eta)^{\alpha-2}} 
         \left(\frac{4 \pi \mu^2}{-p^2-i \eta}\right)^{\eps}
            \nonumber \\ &  & \times \, 
         \left[\frac{\Gamma(\alpha-2+\eps)}{\Gamma(\alpha)}
        \frac{\Gamma(3-\alpha-\eps) \Gamma(2-\eps)}{
              \Gamma(5-\alpha-2 \eps)} (2-\eps)
            \right. \nonumber \\ &  & \left. \quad 
                  +
         \frac{\Gamma(\alpha-1+\eps)}{\Gamma(\alpha)}
        \frac{\Gamma(2-\alpha-\eps) \Gamma(3-\eps)}{
              \Gamma(5-\alpha-2 \eps)} 
               \right] 
                 \, .
%\label{eq:IalphaRes}
\end{eqnarray*}
%\label{eq:Ialpha}
%\end{subequations}
Here, the first fraction 
in the terms containing $\Gamma$ functions
corresponds to the loop-momentum integration 
possibly resulting in UV singularities,
while the second fraction corresponds to the
integration over Feynman parameters and 
consequently, to collinear singularities.
For $\alpha < 2+\tilde{\eps}$, where $\tilde{\eps}=0$ or $\ll$,
only the UV singularities appear, 
while for $\alpha > 2+\tilde{\eps}$, 
only the collinear singularities appear.
The two terms in the bracket %from \req{eq:IalphaRes}  
cancel in both cases, so
$I_{[\alpha\ne2+\tilde{\eps}], \tilde{\eps}\ll}=0$.
However, for $\alpha = 2+\tilde{\eps}$,
both UV and collinear singularities are present,
and the cancellation can occur only if we
abandon distinguishing them.}.
By discriminating between the singularities
of different origin, we obtain
the following expression:
\begin{eqnarray}
  \lefteqn{-i \, (4 \pi)^2 \,
         \left[ \mu^{2 \eps}
             \int \frac{d^D l}{(2 \pi)^D}
        \frac{1}{(l^2 + i \eta )^2} \right]}
         \nonumber \\
             & = &
         \left[ \Gamma_{UV}(\eps) \frac{1}{1-2 \eps} 
                \left( 1 - \frac{\eps}{2} \right)
            \right. \nonumber \\ &  & \left. 
                    +
               \Gamma_{IR}(\eps) \frac{1}{1-2 \eps}
                \left( 1 - \frac{\eps}{2} \right)
               \right] 
         \left(\frac{\mu^2}{\widetilde{Q}^2}\right)^{\eps}
                 \, ,
\label{el:Int2}
\end{eqnarray}
where $\tilde{Q}^2>0$ represents the auxiliary scale.

The NLO contribution $\tilde{\phi}^{(1)}(u,t)$,
to which Feynman diagrams of Fig. \ref{f:DAnlo} contribute,
can then be expressed by
\begin{eqnarray}
\tilde{\phi}^{(1)}(u,t) & = &   
         \left[ \Gamma_{UV}^{(0)}(\eps) \frac{1}{1-2 \eps} 
                \left( 1 - \frac{\eps}{2} \right)
                  K^{(1)}(u,t)
            \right. \nonumber \\ &  & \left. 
                    +
               \Gamma_{IR}^{(0)}(\eps) \frac{1}{1-2 \eps}
                \left( 1 - \frac{\eps}{2} \right)
                  K^{(1)}(u,t)
               \right] 
         \left(\frac{\mu^2}{\widetilde{Q}^2}\right)^{\eps}
             \, ,
             \nonumber \\ &  & 
\label{eq:phi1} 
\end{eqnarray}
where the function $K^{(1)}$ calculated
in the naive-$\gamma_5$ scheme amounts to
\begin{eqnarray}
  K^{(1)} & = & 2\, C_F\,
  \left\{ \frac{u}{t} 
    \left[ (1-\eps) + \frac{1}{t-u} \right] 
         \theta(t-u) 
              \right. \nonumber \\ & & \left.
        + \bigg( \begin{array}{c} u \rightarrow 1-u \\
                                t \rightarrow 1-t
               \end{array} \bigg) \right\}_+
               \, .
  \label{eq:K1}   
\end{eqnarray}
 
\subsubsection{$n_f$-proportional NNLO contributions}
\begin{figure}
  \centerline{\epsfig{file=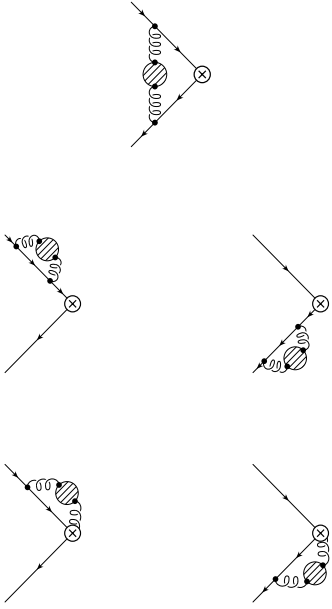,height=240pt,width=135pt,silent=}}
 \caption{Distinct two-loop Feynman diagrams 
          contributing to the $\tilde{\phi}(u,t)$ distribution
          \protect\req{eq:phiOqq}, i.e., \protect\req{eq:phiOqqFr}, 
          which have been obtained
          by inserting the vacuum polarization bubbles in
          the one-loop diagrams of Fig. \protect\ref{f:DAnlo}.}
 \label{f:DAbub}
\end{figure}

By inserting the vacuum polarizations in the NLO 
diagrams of Fig. \ref{f:DAnlo}, we obtain the
NNLO diagrams displayed in Fig. \ref{f:DAbub}.
The $n_f$-proportional contributions% 
\footnote{There are more two-loop diagrams
containing quark loops, but they
contribute to the $n_f$-proportional
NNLO part of the meson singlet
distribution amplitude.}
of these diagrams,
evaluated using Eqs. (\ref{eq:GBUBef}-\ref{eq:nf}), read
 
%%%%%
\begin{eqnarray}
   \lefteqn{\tilde{\phi}_{(X)_{nf}}(u,t)}
         \nonumber \\ &  = & 
      \frac{\alpha_S^2}{(4 \pi)^2} \, 
  \left( - \frac{2}{3} n_f \right) \,
       K_{(X)_{nf}}(u,t) 
              \, (1+ \eps) \;
              f_{n_f}(\eps, \mu^2) \, 
         \nonumber \\ &  & \times \, 
        \left\{\frac{(4 \pi)^2}{i} \, 
          \left[ \mu^{2 \eps}
             \int \frac{d^D l}{(2 \pi)^D}
        \frac{1}{(l^2 + i \eta )^{2+\eps}} \right] \right\}
                \, .
\label{eq:phiXnfab}
\end{eqnarray}
The function
$f_{n_f}(\eps, \mu^2)$ is defined in \req{eq:fnf},
while $K_{(X)_{nf}}$ %and $K_{(X)_{nf}}^{(b)}$
%(obtained from the $g^{\kappa \lambda}l^2$ and 
%$l^{\kappa}l^{\lambda}$ proportional part
%of \req{eq:Inf}, respectively) 
is given by
%\begin{subequations}
\begin{equation}
  K_{(X)_{nf}}(u,t)  = 
        2\, C_F\, \int_0^1 dy \, y^{\eps}\, F_X(u,t; y)  
             \, ,
      \label{eq:KXnfa}
\end{equation}
%\label{eq:KXnfab}
%\end{subequations}
and $F_X(u,t; y)$ and  $K_{(X)_{nf}}(u,t)$ can be read
from Table \ref{t:FXintFX}. 

Similarly to \req{el:Int2}, the $D$ dimensional integral
from Eq. \req{eq:phiXnfab} gives
\begin{eqnarray}
   \lefteqn{-i \, (4 \pi)^2 \,
         \left[ \mu^{2 \eps}
             \int \frac{d^D l}{(2 \pi)^D}
        \frac{1}{(l^2 + i \eta )^{2+\eps}} \right]}
       \nonumber \\
             & = &
         \frac{1}{(-\widetilde{Q}^2 + i \eta)^\eps}
         \left(\frac{\mu^2}{\widetilde{Q}^2}\right)^{\eps}
             \nonumber \\ &  & \times \,
         \left[ \Gamma_{UV}^{(1)}(\eps) 
                \frac{1-\eps}{(1-3 \eps)(1-\frac{3}{2}\eps)} 
                \frac{1 - \frac{\eps}{2}}{1+\eps}
            \right. \nonumber \\ &  & \left. \quad 
                    +
               \Gamma_{IR}^{(1)}(\eps) 
                \frac{1-\eps}{(1-3 \eps)(1-\frac{3}{2}\eps)} 
                \frac{1 - \frac{\eps}{2}}{1+\eps}
               \right] 
                     \, ,
\label{el:Int2e}
\end{eqnarray}
where $\tilde{Q}^2>0$ represents the auxiliary scale.

The $n_f$-proportional NNLO contribution
of the diagrams displayed in Fig. \ref{f:DAbub}
takes the form
\begin{widetext}
\begin{eqnarray}
\tilde{\phi}^{(2, n_f)}(u,t)& = &   
         \left[ \Gamma_{UV}^{(0)}(\eps) \Gamma_{UV}^{(1)}(\eps) 
 \frac{(1-\eps)^2(1 - \frac{\eps}{2})}{(1-\frac{2}{3}\eps)(1-2\eps) 
                                       (1-3 \eps)(1-\frac{3}{2}\eps)} 
                  K^{(2,n_f)}(u,t)
            \right. \nonumber \\ &  & \left. 
                    +
               \Gamma_{UV}^{(0)}(\eps) \Gamma_{IR}^{(1)}(\eps) 
 \frac{(1-\eps)^2(1 - \frac{\eps}{2})}{(1-\frac{2}{3}\eps)(1-2\eps)
                                      (1-3 \eps)(1-\frac{3}{2}\eps)} 
                  K^{(2,n_f)}(u,t)
               \right] 
         \left(\frac{\mu^2}{\widetilde{Q}^2}\right)^{2 \eps}
               \, ,
\label{eq:phi2nf}
\end{eqnarray}
\end{widetext}
where the function $K^{(2,n_f)}$ calculated
in the naive-$\gamma_5$ scheme amounts to
\begin{eqnarray}
 K^{(2,n_f)}(u,t) & = & 2\, C_F\,
  \left\{ \frac{u}{t} 
    \left[ (1-\eps) + \frac{1}{t-u} \right] 
       \left(\frac{u}{t}\right)^{\eps} 
         \theta(t-u)  \right. 
        \nonumber \\ & & \left.
                    + 
        \bigg( \begin{array}{c} u \rightarrow 1-u \\
                                t \rightarrow 1-t
               \end{array} \bigg) \right\}_+
          \, .
  \label{eq:K2nf}
\end{eqnarray}

\subsubsection{The HV scheme results}

The preceding results have been calculated in the naive-$\gamma_5$
scheme.
When the HV scheme is used, only the results for $A$ diagram
from Fig. \ref{f:DAnlo} and the corresponding ``bubble'' diagram
from Fig. \ref{f:DAbub} differ from the naive-$\gamma_5$ results,
and are given by
\begin{eqnarray}
   \tilde{\phi}_A^{HV}(u,t)&=&\tilde{\phi}_A(u,t)+ \Delta \tilde{\phi}_A(u,t)
        \nonumber \\ &=&
     \tilde{\phi}_A(u,t) \left( 1 + 4 \eps 
\frac{1+\frac{\eps}{4}}{(1-\frac{\eps}{2})(1-\eps)} \right) 
         \, , \qquad
\label{eq:phiAbm}
\end{eqnarray}
and
\begin{eqnarray}
   \tilde{\phi}_{(A)_{nf}}^{HV}(u,t)&=&\tilde{\phi}_{(A)_{nf}}(u,t)+
\Delta \tilde{\phi}_{(A)_{nf}}(u,t)
        \nonumber \\ &=&
     \tilde{\phi}_{(A)_{nf}}(u,t) \left( 1 + 4 \eps 
\frac{1}{(1-\frac{\eps}{2})(1-\eps)} \right) 
         \, .
       \nonumber \\ & &
\label{eq:phiAnfbm}
\end{eqnarray}

Hence, when using the HV scheme, the functions $K^{(1)}$
and $K^{(2,n_f)}$ 
in Eqs. \req{eq:phi1} and \req{eq:phi2nf}
get replaced by
\begin{eqnarray}
  K^{(1)}_{HV} & = &  
        2 \, C_F \left\{ 4 \eps  \, 
      \frac{1+\frac{\eps}{4}}{1-\frac{\eps}{2}} 
  \left[ \frac{u}{t} \, \theta(t-u) \right. \right.
         \nonumber \\ & & \left. \left. 
                    + 
        \bigg( \begin{array}{c} u \rightarrow 1-u \\
                                t \rightarrow 1-t
               \end{array} \bigg) \right] \right\}
          + K^{(1)}
             \, ,
\label{eq:K1bm}
\end{eqnarray}
and
\begin{eqnarray}
 K^{(2,n_f)}_{HV}(u,t) & = & 
       2 \, C_F \left\{  4 \eps \, 
      \frac{1}{1-\frac{\eps}{2}} 
  \left[ \frac{u}{t} \, \left(\frac{u}{t}\right)^{\eps} \, \theta(t-u) 
                    \right. \right. 
         \nonumber \\ & & \left. \left.
      +  \bigg( \begin{array}{c} u \rightarrow 1-u \\
                                t \rightarrow 1-t
               \end{array} \bigg) \right] \right\}
          + K^{(2,n_f)} \, ,
         \nonumber \\ & & 
\label{eq:K2nfbm}
\end{eqnarray}
respectively. These results, obviously, these results
bear the signature of chiral symmetry violation.

\section{Renormalization and factorization of collinear
singularities}
\label{ss:ren}

\subsection{General renormalization procedure}
\label{sss:M}

Since in this work we present the calculation
(up to $n_f$-proportional NNLO contributions)
of the hard-scattering amplitude $T(u,Q^2)$
\req{eq:T}, 
as well as of the perturbatively calculable part
of the DA $\tilde{\phi}(u,t)$ 
\req{eq:phiutraz}, 
both containing UV and collinear singularities,
here we outline the general procedure
for the renormalization of UV and the factorization 
of collinear singularities.
\begin{widetext}
We introduce the amplitude ${\cal M}$ 
(having the same form of the perturbative expansion as
the amplitudes $T(u,Q^2)$ and $\tilde{\phi}(u,t)$):
\begin{subequations}
\begin{equation}
 {\cal M} = {\cal M}^{(0)} + \frac{\alpha_S}{4 \pi} {\cal M}^{(1)} 
              + \frac{\alpha_S^2}{(4 \pi)^2}
    \left[ \left(-\frac{2}{3} n_f \right) {\cal M}^{(2,n_f)}
              + \cdots \right] + \cdots
         \, ,
\label{eq:M}
\end{equation}
where
\begin{eqnarray}
 {\cal M}^{(1)} &=& 
          \left[
      \Gamma_{UV}^{(0)}(\eps) 
   \left( a_0^{UV} +  \eps \, a_1^{UV}  + \eps^2 \, a_2^{UV}   
                      + O(\eps^3) \right) 
           \right. \nonumber \\ & & \left. 
      + \, \Gamma_{IR}^{(0)}(\eps) 
   \left( a_0^{IR} +  \eps \, a_1^{IR}  + \eps^2 \, a_2^{IR}   
                      + O(\eps^3) \right) 
          \right]
    \left( \frac{\mu^2}{Q^2} \right)^\eps \, ,
             \label{eq:M1} \\[0.3cm] 
 {\cal M}^{(2, n_f)} &=& 
          \left[ 
      \Gamma_{UV}^{(0)}(\eps) \Gamma_{UV}^{(1)}(\eps)
   \left( b_0^{n_f,UV} +  \eps \, b_1^{n_f,UV}  
                + \eps^2 \, b_2^{n_f,UV}   + O(\eps^3) \right) 
        \right.  \nonumber \\ & & \left. 
      + \, \Gamma_{UV}^{(0)}(\eps) \Gamma_{IR}^{(1)}(\eps) 
   \left( b_0^{n_f,IR} +  \eps \, b_1^{n_f,IR}  
                 + \eps^2 \, b_2^{n_f,IR}  + O(\eps^3) \right) 
          \right] 
      \left( \frac{\mu^2}{Q^2} \right)^{2 \eps}
             \, ,
             \label{eq:M2}
\end{eqnarray}
\label{eq:M12}
\end{subequations}
with $a_i^{UV(IR)}$ and $b_i^{UV(IR)}$ representing
general coefficients in the expansion over $\eps$.
\end{widetext}

As a first step, we perform the coupling constant renormalization
in the $\overline{MS}$ renormalization scheme. 
Note that in the functions $\Gamma_{UV}^{(0)}(\eps)$ and
$\Gamma_{IR}^{(0)}(\eps)$, defined by \req{eq:Gamma0},
%and conveniently rearranged as
%\begin{subequations}
%\begin{eqnarray}
%     \Gamma_{UV}^{(0)}(\eps)
%       &=& \Gamma(\eps)\Gamma(1-\eps)
%         \frac{\Gamma(1-\eps)}{\Gamma(1- 2 \eps)}
%          (4 \pi )^{\eps} 
%         \\ 
%     \Gamma_{IR}^{(0)}(\eps)
%       &=& \Gamma(-\eps)\Gamma(1+\eps)
%         \frac{\Gamma(1-\eps)}{\Gamma(1- 2 \eps)}
%%          (4 \pi )^{\eps} 
%          \, ,
%\end{eqnarray}
%\label{eq:Gamma0n}
%\end{subequations}
the singularities are contained in 
\begin{subequations}
\begin{eqnarray}
\Gamma(\eps)\Gamma(1-\eps)&=&
\displaystyle \frac{\pi}{\sin \pi \, \eps}=
\frac{1}{\eps}+\frac{\pi^2}{6} \eps+ O(\eps^3)
 \nonumber \\ &=&
- \Gamma(-\eps)\Gamma(1+\eps) \, , 
\label{eq:Gsing}
\end{eqnarray}
while the remaining artifacts of dimensional 
re\-gu\-la\-ri\-za\-ti\-on 
%\newpage
can be found in 
\begin{equation}
\displaystyle \frac{\Gamma(1-\eps)}{\Gamma(1- 2 \eps)}
(4 \pi )^{\eps} = 1 + \eps (-\gamma+\ln (4 \pi) ) 
+ O(\eps^2) 
\label{eq:Gnsing}
\end{equation}
\label{eq:Gfact}
\end{subequations}
(and similarly for $\Gamma_{UV}^{(1)}$
and $\Gamma_{IR}^{(1)}$ functions \req{eq:Gamma1}).
By expanding $\Gamma$ functions over $\eps$, 
in relation \req{eq:M12},
an unnecessary complication of keeping track of
various $\gamma$, $\pi^2$, and $\ln 4\pi$ terms,
would be introduced.
Instead, we make use of the freedom in
defining the $\overline{MS}$ scheme beyond $O(\eps^0)$,
which is explained in detail in Appendix \ref{app:alphaS}
(along with some other conventions and ``misconventions''),
and define the bare coupling constant $\alpha_S$
in terms of the running coupling $\alpha_S(\muR)$
by
\begin{eqnarray}
   \alpha_S 
   & = &  \left( \frac{\muR}{\mu^2} \right)^\eps 
                   \left( \eps \,   
           \Gamma_{UV}^{(0)}(\eps) \right)^{-1}
                    \alpha_S(\muR)
        \left( 1 - \frac{\alpha_S(\muR)}{4 \pi} \beta_0
                \frac{1}{\eps} \right)
         \nonumber \\ & & 
                \, ,
\label{eq:renalphaS}
\end{eqnarray}
where $\beta_0=11-2/3 \, n_f$.
The inclusion of the factor
$ \eps  \, \Gamma_{UV}^{(0)}(\eps) $
in \req{eq:renalphaS}
turns out to be very suitable for this type of calculation, in which
both UV and IR singularities are regularized by
the dimensional regularization method. 
The elegance and advantage introduced
in the calculation by the choice \req{eq:renalphaS} becomes clear 
when one notes that
\begin{subequations}
\begin{eqnarray}
    \Gamma_{UV}^{(1)}(\eps) &=& \frac{1}{2} \Gamma_{UV}^{(0)}(\eps) + O(\eps^2)
                 \, ,
          \label{eq:trikUV}  \\
    \Gamma_{IR}^{(1)}(\eps) &=& \frac{1}{2} \Gamma_{IR}^{(0)}(\eps) + O(\eps^2)
          \label{eq:trikIR}
              \, .
\end{eqnarray}
\label{eq:trik}
\end{subequations}
So, one can see that the presence of the factor
$ \eps  \, \Gamma_{UV}^{(0)}(\eps)$
in \req{eq:renalphaS}
is natural in the sense that it contains the combinations
of $\Gamma$'s that naturally emerge in this calculation,
and leads to their cancellation without expanding
the whole result over $\eps$.
That is in contrast to ``artificial'' choices
like
$\exp \left(\eps(-\gamma+\ln 4 \pi)\right) $ 
and $ (4 \pi)^\eps/\Gamma(1-\eps)$ found in the literature
(for example, \cite{Neer,Cat98} and \cite{KuS94}, respectively).

\begin{widetext}
By substituting \req{eq:trik} into \req{eq:M12}, and performing
the coupling constant renormalization according to \req{eq:renalphaS},
one obtains
\begin{subequations}
\begin{equation}
 {\cal M} = {\cal M}^{(0)} 
    + \frac{\alpha_S(\muR)}{4 \pi} \hat{{\cal M}}^{(1)} 
              + \frac{\alpha_S^2(\muR)}{(4 \pi)^2}
    \left[ \left(-\frac{2}{3} n_f \right) \hat{{\cal M}}^{(2)}_{n_f}
              + \cdots \right] + \cdots
         \, ,
\label{eq:Masren}
\end{equation}
where
\begin{eqnarray}
\hat{{\cal M}}^{(1)}&=&
          \left[
          \frac{1}{\eps} \,
   \left( a_0^{UV} +  \eps \, a_1^{UV}  
                      + O(\eps^2) \right) 
      + \, \frac{1}{-\eps} \,
   \left( a_0^{IR} +  \eps \, a_1^{IR}  
                      + O(\eps^2) \right) 
          \right]
    \left( \frac{\muR}{Q^2} \right)^\eps
       \label{eq:M1asren} \\[0.3cm] 
\hat{{\cal M}}^{(2,n_f)}&=&
          \left[ 
         \frac{1}{\eps^2} \, 
   \left( 
       \left( \frac{b_0^{n_f,UV}}{2} - 
     \left(\frac{\muR}{Q^2}\right)^{-\eps} a_0^{UV} \right)
    +  \eps \, 
       \left( \frac{b_1^{n_f,UV}}{2} -  
     \left(\frac{\muR}{Q^2}\right)^{-\eps} a_1^{UV} \right)
         \right. \right.  \nonumber \\ & &
         \left. \left.  \quad \quad
    + \eps^2 \, 
      \left( \frac{b_2^{n_f,UV}}{2} -
     \left(\frac{\muR}{Q^2}\right)^{-\eps} a_2^{UV} \right)
    + O'(\eps^3) \right) 
        \right.  \nonumber \\ & &  \left. 
      + \, \frac{1}{-\eps^2} 
   \left( 
       \left( \frac{b_0^{n_f,IR}}{2} - 
     \left(\frac{\muR}{Q^2}\right)^{-\eps} a_0^{IR} \right)
    +  \eps \, 
       \left( \frac{b_1^{n_f,IR}}{2} -  
     \left(\frac{\muR}{Q^2}\right)^{-\eps} a_1^{IR} \right)
         \right. \right.  \nonumber \\ & &
         \left. \left.  \quad  \quad
    + \eps^2 \, 
      \left( \frac{b_2^{n_f,IR}}{2} -
     \left(\frac{\muR}{Q^2}\right)^{-\eps} a_2^{IR} \right)
    + O'(\eps^3)\right) 
          \right] 
         \left( \frac{\muR}{Q^2} \right)^{2 \eps}
       \, .
\label{eq:M2asren}
\end{eqnarray}
\label{eq:M12asren}
\end{subequations}
\end{widetext}
Note that the only artifact of dimensional regularization
we are left with is the dimensional parameter $\eps$.
The result \req{eq:M12asren} is given in a simple and compact form
in which all terms in the expansion over $\eps$ are still
retained.
Also, the distinction between the singularities of UV and IR
origin is still preserved.

If the coefficients of the $1/\eps$ and $1/\eps^2$ poles
of UV origin are different from zero, the additional 
renormalization should be performed, 
as in the case of the multiplicatively 
renormalizable composite operator from 
$\tilde{\phi}(u,t)$ \req{eq:phiOqq}.
The UV singularities are then factorized in the renormalization
constant $Z_{{\cal M},ren}$.
After all UV divergences are properly renormalized,
the remaining $1/(-\eps)$ and $1/(-\eps^2)$ 
collinear poles should be, 
at some factorization scale,
factorized in $Z_{{\cal M},col}$.

\subsection{Renormalization of the hard-scattering amplitude $T$}
\label{ss:Tren}

%%%%%
\renewcommand{\arraystretch}{1.7}
\begin{table*}
\caption{
The NLO coefficients $a_j^{UV(IR)}$
defined in \protect\req{eq:M1}
and determined for 
${\cal M}^{(1)} \equiv T^{(1)}(u)$
\protect\req{eq:T1all}.
The $n_f$-proportional NNLO coefficients 
$b_j^{nf,UV (IR)}$ defined in \protect\req{eq:M2}
and determined  for 
${\cal M}^{(2, n_f)} \equiv T^{(2,n_f)}(u)$
\protect\req{eq:T2nfall}.} 
%\small
 
        \begin{ruledtabular} \begin{tabular}     {ll}
$a_0^{UV}$ & $0$ \\[0.3cm]
$a_1^{UV}$ & $\displaystyle C_F \, \frac{1}{1-u}
 \left( \frac{1}{2} + \frac{1}{u} \ln (1-u) \right) 
     + (u \rightarrow 1-u) $ \\[0.3cm]
$a_2^{UV}$ & $\displaystyle C_F \, \frac{1}{1-u}
 \left( \frac{1}{2} -  \left( \frac{1}{2}-\frac{1}{u} \right)  \ln (1-u) 
       - \frac{1}{2 u} \ln^2 (1-u) \right) 
     + (u \rightarrow 1-u) $ \\[0.3cm]
\hline
$a_0^{IR}$ & $\displaystyle C_F \, \frac{1}{1-u}
 \left( 3 + 2 \ln (1-u) \right)  
     + (u \rightarrow 1-u) $\\[0.3cm]
$a_1^{IR}$ & $\displaystyle C_F \, \frac{1}{1-u}
 \left( \frac{19}{2} 
   + \left( -(8 \delta-7) + \frac{8 \delta -6}{u} \right) \ln (1-u) 
       -  \ln^2 (1-u) \right)  
     + (u \rightarrow 1-u) $\\[0.3cm]
$a_2^{IR}$ & $\displaystyle C_F \, \frac{1}{1-u}
 \left( \frac{37}{2} 
   + \left( -\frac{40 \delta-29}{2} + \frac{20 \delta -15}{u} \right) \ln (1-u) 
 + \left( \frac{8 \delta-7}{2} - \frac{8 \delta -6}{2 u} \right) \ln^2 (1-u) 
     \right. $ \\[0.1cm] 
  & $ \displaystyle \left. 
       + \frac{1}{3} \ln^3 (1-u) \right) 
     + (u \rightarrow 1-u) $ \\[0.3cm]
\hline
$b_0^{n_f,UV}$ & $0$ \\[0.3cm]
$b_1^{n_f,UV}$ & $\displaystyle C_F \, \frac{1}{1-u}
 \left( 1 + \frac{2}{u} \ln (1-u) \right) 
     + (u \rightarrow 1-u) $ \\[0.3cm]
$b_2^{n_f,UV}$ & $\displaystyle C_F \, \frac{1}{1-u}
 \left( \frac{11}{3} -  \left( 2-\frac{25}{3 u} \right)  \ln (1-u) 
       - \frac{2}{ u} \ln^2 (1-u) + \frac{2}{u} \text{Li}_2(u) \right) 
     + (u \rightarrow 1-u) $ \\[0.3cm]
\hline
$b_0^{n_f,IR}$ & $\displaystyle C_F \, \frac{1}{1-u}
 \left( 3 + 2 \ln (1-u) \right)  
     + (u \rightarrow 1-u) $\\[0.3cm]
$b_1^{n_f,IR}$ & $\displaystyle C_F \, \frac{1}{1-u}
 \left( \frac{39}{2} 
   + \left( -\frac{24 \delta-34}{3} 
      + \frac{24 \delta -18}{3 u} \right) \ln (1-u) 
       - 2  \ln^2 (1-u)  + 2 \text{Li}_2(u) \right)  
      $ \\[0.1cm] & $ \displaystyle 
     + (u \rightarrow 1-u) $\\[0.3cm]
$b_2^{n_f,IR}$ & $\displaystyle C_F \, \frac{1}{1-u}
 \left( \frac{311}{4} 
   + \left( -\frac{408 \delta-347}{9} + \frac{408 \delta -270}{9 u} \right) 
                 \ln (1-u) 
     \right. $ \\[0.1cm] 
  & $ \displaystyle \left. 
 + \left( \frac{24 \delta-34}{3} - \frac{24 \delta -18}{3 u} \right) 
     \ln^2 (1-u) 
       + \frac{4}{3} \ln^3 (1-u) 
     \right. $ \\[0.1cm] 
  & $ \displaystyle \left. 
   + \left( -\frac{24 \delta-52}{3} 
      + \frac{24 \delta -18}{3 u} \right) \text{Li}_2(u)
      -2 \text{Li}_3(u) + 4 S_{1,2}(u)
    \right) 
     + (u \rightarrow 1-u) $ \\[0.3cm]
\end{tabular} \end{ruledtabular}             
 
%\normalsize
\label{t:abUVIR}
\end{table*}
\renewcommand{\arraystretch}{1}

We shall now apply the results of the preceding subsection
to the renormalization of our results
for the hard-scattering amplitude $T$ given by
Eqs. \req{eq:T}, \req{eq:T0}, \req{eq:T1all}, and, \req{eq:T2nfall}.
By comparing \req{eq:T} and \req{eq:M},
we can identify
${\cal M} \equiv T(u,Q^2) / (N_T/Q^2) $ 
and 
${\cal M}^{(i)} \equiv T^{(i)}(u)$.

The NLO coefficients $a_j^{UV(IR)}$
appearing in \req{eq:M1}
are determined from \req{eq:T1all} 
by expanding the coefficients of 
$\Gamma^{(0)}_{UV,IR}(\eps)$ over $\eps$,
while
the $n_f$-proportional NNLO coefficients 
$b_j^{nf,UV (IR)}$ in \req{eq:M2}
are obtained  from \req{eq:T2nfall} 
by expanding the coefficients of 
$\Gamma^{(0)}_{UV}(\eps)\Gamma^{(1)}_{UV,IR}(\eps)$ 
over $\eps$.
The special cases of the generalized Nielsen polylogarithms
%\footnote{ 
%\begin{displaymath}
%  S_{n,p}(u)=\frac{(-1)^{n+p-1}}{(n-1)! \, p!}
%    \int_0^1 dx \, \frac{\ln^{n-1}(x) \ln^p(1-x u)}{x}  
%%\label{eq:Spencef}
%\end{displaymath}
%}
%%%%%
\begin{eqnarray}
  \text{Li}_2(u) &=&S_{1,1}(u) =
    - \int_0^u dx \, \frac{\ln(1-x)}{x} 
       \, , \nonumber \\
  \text{Li}_3(u) &=&S_{2,1}(u)=
     \int_0^u dx \, \frac{\text{Li}_2(x)}{x} 
       \, , \nonumber \\
  S_{1,2}(u) &=&
    \frac{1}{2} \, \int_0^u dx \, \frac{\ln^2(1-x)}{x} 
       \, ,
\label{eq:Li23S12}
\end{eqnarray}
appear in these results,
and the useful identity is
\begin{eqnarray}
  S_{1,2}(u) &=& -\text{Li}_3(1-u) - \ln(1-u) \text{Li}_2(u)
              \nonumber \\ & &
         -\frac{1}{2} \ln(u) \ln^2 (1-u) 
         + \frac{\pi^2}{6} \ln (1-u) + \xi (3)
             \, . \nonumber \\
\label{eq:S12toLi}
\end{eqnarray}

%\newpage 
So, after the coupling constant renormalization
has been performed,
the hard-scattering amplitude 
$T/(N_T/Q^2)$%\equiv{\cal M}$ 
takes the form given by \req{eq:M12asren}, where
the LO contribution $T^{(0)}$ is given by
\req{eq:T0} and the coefficients
$a_j^{UV(IR)}$ and $b_j^{nf,UV (IR)}$ 
are listed in Table \ref{t:abUVIR}.
As expected, 
the coefficients of the UV poles
in \req{eq:M12asren} vanish, since all UV singularities get 
removed by the coupling constant renormalization.
According to Eq. \req{eq:TTHZ}, i.e., 
\begin{displaymath}
    T(u,Q^2) = T_H(x, Q^2, \muF) \, \otimes \, Z_{T,col}(x, u; \muF)
           \, ,
%\label{eq:TTHZ}
\end{displaymath}
the remaining singularities of the collinear type
factorize at the factorization scale $\muF$
in 
\begin{eqnarray}
  \lefteqn{Z_{T,col}(x,u; \muF)} \nonumber \\
    &=& \delta(x-u) 
    +\frac{\alpha_S(\muR)}{4 \pi} 
    \left( \frac{\muR}{\muF} \right)^{\eps}
    \frac{1}{-\eps}
     \tilde{a}_0^{IR}(x,u)
     \nonumber \\ & & 
    +\,\frac{\alpha_S^2(\muR)}{(4 \pi)^2} 
    \left( \frac{\muR}{\muF} \right)^{2 \eps}
    \frac{1}{-\eps^2}
    \left\{ \left( - \frac{2}{3} n_f \right)
     \right. \nonumber \\ & & \left. \quad \times
    \left[ \left(
       \frac{\tilde{b}_0^{n_f,IR}(x,u)}{2}-
    \left( \frac{\muR}{\muF} \right)^{-\eps}
     \tilde{a}_0^{IR}(x,u)
            \right)
     \right. \right. \nonumber \\ & & \left. \left. \qquad 
    +\eps
        \left(
       \frac{\tilde{b}_1^{n_f,IR}(x,u)}{2}-
     \tilde{a}_1^{IR}(x,u)
        \right) 
       \right] + \cdots \right\} + \cdots
         \nonumber \\ & &
             \, ,
\label{eq:ZTcol1} 
\end{eqnarray}
where the coefficients 
$\tilde{a}_i^{IR}(x,u)$ and $\tilde{b}_i^{n_f,IR}(x,u)$
satisfy the relations
\begin{eqnarray*}
T^{(0)}(x) \otimes \tilde{a}_i^{IR}(x,u) &=&a_i^{IR}(u) \, , \\
T^{(0)}(x) \otimes \tilde{b}_i^{n_f,IR}(x,u)& =&b_i^{n_f,IR}(u) \, . 
\end{eqnarray*}

With the help of
\pagebreak
\begin{eqnarray}
 \lefteqn{\alpha_S(\muR)}  
        \nonumber \\ &=&   \left( \frac{\muF}{\muR} \right)^\eps 
            \alpha_S(\muF)
     \left[ 1 + \frac{\alpha_S(\muF)}{4 \pi} \beta_0
           \frac{1}{\eps} 
         \left( \left(\frac{\muF}{\muR} \right)^\eps
                -1 \right) \right]
              \, ,
          \nonumber \\ & & 
\label{eq:chalmuRF}
\end{eqnarray}
one can easily demonstrate that $Z_{T,col}(x,u; \muF)$
is indeed independent of the hard-scattering renormalization
scale $\muR$:
\begin{eqnarray}
  \lefteqn{Z_{T,col}(x,u; \muF)} \nonumber \\
    &=& \delta(x-u) 
    +\frac{\alpha_S(\muF)}{4 \pi} 
    \frac{1}{-\eps}
     \tilde{a}_0^{IR}(x,u)
     \nonumber \\ & & 
    +\,\frac{\alpha_S^2(\muF)}{(4 \pi)^2} 
    \frac{1}{-\eps^2}
    \left\{ \left( - \frac{2}{3} n_f \right)
     \right. \nonumber \\ & & \left. \quad \times
    \left[ \left(
       \frac{\tilde{b}_0^{n_f,IR}(x,u)}{2}-
     \tilde{a}_0^{IR}(x,u)
            \right)
     \right. \right. \nonumber \\ & & \left. \left. \qquad 
    +\eps
        \left(
       \frac{\tilde{b}_1^{n_f,IR}(x,u)}{2}-
     \tilde{a}_1^{IR}(x,u)
        \right) 
       \right] + \cdots \right\} + \cdots
             \, .
         \nonumber \\ & &
\label{eq:ZTcol} 
\end{eqnarray}

\begin{widetext}
After factorizing the collinear singularities 
from \req{eq:M12asren} by \req{eq:ZTcol1}, and taking into
account that $b_0^{n_f,IR} = a_0^{IR}$,
we obtain 
\begin{eqnarray}
\frac{T_H(x, Q^2, \muF)}{N_T/Q^2} &=& 
  T^{(0)}(x) 
  + \frac{\alpha_S(\muR)}{4 \pi} 
  \left[ \left( a_1^{UV}(x) - a_1^{IR}(x) \right)
       - a_0^{IR}(x) \ln \frac{\muF}{Q^2}
         + O(\eps) \right]
     \nonumber \\ & & 
  + \frac{\alpha_S^2(\muR)}{(4 \pi)^2} 
  \left\{ 
   \left( - \frac{2}{3} n_f \right)
   \left[
 \left( \frac{b_2^{n_f,UV}(x)}{2} -
        \frac{b_2^{n_f,IR}(x)}{2} \right) 
  - \left( a_2^{UV}(x) - a_2^{IR}(x) \right)  
        \right. \right. \nonumber \\ & &
       \left. \left. 
  + \left( a_1^{UV}(x) - a_1^{IR}(x)
       - a_0^{IR}(x) \ln \frac{\muF}{Q^2}
    \right) \ln \frac{\muR}{Q^2}
        \right. \right. \nonumber \\ & &
        \left. \left. 
  - \left( b_1^{n_f, IR}(x) - 2 a_1^{IR}(x) \right)
      \ln \frac{\muF}{Q^2}
  + \frac{1}{2} a_0^{IR}(x) \ln^2 \frac{\muF}{Q^2}
   + O(\eps) \right] + \cdots \right\}
         \nonumber \\ & & 
    +\cdots
        \, ,
\label{eq:THsh}
\end{eqnarray}
where the $O(\eps)$ terms can now be safely neglected
(notice that we have kept all $\eps^n$ terms till
the end of calculation).

Finally, having evaluated all the necessary terms,
we summarize our result
for the hard-scattering amplitude $T_H(x,Q^2,\muF)$
in the form
\begin{eqnarray}
   T_H(x, Q^2, \muF) &=& 
        T_H^{(0)}(x,Q^2) 
        + \frac{\alpha_S(\muR)}{4 \pi}\,  \, 
          T_H^{(1)}(x,Q^2,\muF)
              \nonumber \\ & &
        + \frac{\alpha_S^2(\muR)}{(4 \pi)^2} 
          \left[  \left( -\frac{2}{3} n_f \right) \, 
           T_{H}^{(2,n_f)}(x,Q^2,\muR,\muF)
              + \cdots \right] + \cdots 
         \, ,
\label{eq:THres}
\end{eqnarray}
where
\begin{subequations}
\begin{eqnarray}
   T_H^{(0)}(x,Q^2)&=&\frac{N_T}{Q^2} \,
       A^{(0)}(x) + (x \rightarrow 1-x) 
        \, ,  \label{eq:TH0}  \\[0.2cm] 
   T_H^{(1)}(x,Q^2,\muF)&=&\frac{N_T}{Q^2} \,
      \left( 
          A^{(1)}(x) \, 
        - A^{(1)}_{col}(x) \, \ln \frac{\muF}{Q^2} \right)
               +  (x \rightarrow 1-x)
         \, \label{eq:TH1} \\[0.2cm] 
   T_H^{(2,n_f)}(x,Q^2,\muR,\muF)&=&\frac{N_T}{Q^2} \,
      \left[ 
         A^{(2,n_f)}(x) \,
    +  \left( A^{(1)}(x) 
    -  A^{(1)}_{col}(x) \,
           \ln \frac{\muF}{Q^2} \right) \ln \frac{\muR}{Q^2}
             \right. \nonumber \\ & &  \left.
    -  A^{(2,n_f)}_{col}(x) \,\ln \frac{\muF}{Q^2}
    +  \frac{1}{2} A^{(1)}_{col}(x) \, \ln^2 \frac{\muF}{Q^2} 
              \right] 
               +  (x \rightarrow 1-x)
        \, .  \nonumber \\
\label{eq:TH2nf}
\end{eqnarray}
\label{eq:TH012nf}
\end{subequations}
We have introduced the functions $A^{(i)}$ and $A^{(i)}_{col}$,
which are given by
\begin{subequations}
\begin{equation}
    A^{(0)}(x) = \frac{1}{1-x}
         \label{eq:A0}  \, ,  
\end{equation} 
and
\begin{eqnarray}
    A^{(1)}(x) &=& 
           C_F \, \frac{1}{1-x} 
      \left[ -9 - (8 \delta -7 ) \frac{1-x}{x} \ln(1-x) + 
             \ln^2 (1-x) \right]
         \label{eq:A1} \, ,   \\[0.2cm] 
    A^{(2,n_f)}(x) &=&
           C_F \, \frac{1}{1-x} 
       \left[
    -\frac{457}{24} 
    + \left( \frac{(48 \delta - 95)}{18} +
             \frac{(-16 \delta + 19)}{6 x} \right) \ln (1-x) 
              \right. \nonumber \\ & &  \left.
    + \left( \frac{13}{6} -
             \frac{1}{2 x} \right) \ln^2 (1-x) 
    - \frac{1}{3} \ln^3 (1-x) 
             \right.  \nonumber \\ & &  \left.
    + \left( \frac{(12 \delta - 26)}{3} -
             \frac{(4 \delta - 4)}{x} \right) \text{Li}_2(x)
    + \text{Li}_3(x) - 2 S_{1,2}(x) \right]
      \label{eq:A2nf} \, ,
\end{eqnarray}
\label{eq:A}
\end{subequations}
while
\begin{subequations}
\begin{eqnarray}
    A^{(1)}_{col}(x) &=& 
           C_F \, \frac{1}{1-x} 
        \left(3 + 2 \ln(1-x) \right)
        \label{eq:Acol1} \, ,     \\[0.2cm] 
    A^{(2,n_f)}_{col}(x) &=&
           C_F \, \frac{1}{1-x} 
      \left[ \frac{1}{2} 
       + \left( \frac{(24 \delta -8)}{3} +
                \frac{(-8 \delta +6)}{x} \right) \ln (1-x)
       + 2 \text{Li}_2(x) \right] 
      \label{eq:Acol2nf} \, .
\end{eqnarray}
\label{eq:Acol}
\end{subequations}
\end{widetext}

The collinearly singular terms removed  from \req{eq:M12asren}
by \req{eq:ZTcol1}
correspond to
\begin{eqnarray}
 \lefteqn{ T_H^{(0)}(x,Q^2) \, \otimes \, Z_{T,col}(x,u; \muF)} 
      \nonumber \\ \; &=& 
    \frac{N_T}{Q^2} \Biggl\{
     A^{(0)}(u)
    +\frac{\alpha_S(\muF)}{4 \pi}   
    \frac{1}{-\eps} 
       A^{(1)}_{col}(u) 
    +\,\frac{\alpha_S^2(\muF)}{(4 \pi)^2} 
    \frac{1}{-\eps^2}
            \nonumber \\ & & \: \times 
    \left[ \left( - \frac{2}{3} n_f \right)
    \left( 
    - \frac{1}{2} A^{(1)}_{col}(u) 
         +\eps \,  
      \frac{1}{2} A^{(2,n_f)}_{col}(u) 
        \right) 
       + \cdots \right] 
            \nonumber \\ & & \,
       + \cdots \Biggr\}
               +  (u \rightarrow 1-u)
         \, .
\label{eq:TH0ZTcol} 
\end{eqnarray}
The functions $A^{(i)}_{col}$ \req{eq:Acol}, 
which appear in 
\req{eq:TH0ZTcol} and as coefficients of 
$\ln^n (\muF/Q^2)$ in \req{eq:TH012nf},
are obviously connected to collinear singularities
of the hard-scattering amplitude $T$. 

\subsection{Renormalization of the perturbatively calculable DA part}
\label{ss:DAren}

\subsubsection{General analysis}

Next, we proceed to renormalize the $\tilde{\phi}(u,t)$
following the procedure outlined in 
subsection \ref{sss:M}.
By comparing \req{eq:M12} with 
\req{eq:phiutraz}
we identify ${\cal M}\equiv\tilde{\phi}(u,t)$ and
${\cal M}^{(i)}\equiv\tilde{\phi}^{(i)}(u,t)$,
while the $Q^2$ scale corresponds
to the scale $\tilde{Q}^2$. 
The coefficients 
\begin{eqnarray}
a_i^{UV}&=&a_i^{IR}\equiv a_i(u,t) 
        \nonumber \\
b_i^{n_f,UV}&=&b_i^{n_f,IR}\equiv b_i^{n_f}(u,t)
\label{eq:abnf}
\end{eqnarray}
are determined from Eqs.
\req{eq:phi1} and \req{eq:K1}
by expanding the coefficients of 
$\Gamma^{(0)}_{UV,IR}(\eps)$ over $\eps$,
and from Eqs.
\req{eq:phi2nf} and \req{eq:K2nf} 
by expanding the coefficients of 
$\Gamma^{(0)}_{UV}(\eps)\Gamma^{(1)}_{UV,IR}(\eps)$ 
over $\eps$, respectively.
Although in this work only the
$n_f$-proportional part of the NNLO contribution
$\tilde{\phi}^{(2)}_{n_f}(u,t)$
has been determined, 
our symbolical analysis can be extended to
the whole NNLO contribution $\tilde{\phi}^{(2)}(u,t)$.
In this case 
the general coefficients
of the order $\alpha_S^2$ term
$\tilde{\phi}^{(2)}(u,t)\equiv{\cal M}^{(2)}$ 
\begin{equation}
b_i^{UV}=b_i^{IR}\equiv b_i(u,t) 
\label{eq:b}
\end{equation}
appear.
After the coupling constant renormalization,
the distribution amplitude 
$\tilde{\phi}(u,t)\equiv {\cal M}$
is given by the expression \req{eq:M12asren}
with the renormalization scale denoted by
$\tilde{\mu}_R^2$.
The complete order $\alpha_S^2(\tilde{\mu}_R^2)$
coefficient $\hat{{\cal M}}^{(2)}$ reads
\begin{widetext}
\begin{eqnarray}
\hat{{\cal M}}^{(2)}&=&
          \left[ 
         \frac{1}{\eps^2} \, 
   \left( 
       \left( \frac{b_0}{2} - \beta_0
     \left(\frac{\tilde{\mu}_R^2}{\tilde{Q}^2}\right)^{-\eps} a_0 \right)
    +  \eps \, 
       \left( \frac{b_1}{2} - \beta_0 
     \left(\frac{\tilde{\mu}_R^2}{\tilde{Q}^2}\right)^{-\eps} a_1 \right)
         \right. \right.  \nonumber \\ & &
         \left. \left.  \quad \quad
    + \eps^2 \, 
      \left( \frac{b_2}{2} - \beta_0
     \left(\frac{\tilde{\mu}_R^2}{\tilde{Q}^2}\right)^{-\eps} a_2 \right)
    + O'(\eps^3) \right) 
        \right.  \nonumber \\ & &  \left. 
      + \, \frac{1}{-\eps^2} 
   \left( 
       \left( \frac{b_0}{2} - \beta_0
     \left(\frac{\tilde{\mu}_R^2}{\tilde{Q}^2}\right)^{-\eps} a_0 \right)
    +  \eps \, 
       \left( \frac{b_1}{2} - \beta_0 
     \left(\frac{\tilde{\mu}_R^2}{\tilde{Q}^2}\right)^{-\eps} a_1 \right)
         \right. \right.  \nonumber \\ & &
         \left. \left.  \quad  \quad
    + \eps^2 \, 
      \left( \frac{b_2}{2} - \beta_0
     \left(\frac{\tilde{\mu}_R^2}{\tilde{Q}^2}\right)^{-\eps} a_2 \right)
    + O'(\eps^3)\right) 
          \right] 
         \left( \frac{\tilde{\mu}_R^2}{\tilde{Q}^2} \right)^{2 \eps}
       \, ,
\label{eq:comM2asren}
\end{eqnarray}
\end{widetext}
where Eqs. (\ref{eq:abnf}-\ref{eq:b}) have
already been taken into account.

As denoted in \req{eq:ZfVZ}
\begin{eqnarray*}
   \lefteqn{\tilde{\phi}(u,t)} \nonumber \\
& = & Z_{\phi,ren}(u,v; \tilde{\mu}_R^2) \otimes 
              \phi_V(v,s; \tilde{\mu}_R^2, \mu_0^2) \otimes
               Z_{\phi,col}(s,t; \mu_0^2)
                 \, .
           \nonumber \\ 
%\label{eq:ZfVZ}
\end{eqnarray*}
the remaining UV singularities are multiplicatively
renormalizable and factorize in
the renormalization constant $Z_{\phi,ren}(u,v; \tilde{\mu}_R^2)$
given by 
\begin{eqnarray}
Z_{\phi,ren}& =&
   \openone  
 + \frac{\alpha_S(\tilde{\mu}_R^2)}{4 \pi}
 \, \frac{1}{\eps} \, a_0 
 + \frac{\alpha_S^2(\tilde{\mu}_R^2)}{(4 \pi)^2}
 \, \frac{1}{\eps^2} 
    \nonumber \\ & & \times
 \left[ \left( \frac{b_0}{2} - \beta_0 \, a_0 \right)
   + \eps \left( \frac{b_1}{2} - \beta_0 \, a_1  
              - a_0\, a_1 \right) \right] 
     \nonumber \\ & & 
     + \cdots \, ,
\label{eq:Zfren}
\end{eqnarray}
with
\begin{equation}
b_0-\beta_0 \, a_0 - a_0^2 =0
\label{eq:mrcon} 
\end{equation}
(i.e.,
$b_0(x,y) -\beta_0 \, a_0(x,y) - a_0(x,u) \otimes a_0(u,y) =0$)
emerging
as the 
condition of multiplicative 
renormalizability.
As for the collinear singularities, they factorize
at the factorization scale $\mu_0^2$
in $Z_{\phi,col}(s,t; \mu_0^2)$ given by
\begin{eqnarray}
\lefteqn{Z_{\phi,col}} \nonumber \\ & =&
   \openone
 + \frac{\alpha_S(\mu_0^2)}{4 \pi}
 \, \frac{1}{-\eps} \, a_0 
 + \frac{\alpha_S^2(\mu_0^2)}{(4 \pi)^2}
 \, \frac{1}{-\eps^2} 
    \nonumber \\ & & \times
 \left[ \left( \frac{b_0}{2} - \beta_0 \, a_0 - a_0^2 \right)
   + \eps \left( \frac{b_1}{2} - \beta_0 \, a_1  
              - a_0\, a_1 \right) \right] 
     \nonumber \\ & & 
 + \cdots
          \, .
\label{eq:Zfcol}
\end{eqnarray}
Finally, based on Eqs.\req{eq:ZfVZ} and
(\ref{eq:Zfren}-\ref{eq:Zfcol}),
the function $\phi_V(v,s; \tilde{\mu}_R^2, \mu_0^2)$
is obtained.
It is free of singularities,
and after the $\eps \rightarrow 0$ limit is taken,
it takes the form
\begin{eqnarray}
\phi_V &=& \openone
 + \frac{\alpha_S(\tilde{\mu}_R^2)}{4 \pi} 
   a_0 \ln \frac{\tilde{\mu}_R^2}{\mu_0^2}
 + \frac{\alpha_S^2(\tilde{\mu}_R^2)}{(4 \pi)^2} 
    \nonumber \\ & & \times
   \left[ \frac{b_0}{2} \ln^2 \frac{\tilde{\mu}_R^2}{\mu_0^2}
      + \Big( b_1-2 \beta_0 \, a_1 - 2 a_0 \, a_1 \Big)
    \ln \frac{\tilde{\mu}_R^2}{\mu_0^2} \right] 
     \nonumber \\ & & 
    + \cdots
          \, .
\label{eq:phiV}
\end{eqnarray}
Note that the auxiliary scale $\tilde{Q}^2$ has disappeared
after renormalization and factorization of collinear 
singularities. 
We can make a distinction between the scale $\tilde{\mu}_{R,1}^2$
introduced by the coupling constant renormalization
and the scale $\tilde{\mu}_{R,2}^2$ at which the remaining
UV singularities are factorized in the renormalization constant 
$Z_{\phi,ren}$. It can be easily shown that the scale
$\tilde{\mu}_{R,1}^2$ vanishes from the end results, i.e.,
that $Z_{\phi,ren}$ and $\phi_V$ depend only on the
scale $\tilde{\mu}_{R,2}^2$. 
Hence, $\tilde{\mu}_R \equiv \tilde{\mu}_{R,2}^2$
and $\mu_0^2$ are the only relevant scales.
Also, note that
$Z_{\phi,col}(\mu^2)=Z_{\phi,ren}^{-1}(\mu^2)$,
which is expected, 
since, in dimensional regularization, $\tilde{\phi}=\openone $ 
when the distinction between UV and collinear singularities
is abandoned.

\subsubsection{Remarks on the evolutional part of the DA}

As explained in Sec. \ref{s:formalism},
the function $\tilde{\phi}(u,t)$ represents
a perturbatively calculable part of the
unrenormalized pion distribution amplitude
$\Phi(u)$. 
By taking into account Eqs. \req{eq:Phitphirest}
and \req{eq:ZfVZ}
the distribution
$\Phi(u)$ can be expressed by \req{eq:PhiZfVPhi}
\begin{displaymath}
   \Phi(u)=
         Z_{\phi,ren}(u,v; \tilde{\mu}_R^2) \otimes
         \phi_V(v,s; \tilde{\mu}_R^2, \mu_0^2) \otimes
         \Phi(s, \mu_0^2)
           \, ,
%\label{eq:PhiZfVPhi}
\end{displaymath}
where $\Phi(s, \mu_0^2)$ represents the pion distribution
amplitude determined at the scale $\mu_0^2$.
Its evolution to the scale $\tilde{\mu}_R^2$
is determined by
$\phi_V(v,s; \tilde{\mu}_R^2, \mu_0^2)$
and given by  \req{eq:PhifVPhi}:
%The function
%$\phi_V(v,s; \tilde{\mu}_R^2, \mu_0^2)$
%evolutes the pion DA 
%determined at the scale $\mu_0^2$
%to the scale $\tilde{\mu}_R^2$ as denoted by \req{eq:PhifVPhi}:
\begin{displaymath}
   \Phi(v,\tilde{\mu}_R^2)=
         \phi_V(v,s; \tilde{\mu}_R^2, \mu_0^2) \otimes
         \Phi(s, \mu_0^2)
           \, .
%\label{eq:PhifVPhi}
\end{displaymath}

The evolution potential $V$ defined in \req{eq:eveq}
can be obtained
from \req{eq:VZ} 
\begin{displaymath}
   V = -Z_{\phi, ren}^{-1} \left( \tilde{\mu}_R^2 
       \frac{\partial}{\partial \tilde{\mu}_R^2} Z_{\phi, ren}
              \right)
            \, ,
\end{displaymath}
using \req{eq:Zfren}, 
and it reads
\begin{subequations}
\begin{equation}
   V    = \frac{\alpha_S(\tilde{\mu}_R^2)}{4 \pi} \, V_1
         + \frac{\alpha_S^2(\tilde{\mu}_R^2)}{(4 \pi)^2}
            \, V_2 + \cdots 
         \, ,
\label{eq:V}
\end{equation}
where
\begin{eqnarray}
     V_1 &=& a_0 \nonumber \\
     V_2 &=&
      b_1-2 \beta_0 \, a_1 - 2 a_0 \, a_1 
       \, .
\label{eq:V1V2}
\end{eqnarray}
\label{eq:VV1V2}
\end{subequations}

By noting that
\begin{equation}
\frac{\alpha_S(\tilde{\mu}_R^2)}{4 \pi} 
  \ln \frac{\tilde{\mu}_R^2}{\mu_0^2} 
  =\frac{1}{\beta_0} \left( 1 - 
\frac{\alpha_S(\tilde{\mu}_R^2)}{\alpha_S(\mu_0^2)} \right)
 = O(\alpha_S^0)
\label{eq:alphaln}
\end{equation}
and by employing the multiplicative renormalizability
condition \req{eq:mrcon} as well as the results \req{eq:V1V2},
the $\alpha_S$ expansion of $\phi_V$ given in \req{eq:phiV} 
can be reorganized and written in the form
\begin{subequations}
\begin{equation}
\phi_V = 
      \phi_{V}^{LO} 
 + \frac{\alpha_S(\tilde{\mu}_R^2)}{4 \pi} 
      \phi_{V}^{NLO} + \cdots
             \, ,
\label{eq:phiVevLONLO}
\end{equation}
where 
\begin{eqnarray}
 \phi_{V}^{LO} &=&
     \openone
 + \frac{\alpha_S(\tilde{\mu}_R^2)}{4 \pi} 
    \ln \frac{\tilde{\mu}_R^2}{\mu_0^2} \, V_1
      \nonumber \\ & &
 + \frac{\alpha_S^2(\tilde{\mu}_R^2)}{(4 \pi)^2} 
    \ln^2 \frac{\tilde{\mu}_R^2}{\mu_0^2}\, 
   \frac{1}{2} \, (V_1^2 + \beta_0 \, V_1) 
 + \cdots \qquad \quad
    \label{eq:phiVevLO}
\end{eqnarray}
and
\begin{eqnarray}
\phi_{V}^{NLO} &=&
  \frac{\alpha_S(\tilde{\mu}_R^2)}{4 \pi} 
    \ln \frac{\tilde{\mu}_R^2}{\mu_0^2} 
           \, V_2 \,
      + \cdots 
           \, 
    \label{eq:phiVevNLO} 
\end{eqnarray}
\label{eq:phiVev}
\end{subequations}
denote
the LO and NLO part, respectively.
By substituting \req{eq:phiVevLONLO}
into \req{eq:PhifVPhi} 
one obtains 
\begin{eqnarray}
   \lefteqn{\Phi(v, \tilde{\mu}_R^2)} \nonumber \\ & =&
       \phi_{V}^{LO}(v,s; \tilde{\mu}_R^2, \mu_0^2)
       \otimes \Phi(s, \mu_0^2) 
           \nonumber \\ & &
 + \frac{\alpha_S(\tilde{\mu}_R^2)}{4 \pi} \;
       \phi_{V}^{NLO}(v,s; \tilde{\mu}_R^2, \mu_0^2)
       \otimes \Phi(s, \mu_0^2) + \cdots
        \nonumber \\
          & = & 
   \Phi^{LO}(v, \tilde{\mu}_R^2)
 + \frac{\alpha_S(\tilde{\mu}_R^2)}{4 \pi} \;
   \, \Phi^{NLO}(v, \tilde{\mu}_R^2) + \cdots
          \, . \qquad
\label{eq:phiVPhi}
\end{eqnarray}
As it is seen from \req{eq:phiVevLO} and \req{eq:phiVevNLO},
the results of the two-loop calculation
correspond to the first terms of the LO 
and NLO contributions to the $\phi_V$ function. 

The complete LO and NLO behavior of $\phi_V(v,s; \tilde{\mu}_R^2)$
and, consequently, of $\Phi(v,\tilde{\mu}_R^2)$
can be determined by solving the
evolution equation \req{eq:eveq}, or equivalently 
\req{eq:VfV}
\begin{eqnarray*}
       \lefteqn{\tilde{\mu}_R^2 
       \frac{\partial}{\partial \tilde{\mu}_R^2} 
                        \phi_V (v,s,\tilde{\mu}_R^2,\muO)}
           \nonumber \\  
        \qquad & \qquad =& V(v,s',\tilde{\mu}_R^2) \, \otimes \, 
                           \phi_V(s',s,\tilde{\mu}_R^2,\muO)
             \, . \qquad 
%\label{eq:VfV}
\end{eqnarray*}
The LO result is of the form
\begin{eqnarray}
  \phi_V^{LO}(v,s; \tilde{\mu}_R^2)& = &
      \sum_{n=0}^{\infty} {}' \frac{v (1-v)}{N_n}
        C_n^{3/2}(2 v-1) 
     \nonumber \\ & & \times \, C_n^{3/2}(2 s -1)
        \left( 
    \frac{\alpha_S(\tilde{\mu}_R^2)}{\alpha_S(\mu_0^2)}
        \right)^{-\gamma_n^{(0)}/\beta_0}
          \, ,
       \nonumber \\ & &
\label{eq:phiVLOcom}
\end{eqnarray}
where $N_n=(n+1)(n+2)/(4 (2 n+3) )$,
and $C_n^{3/2}(2 x -1)$ are the Gegenbauer polynomials
representing the eigenfunctions of the LO kernel $V_1$
with the corresponding eigenvalues
\begin{equation}
  \gamma_n^{(0)}  =  C_F \left[
            3 + \frac{2}{(n+1) (n+2)}
           - 4 \sum_{i=1}^{n+1} \frac{1}{i} \right] 
             \, .
\label{eq:gamman0}
\end{equation}
One can show the agreement between the 
complete LO prediction given above and the expansion \req{eq:phiVevLO}.
The complete formal solution of the NLO evolution equation
was obtained in \cite{Mu94etc} making use of conformal constraints,
and the form of $\phi_V^{NLO}$ can
be extracted from the results listed in \cite{MNP99}.

\subsubsection{Analytical results up to $n_f$-proportional
NNLO terms (obtained using the naive-$\gamma_5$ scheme)}

After this lengthy general analysis
we now turn to displaying the results.
Using the multiplicative renormalizability condition
\req{eq:mrcon} and the notation \req{eq:V1V2},
the renormalization constant
$Z_{\phi,ren}(u,v; \tilde{\mu}_R^2)$
from Eq. \req{eq:Zfren} is expressed by
\begin{eqnarray}
\lefteqn{Z_{\phi,ren}(u,v; \tilde{\mu}_R^2)} \nonumber \\
  & =&
   \delta(u-v)
 + \frac{\alpha_S(\tilde{\mu}_R^2)}{4 \pi}
 \, \frac{1}{\eps} \, V_1(u,v) 
 + \frac{\alpha_S^2(\tilde{\mu}_R^2)}{(4 \pi)^2}
 \, \frac{1}{\eps^2} 
    \nonumber \\ & & \times
 \left[ \left(-\frac{2}{3} n_f \right)
    \left( \frac{-V_1(u,v)}{2}
   + \eps \frac{V_2^{n_f}(u,v)}{2}
              \right) + \cdots \right] 
          \nonumber \\ & &
  + \cdots
               \, .
\label{eq:ZfrenNF}
\end{eqnarray}
Here we list only the relevant 
combinations of $a_i$, $b_i$ coefficients:
\begin{subequations}
\begin{eqnarray}
   \lefteqn{V_1(u,t) = a_0(u,t)} 
     \nonumber \\ & = & 
   2\, C_F\,
  \left\{ \frac{u}{t} 
    \left[ 1 + \frac{1}{t-u} \right] 
         \theta(t-u) 
%     \right. \nonumber \\ &  & \left. \;
                    + 
        \bigg( \begin{array}{c} u \rightarrow 1-u \\
                                t \rightarrow 1-t
               \end{array} \bigg) \right\}_+
          \, ,
      \nonumber \\  & &
\label{eq:V1}  
\end{eqnarray}
\begin{eqnarray}
   \lefteqn{V_2^{n_f}(u,t) = b_1^{n_f}(u,t)-2 a_1(u,t)}
         \nonumber \\ &= &
   2\, C_F\,
  \left\{ \frac{u}{t} 
    \left[ 1 + \left(1+\frac{1}{t-u}\right)
               \left(\frac{5}{3}+\ln \frac{u}{t}\right) \right] 
         \theta(t-u) 
         \right. \nonumber \\ & & \left. \;
                    + 
        \bigg( \begin{array}{c} u \rightarrow 1-u \\
                                t \rightarrow 1-t
               \end{array} \bigg) \right\}_+
           \, .
\label{eq:V2nf}
\end{eqnarray}
\label{eq:V1V2nf}
\end{subequations}
Our results confirm the well-known form of the
one-loop kernel $V_1$ \cite{LeBr80} and 
the two-loop $n_f$-proportional kernel $V_2^{n_f}$ 
\cite{DiR84etc,Ka85etc,MiR85}.
For later use
we also specify the convolution
of the functions given in \req{eq:V1V2nf}
with the frequently encountered $1/(1-x)$
term:
\begin{subequations}
\begin{eqnarray}
  \frac{1}{1-x} 
  \, \otimes \, V_1(x,u) & = &
   C_F \,\frac{1}{1-u} \left( 3 + 2 \ln (1-u) \right) 
         \nonumber \\ & &
\label{eq:1xV1} 
\end{eqnarray}
and
\begin{eqnarray}
  \lefteqn{\frac{1}{1-x} 
  \, \otimes \, V_2^{n_f}(x,u)}
      \nonumber \\ & = &
   C_F \,\frac{1}{1-u} \left[ 
      \frac{1}{2} 
  + \left( \frac{16}{3} - \frac{2}{u} \right) \ln (1-u)
  + 2 \, \text{Li}_2(u) \right]
            \, .
         \nonumber \\ & &
\label{eq:1xV2nf}
\end{eqnarray}
\label{eq:1xV1V2nf}
\end{subequations}

\subsubsection{The results obtained using the HV scheme}

In the following we present the results obtained
using the HV scheme.

The function $\tilde{\phi}$
calculated in the naive-$\gamma_5$ scheme
and the function $\tilde{\phi}^{HV}$
obtained in the HV scheme are related by
\begin{equation}
    \tilde{\phi} = {\cal Z}_{HV,UV}^{-1} \; \tilde{\phi}^{HV}
                 \;{\cal Z}_{HV,col}^{-1} 
        \, .
\label{eq:ffBM}
\end{equation}
The factors 
${\cal Z}_{HV,UV}^{-1}$ and 
${\cal Z}_{HV,col}^{-1}$ remove
the ``spurious'' anomalous terms 
introduced by the presence of dimensionally regulated 
UV and collinear singularities, respectively.

The renormalization 
of the $\tilde{\phi}^{HV}$ function
proceeds analogously to 
the renormalization of $\tilde{\phi}$
described in preceding subsection.
While
\begin{equation}
a_0^{HV}=a_0 \, , \quad b_0^{HV}=b_0,
\end{equation}
the coefficients $a_i$, $b_i$ for $i\geq1$ 
get replaced by
\begin{eqnarray}
a_i^{HV}&=&a_i + \Delta a_i 
  \nonumber \\ 
b_i^{HV}&=&b_i + \Delta b_i
    \, ,
\label{eq:abBM}
\end{eqnarray}
and, according to \req{eq:ZfVZ}, the UV and collinear
singularities are factorized 
%in $Z_{\phi,ren}^{HV}$ and $Z_{\phi,col}^{HV}$
\begin{equation}
   \tilde{\phi}^{HV} =  
  Z_{\phi,ren}^{HV} \; \phi_V^{HV} \; Z_{\phi,col}^{HV}
         \, .
\label{eq:ZfZBM}
\end{equation}
By comparing $Z_{\phi,ren}^{HV}$, $Z_{\phi,col}^{HV}$,
and $\phi_V^{HV}$ with the results 
(\ref{eq:Zfren}-\ref{eq:phiV}) obtained
using the naive-$\gamma_5$ scheme,
Eq. \req{eq:ZfZBM} takes the form
\begin{eqnarray}
   \tilde{\phi}^{HV} & = &  
\left( Z_{\phi,ren} \;  {\cal Z}_{HV,UV}^{div} \right)
\left( {\cal Z}_{HV,UV}^{fin} \; \phi_V \;
{\cal Z}_{HV,col}^{fin} \right ) \; 
     \nonumber \\ & & \times \,
\left( {\cal Z}_{HV,col}^{div} Z_{\phi,col} \right)
  \, ,
\label{eq:ZZBMetc}
\end{eqnarray}
where
\begin{eqnarray}
\lefteqn{{\cal Z}_{HV}^{div} \equiv
{\cal Z}_{HV,UV}^{div} } \nonumber \\ &=&
   \openone 
 + \frac{\alpha_S^2(\tilde{\mu}_R^2)}{(4 \pi)^2}
 \, \frac{1}{\eps} 
  \left( \frac{\Delta b_1}{2} -\beta_0 \Delta a_1 -
        a_0 \, \Delta a_1 \right) 
     \nonumber \\ & & 
  + \cdots
        \,  
\label{eq:ZBMdiv}
\end{eqnarray}
and
\begin{eqnarray}
\lefteqn{{\cal Z}_{HV}^{fin} \equiv
{\cal Z}_{HV,UV}^{fin} } \nonumber \\ & =&
     \openone
 + \frac{\alpha_S(\tilde{\mu}_R^2)}{4 \pi}
 \, \Big( \Delta a_1 + O(\eps) \Big) 
 + \frac{\alpha_S^2(\tilde{\mu}_R^2)}{(4 \pi)^2}
    \nonumber \\ & & \times
    \left( \frac{\Delta b_1}{2} - \beta_0 \Delta a_2 
           - a_0 \, \Delta a_2 - a_1 \Delta a_1  
           + O(\eps) \right)
     \nonumber \\ & & 
               + \cdots
         \, ,
\label{eq:ZBMfin}
\end{eqnarray}
while 
${\cal Z}_{HV,col}^{div,fin}(\mu^2)=({\cal Z}_{HV,UV}^{div,fin})^{-1}(\mu^2)$.
The condition of multiplicative renormalizability
of the ``spurious'' anomalous terms introduced 
by the HV scheme reads
\begin{equation}
 \Delta b_1 - \beta_0 \Delta a_1 - 2 a_0 \Delta a_1=0
        \, .
\label{eq:mrconBM}
\end{equation}
   
Finally, we list the results obtained
by substituting
$K_{HV}^{(1)}$ and
$K_{n_f,HV}^{(2)}$ (\ref{eq:K1bm}-\ref{eq:K2nfbm}) in place
of $K^{(1)}$ and $K^{(2)}$ in Eqs. 
(\ref{eq:phiutraz}-\ref{eq:phi2nf}).
The combinations of the $\Delta a_i$
and $\Delta b_i$ coefficients,
which appear in (\ref{eq:ZBMdiv}-\ref{eq:ZBM})
after \req{eq:mrconBM} is taken into account,
read
\begin{subequations}
\begin{eqnarray}
  \Delta a_1(u,t) &=& 
   2\, C_F\,
  \left\{ 4 \, \frac{u}{t} 
         \theta(t-u) 
                    + 
        \bigg( \begin{array}{c} u \rightarrow 1-u \\
                                t \rightarrow 1-t
               \end{array} \bigg) \right\}
                   \, ,
        \nonumber \\ & &
\label{eq:deltaa1} 
\end{eqnarray}
\begin{eqnarray}
  \lefteqn{\Delta b_1^{n_f}(u,t) - 2 \Delta a_2(u,t)}
          \nonumber \\
           & = & 
   2\, C_F\,
  \left\{  \frac{u}{t} 
    \left( \frac{8}{3} + 4 \ln \frac{u}{t} \right)
         \theta(t-u) 
                    + 
        \bigg( \begin{array}{c} u \rightarrow 1-u \\
                                t \rightarrow 1-t
               \end{array} \bigg) \right\}
           \, .
        \nonumber \\ & &
\label{eq:deltab1a2}
\end{eqnarray}
\label{eq:deltaab}
\end{subequations}
The complete renormalization constant
${\cal Z}_{HV} \equiv {\cal Z}_{HV,UV}$
from Eq. \req{eq:ffBM} is then given by
\begin{equation}
 {\cal Z}_{HV}(u,v; \tilde{\mu}_R^2) =   
 {\cal Z}_{HV}^{div}(u,w; \tilde{\mu}_R^2) \, \otimes \,   
 {\cal Z}_{HV}^{fin}(w,v; \tilde{\mu}_R^2)   
      \, .
\label{eq:ZBM}
\end{equation}
By utilizing 
the condition
of multiplicative renormalizability of the
``spurious'' terms
\req{eq:mrconBM}, 
it takes the form
\begin{widetext}
\begin{eqnarray}
 {\cal Z}_{HV}(u,v; \tilde{\mu}_R^2)& =&   
   \delta(u-v)
 + \frac{\alpha_S(\tilde{\mu}_R^2)}{4 \pi}
 \, \Big( \Delta a_1(u,v) + O(\eps) \Big)
    \nonumber \\ & & 
 + \, \frac{\alpha_S^2(\tilde{\mu}_R^2)}{(4 \pi)^2}
 \left[ \left(-\frac{2}{3} n_f \right)
    \left( \frac{1}{\eps} \,
    \frac{-\Delta a_1(u,v)}{2} + 
    \frac{\Delta b_1^{n_f}(u,v) - 2 \Delta a_2(u,v)}{2}  + O(\eps) \right) 
    + \cdots \right]
         \nonumber \\ &  &
        + \cdots
            \, .
\label{eq:ZBMour}
\end{eqnarray}
\end{widetext}
It is interesting to note that by using the ``reduction''
formulas, which relate 
the exclusive (``nonforward'') and inclusive (``forward'')
kernels \cite{MiR85}, the agreement between
the renormalization constant ${\cal Z}_{HV}$ given above
and an analogous ``HV'' renormalization constant 
for the longitudinal spin structure function
$g_1$ \cite{MSNe98} is established (up to $O(\eps)$ terms).

For later use we specify  
the convolution of functions defined in Eq. \req{eq:deltaab}
with the $1/(1-x)$ term:
\begin{subequations}
\begin{eqnarray}
  \frac{1}{1-x} 
  \, \otimes \, \Delta a_1(x,u) & = &
   C_F \,\frac{1}{1-u} \left( - 8 \frac{1-u}{u} \ln(1-u) \right) 
          \, ,
        \nonumber \\ & &
\label{eq:1xdV1}
\end{eqnarray}
and
\begin{eqnarray}
  \lefteqn{\frac{1}{1-x} 
  \, \otimes \, 
 \left(\Delta b_1^{n_f}(x,u) - 2 \Delta a_2(x,u)\right)}
    \nonumber \\  & = &
   C_F \,\frac{1}{1-u} \left( 
  - \frac{16}{3} \frac{1-u}{u} \ln (1-u)
  -8 \frac{1-u}{u} \text{Li}_2(u) \right)
         \, .
    \nonumber \\  &  &
\label{eq:1xdV2nf}
\end{eqnarray}
\label{eq:1xdV1V2nf}
\end{subequations}

\section{The expression for the pion transition form factor
         up to \lowercase{$n_f$}-proportional NNLO terms}
\label{s:res}

We now combine the results 
for the hard-scattering amplitude
and the DA obtained in the preceding sections.
After resolving the $\gamma_5$ problem and discussing
the dependence of the prediction on the factorization
scale $\muF$,
we finally present
the expression for the pion transition form factor
up to $n_f$-proportional NNLO terms.

\subsection{Resolving the $\gamma_5$ ambiguity}

In subsections \ref{ss:Tren} and \ref{ss:DAren}, we have presented
the results of the perturbative treatment of
the hard-scattering amplitude $T(u,Q^2)$
and the distribution amplitude $\Phi(u)$, respectively. 
Along the lines outlined in Sec. \ref{s:cprocedure},
we  now proceed to combine these results 
to obtain the finite and $\gamma_5$-scheme independent
expression for the pion transition form factor
$F_{\gamma^* \gamma \pi}(Q^2)$, up to the
$n_f$-proportional NNLO contributions.

The lack of the ambiguity in the DA results
along with the fact that the prediction for the pion transition form factor
should not depend on the choice of the $\gamma_5$-scheme
make it possible to
resolve the ambiguity of the $\gamma_5$
treatment in the hard-scattering calculation.

\subsubsection{Naive $\gamma_5$-scheme}

The appearance of two $\gamma_5$ matrices imposes
the use of the naive-$\gamma_5$ scheme
in the DA calculation 
and the corresponding results are presented
in (\ref{eq:phiVev}--\ref{eq:V1V2nf}).
The $\gamma_5$ matrix present in
the hard-scattering amplitude can also
be treated in the naive-$\gamma_5$ scheme
in which case a number of results emerge.
After the Ward identities of QED are 
taken into account, the remaining
ambiguity in the hard-scattering amplitude result
(\ref{eq:THres}-\ref{eq:TH0ZTcol})  
is parameterized by the parameter
$\delta$
(as explained in Appendix \ref{app:gamma5} and 
Subsec. \ref{ss:Tfd}). 

Matching Eqs.
\req{eq:Acol} and \req{eq:1xV1V2nf}
one observes that
\begin{subequations}
\begin{eqnarray}
  \frac{1}{1-x} 
  \, \otimes \, V_1(x,u) & = &
         A^{(1)}_{col}(u)
\label{eq:1xV1A}
       \\
  \frac{1}{1-x} 
  \, \otimes \, V_2^{n_f}(x,u) & = &
         A^{(2, n_f)}_{col}(u) |_{\delta=1}
          \, .
\label{eq:1xV2nfA}
\end{eqnarray}
\label{eq:1xV1V2nfA}
\end{subequations}
If these relations are taken into account in Eq. \req{eq:TH0ZTcol},
then a comparison with \req{eq:ZfrenNF}, 
for $\tilde{\mu}_R^2=\muF$
(i.e., for the DA $\Phi(u)$ renormalized at the $\muF$ scale),
gives 
\begin{equation}
  Z_{T,col}|_{\delta=1} 
   = Z_{\phi,ren}^{-1} 
          \, ,
\label{eq:ZTcolZfren}
\end{equation}
i.e., the relation \req{eq:ZTZf} is satisfied
and the singularities in \req{eq:Fpiur1} 
%(or equivalently \req{eq:Fpiur2})
cancel for the $\delta$ parameter 
taking the value $1$.
Hence, we obtain
\begin{equation}
  Z_{T,col}(x,u;\muF) \equiv Z_{T,col}(x,u;\muF)|_{\delta=1} 
          \, ,
\label{eq:ZTcoldef} 
\end{equation}
and, consequently,
\begin{equation}
  T_H(x, Q^2, \muF) \equiv T_H(x, Q^2, \muF)|_{\delta=1} 
         \, ,
\label{eq:THdef}
\end{equation}
where Eq. 
\req{eq:ZTcolZfren}, together with \req{eq:ZfrenNF},
determines
$Z_{T,col}(x,u;\muF)|_{\delta=1}$,
while
taking $\delta=1$ in Eqs.
(\ref{eq:THres}-\ref{eq:Acol}) 
gives
$T_H(x, Q^2, \muF)|_{\delta=1}$. 
Hereby, we have
confirmed the $\gamma_5$ prescription  
employed in \cite{Bra83}
(see Appendix \ref{app:gamma5} for 
further details). 

In preceding consideration,
we have resolved the $\gamma_5$ ambiguity 
of the hard-scattering prediction by adopting the naive
$\gamma_5$-scheme and by using the unambiguous DA results
(along with the QED Ward identities)
to single out the correct prediction. 
%from the set of results.

\subsubsection{HV scheme}

Let us now present the calculation
performed in the HV scheme.
Continuing to $D$ dimensions and adopting
the HV scheme leads to unique results, but 
the ``spurious'' anomalous terms,
which violate chiral symmetry, appear and
additional renormalization is required,
both for the DA and the hard-scattering amplitude.

The results for the hard-scattering amplitude
obtained in the HV scheme
correspond to the $\delta =0$ choice in
(\ref{eq:THres}-\ref{eq:TH0ZTcol}),  
and the notation
$T^{HV}$, $T_H^{HV}$, $Z_{T,col}^{HV}$
has been introduced.
The fact that the UV singularities appearing
in the hard-scattering amplitude get completely
renormalized by the coupling constant renormalization
indicates that, contrary to the DA case,
only the presence of collinear singularities
along with the nonanticommuting nature of $\gamma_5$
matrix introduces ``spurious'' anomalous terms.

The corresponding renormalization
constant for the DA denoted by
${\cal Z}_{HV}={\cal Z}_{HV}^{div} \, {\cal Z}_{HV}^{fin}$
and displayed in 
(\ref{eq:ZBMdiv}-\ref{eq:ZBMour})
has been determined by comparing the ``correct'' results 
obtained in the naive-$\gamma_5$  with
the corresponding results obtained using the HV scheme.
As a result, one finds that
the unrenormalized DA in the HV scheme, $\Phi^{HV}(u')$,
and the unrenormalized DA in the
naive-$\gamma_5$ scheme, $\Phi(u)$,
are related by
\begin{equation}
   \Phi={\cal Z}_{HV}^{-1} \; \Phi^{HV}
        \, .
\label{eq:PhiZBMPhiBM}
\end{equation}
Similarly, 
the renormalization constants
$Z_{\phi,ren}$ and $Z_{\phi,ren}^{HV}$ determined in
the naive-$\gamma_5$ and the HV schemes, respectively,
are related by
\begin{equation}
   Z_{\phi,ren} =
       {\cal Z}_{HV}^{div}{}^{-1} \;
          Z_{\phi,ren}^{HV}
        \, ,
\label{eq:ZrenZBMdivZrenBM}
\end{equation}
while the additional finite ``HV'' renormalization
of the  
renormalized distribution
$\Phi^{HV}(v',\tilde{\mu}_R^2)$ calculated in the HV scheme,
is needed to obtain 
the renormalized DA 
$\Phi(v,\tilde{\mu}_R^2)$ free of ``spurious''
anomalies: 
\begin{equation}
   \Phi={\cal Z}_{HV}^{fin}{}^{-1}
           \; \Phi^{HV}
        \, .
\label{eq:PhiZBMfinPhiBM}
\end{equation}

The prediction for the pion transition form
factor cannot depend on the choice of the scheme
and chiral symmetry is restored for the 
complete result,
i.e.,
\begin{eqnarray}
  F_{\gamma^* \gamma \pi } &=& 
         T \; \Phi^{\dagger} =
         T \; {\cal Z}_{HV}^{-1} \; 
          {\cal Z}_{HV} \;\Phi^{\dagger}
             \nonumber \\ &=&
         T^{HV}  \; \Phi^{HV}{}^{\dagger}
             \, ,
\label{eq:Fpibm}
\end{eqnarray}
or equivalently
\begin{eqnarray}
  F_{\gamma^* \gamma \pi } &=& 
         T_H \; \Phi^* =
         T_H \; \left( {\cal Z}_{HV}^{fin}\right)^{-1} \; 
          {\cal Z}_{HV}^{fin} \;\Phi^*
             \nonumber \\ &=&
         T_H^{HV}  \; \Phi^{HV}{}^*
                 \, .
\label{eq:Fpibm1}
\end{eqnarray}

On the basis of Eqs.\req{eq:1xdV1V2nf} and 
(\ref{eq:ZBMdiv}-\ref{eq:ZBMour}),
along with  Eqs.
(\ref{eq:THres}-\ref{eq:TH0ZTcol}),  
it can easily be shown that 
\begin{equation}
      T = T^{HV} \, {\cal Z}_{HV}
            \, ,
\end{equation}
while 
\begin{equation}
      T_H = T_H^{HV} \, {\cal Z}_{HV}^{fin}
            \, ,
\end{equation}
and
\begin{equation}
      Z_{T,col} = Z_{T,col}^{HV} \, {\cal Z}_{HV}^{div}
            \, ,
\end{equation}
with $T_H$ and $Z_{T,col}$ being given by 
Eqs. \req{eq:THdef} and \req{eq:ZTcoldef}, respectively,
and $T=T_H \, Z_{T,col}$.

Hence, we have resolved the $\gamma_5$ ambiguity
appearing in the hard-scattering amplitude calculation
by  consistently treating, in either the naive-$\gamma_5$ or
the HV scheme, both the hard-scattering amplitude
and the distribution amplitude (which, 
actually, is free of the $\gamma_5$ ambiguity).

\subsection{
Discussing the factorization scale dependence}

After the $\gamma_5$ ambiguity is resolved,
and the collinear singularities present in the $T(u,Q^2)$
and $\Phi(u)$ amplitudes cancel, we are left with
the finite prediction for the pion
transition form factor
\begin{equation}
    F_{\gamma^* \gamma \pi}(Q^{2})= 
      T_{H}(x,Q^{2},\muF)  \, \otimes \,  \Phi^{*}(x,\muF)
                    \,. 
\label{eq:tffcf1}
\end{equation}
The hard-scattering amplitude
$T_{H}(x,Q^{2},\muF)$, evaluated
up to $n_f$-proportional NNLO terms, is given by 
Eqs. (\ref{eq:THres}-\ref{eq:Acol}) with $\delta=1$.

The distribution amplitude
$\Phi^{*}(x,\muF)$ is determined
by evoluting $\Phi^{*}(x,\mu_0^2)$
(obtained at the scale $\mu_0^2$ using 
some nonperturbative method)
to the scale $\muF$ according to 
\req{eq:PhifVPhi} 
\begin{displaymath}
   \Phi(v,\tilde{\mu}_R^2)=
         \phi_V(v,s; \muF, \mu_0^2) \otimes
         \Phi(s, \mu_0^2)
           \, ,
%\label{eq:PhifVPhi}
\end{displaymath}
i.e., \req{eq:phiVPhi}
\begin{eqnarray*}
   \lefteqn{\Phi(v, \muF)} \nonumber \\ & =&
       \phi_{V}^{LO}(v,s; \muF, \mu_0^2)
       \otimes \Phi(s, \mu_0^2) 
           \nonumber \\ & &
 + \frac{\alpha_S(\muF)}{4 \pi} \;
       \phi_{V}^{NLO}(v,s; \muF, \mu_0^2)
       \otimes \Phi(s, \mu_0^2) + \cdots
          \, . \qquad
%\label{eq:phiVPhi}
\end{eqnarray*}
%according to \req{eq:phiVPhi},
%with \req{eq:phiVev} taken into account.
In subsection \ref{ss:DAren} 
we have analysed in detail the evolutional
part $\phi_V$,
and as noted there, 
the two-loop DA calculation explicitly gives only the
first few terms of the LO and NLO contributions,
while the complete LO and NLO behavior of the DA is determined
by solving the evolution equation \req{eq:eveq},
%yielding  (\ref{eq:Phiphi}-\ref{eq:phiNLOexp}).
or equivalently \req{eq:VfV}
\begin{eqnarray*}
       \lefteqn{\muF 
       \frac{\partial}{\partial \muF} 
                        \phi_V (v,s,\muF,\muO)}
           \nonumber \\  
        \qquad & \qquad =& V(v,s',\muF) \, \otimes \, 
                           \phi_V(s',s,\muF,\muO)
             \, . \qquad 
%\label{eq:VfV}
\end{eqnarray*}

The dependence of $T_{H}(x,Q^{2},\muF)$
on the factorization scale $\muF$ 
can be determined analogously to the 
%previously examined 
$\muF$ dependence of the DA.
Thus, by differentiating
Eq. \req{eq:TTHZ} with respect to $\muF$
and taking into account Eqs. \req{eq:VZ} and \req{eq:ZTZf}
one finds that $T_H(x, Q^2, \muF)$ 
satisfies the equation
\begin{equation}
  \muF \frac{\partial}{\partial \muF} 
        T_H(x, Q^2, \muF) = -
        T_H(y, Q^2, \muF)  \, \otimes \,
            V(y,x;\muF) 
         \, , 
\label{eq:EvEqT}
\end{equation}
which, as it is seen, is analogous to the DA evolution
equation \req{eq:eveq}.
Therefore, just as in the case of the DA,
any finite order solution of \req{eq:EvEqT}
contains the complete $\muF$ dependence of
$T_H(x, Q^2, \muF)$ 
which is not the case with the expansion
%the results given by
\req{eq:THres}
%(\ref{eq:TH012nf}-\ref{eq:Acol})
truncated at the same order
(and not taking into account \req{eq:alphaln}).

%Similarly to \req{eq:PhifVPhi}, 
The hard-scattering amplitude $T_H(x, Q^2, \muF)$
can be written in the factorized form 
\begin{equation}
  T_H(x,Q^2,\muF) =  T_H(y,Q^2,\muF=Q^2) \otimes
                   \phi_V(y,x,Q^2,\muF)
            \, ,
\label{eq:tTHfV}
\end{equation}
with $\phi_V(y,x,Q^2,\muF)$ containing all its
$\muF$ dependence.
This can  easily be demonstrated
if use is made of  $T_H$ and $\phi_V$ determined
up to $n_f$-proportional NNLO terms (see Eqs. 
\req{eq:phiVev}, \req{eq:1xV1V2nf}, 
and \req{eq:chscLO}). 
On the other hand, using Eq. \req{eq:VfV} 
one can show by partial integration that
\req{eq:tTHfV} indeed represents the solution of
the evolution equation \req{eq:EvEqT}.

When calculating to any finite order in $\alpha_S$,
it is inappropriate to convolute  the
$\Phi(x,\muF)$ distribution obtained by solving 
the evolution equation \req{eq:eveq} 
(i.e., given by
(\ref{eq:Phiphi}-\ref{eq:phiNLOexp}))
with $T_H(x,Q^2,\muF)$ obtained by the
truncation of the expansion \req{eq:THres}. 
Namely, in the latter function, only the partial dependence
on $\muF$ is included,
in contrast to the former.
%To illustrate,
%this would correspond to convoluting
%the complete $\phi_V^{LO}$ form
%of \req{eq:phiVLOcom} with the  
%$\phi_V^{LO}$ given by terms explicitly displayed \req{eq:phiVevLO}.
Notice that
when the complete  dependence of $T_H$ on $\muF$
is taken into account
even the LO term in \req{eq:tTHfV} is $\muF$ dependent,
in contrast to $T^{(0)}$ given in \req{eq:TH},
and this leads to $\muF$ independent 
LO prediction for the pion transition
form factor.

Substituting Eqs. \req{eq:tTHfV} and \req{eq:PhifVPhi}
into Eq. \req{eq:tffcf1} 
and taking into account that
\begin{equation}
   \phi_V(y,x, Q^2, \muF)\, \otimes \phi_V(x,s,\muF,\mu_0^2) 
          = \phi_V(y,s, Q^2, \mu_0^2) 
               \, ,
\label{eq:fvfvfv}
\end{equation}
one obtains
\begin{equation}
  F_{\gamma^* \gamma \pi}(Q^2) = 
   T_H(y, Q^2,Q^2) 
          \, \otimes \phi_V(y,s, Q^2, \mu_0^2) 
          \, \otimes \,\Phi^*(s, \mu_0^2)
                 \, .
\label{eq:Fpi2q2}
\end{equation}
The relation \req{eq:fvfvfv} is valid at 
every order of perturbative calculation
(to NLO  this can easily be checked  by  
substituting
\req{eq:phiVLOcom} and the NLO results of Ref. \cite{Mu94etc}
into \req{eq:fvfvfv}). 
It represents the resummation of the 
$\ln(Q^2/\mu_0^2)$ logarithms over the intermediate $\muF$ scale, 
performed in such a way that, first, 
the logarithms $\ln(\muF/\mu_0^2)$ originating from the 
perturbative part of the DA are resummed, and then the summation of 
$\ln(Q^2/\muF)$ logarithms from the hard-scattering part is performed. 
Therefore, the summations of the $\muF$ logarithms can be accomplished with 
any choice of $\muF$, because the effect in the final prediction,
at every order, 
is the same as if 
the complete renormalization-group resummation of $\ln(Q^2/\mu_0^2)$ 
logarithms has been performed. 

Consequently, the $F_{\gamma^* \gamma \pi}$
prediction \req{eq:tffcf} (as well as the prediction for any
other exclusive quantity obtained in
the standard hard-scattering picture)
is independent 
of the factorization scale $\muF$  
at every order in $\alpha_S$,
provided both $T_H$ and $\Phi$ are 
consistently treated regarding the $\muF$ dependence. 
The intermediate 
scale at which the short- and long-distance dynamics separate, 
the factorization scale $\muF$, 
disappears from the final prediction at every order in $\alpha_S$
and therefore does not introduce any theoretical uncertainty into the 
PQCD calculation for exclusive processes. 

We would like to point out here that
by adopting the common choice $\mu_F^2 = Q^2$,
one avoids the need
for the resummation of the $\ln(Q^2/\muF)$ logarithms in 
the hard-scattering part, making the calculation simpler 
and hence, for practical purposes, the preferable form of
$F_{\gamma^* \gamma \pi}(Q^{2})$ is given by
\begin{equation}
    F_{\gamma^* \gamma \pi}(Q^{2})= 
   T_H(x, Q^2, Q^2) 
        \, \otimes \,  \Phi^{*}(x,Q^2)
                    \,. 
\label{eq:tffcfN}
\end{equation}

\begin{widetext}
\subsection{Presenting the final results}

Finally, we summarize. 
Taking into account 
Eqs. (\ref{eq:THres}-\ref{eq:Acol}) and the results
of the preceding subsections,
the hard-scattering amplitude
$T_{H}(x,Q^{2},\muF=Q^2)$, free of 
all effects of collinear singularities 
(the terms containing functions 
$A_{col}^{(n)} \ln^n \muF/Q^2$ factorized in $\phi_V$),
takes the form
\begin{eqnarray}
  T_H(x, Q^2, \muF=Q^2) 
     & =& 
        T_H^{(0)}(x,Q^2) 
        + \frac{\alpha_S(\muR)}{4 \pi}\,  \, T_H^{(1)}(x,Q^2,\muF=Q^2)
             \nonumber \\ & &
        + \frac{\alpha_S^2(\muR)}{(4 \pi)^2} 
          \left[  \left( -\frac{2}{3} n_f \right) 
              \,  T^{(2,n_f)}_{H}(x,Q^2,\muF=Q^2,\muR)
              + \cdots \right] + \cdots 
             \, ,
         \nonumber \\ & &
\label{eq:tTHres}
\end{eqnarray}
where
\begin{subequations}
\begin{eqnarray}
   T_H^{(0)}(x,Q^2)&=&\frac{N_T}{Q^2} \,
       A^{(0)}(x) + (x \rightarrow 1-x) \, ,
\label{eq:tTH0}  \\[0.2cm] 
   T_H^{(1)}(x,Q^2,\muF=Q^2)&=&\frac{N_T}{Q^2} \,
          A^{(1)}(x) 
               +  (x \rightarrow 1-x) \, ,
\label{eq:tTH1}  \\[0.2cm] 
   T_H^{(2,n_f)}(x,Q^2,\muF=Q^2,\muR)&=&\frac{N_T}{Q^2} \,
      \left( 
         A^{(2,n_f)}(x) \,
    +   A^{(1)}(x) \ln \frac{\muR}{Q^2}
              \right) 
               +  (x \rightarrow 1-x)
          \, ,
\label{eq:tTH2nf}
\end{eqnarray}
\label{eq:tTH012nf}
\end{subequations}
and
\begin{subequations}
\begin{eqnarray}
    A^{(0)}(x) &=& \frac{1}{1-x} \, ,
\label{eq:A0res} \\[0.2cm] 
    A^{(1)}(x) &=& 
           C_F \, \frac{1}{1-x} 
      \left[ -9 -  \frac{1-x}{x} \ln(1-x) + 
             \ln^2 (1-x) \right] \, ,
\label{eq:A1res}  \\[0.2cm] 
    A^{(2,n_f)}(x) &=&
           C_F \, \frac{1}{1-x} 
       \left[
    -\frac{457}{24} 
    - \left( \frac{47}{18} -
             \frac{1}{2 x} \right) \ln (1-x) 
    + \left( \frac{13}{6} -
             \frac{1}{2 x} \right) \ln^2 (1-x) 
             \right.  \nonumber \\ & &  \left.
    - \frac{1}{3} \ln^3 (1-x) 
    - \frac{14}{3} \text{Li}_2(x)
    + \text{Li}_3(x) - 2 S_{1,2}(x) \right]
       \, .
\label{eq:A2nfres}
\end{eqnarray}
\label{eq:Ares}
\end{subequations}
\end{widetext}

%\begin{eqnarray}
%     \lefteqn{
%  F_{\gamma^* \gamma \pi}(Q^2)
%     } \nonumber \\[0.3cm]
%        & =&
%    \tilde{T}_H^{(0)}(x,Q^2) \otimes \Phi^{LO}(x,Q^2) 
%       \nonumber \\ & &
%    +\left( \frac{\alpha_S(\muR)}{4 \pi}
%    \tilde{T}_H^{(1)}(x,Q^2) \otimes \Phi^{LO}(x,Q^2) + 
%    \frac{\alpha_S(Q^2)}{4 \pi}
%    \tilde{T}_H^{(0)}(x,Q^2) \otimes \Phi^{NLO}(x,Q^2) \right) 
%       \nonumber \\ & &
%    +\frac{\alpha_S^2(\muR)}{(4 \pi)^2}
%      \left[ \left(-\frac{2}{3} n_f \right)
%        \tilde{T}_H^{(2, n_f)}(x,Q^2,\muR) \, \otimes
%                          \Phi^{LO}(x,Q^2)
%         + \cdots \right]
%      + \cdots
%           \, .
%\label{eq:anres}
%\end{eqnarray}

To obtain 
the distribution amplitude $\Phi(x,Q^2)$
one evolutes $\Phi(x,\mu_0^2)$,
determined using some nonperturbative methods
at the scale $\mu_0^2$, to the scale $Q^2$
according to
(\ref{eq:Phiphi}-\ref{eq:phiNLOexp}).

By substituting Eqs. \req{eq:tTHres}, 
\req{eq:Phiphi}, and \req{eq:phiexp} into
\req{eq:tffcfN} and taking 
\req{eq:chscLO} into account
the pion transition form factor $F_{\gamma^* \gamma \pi}$
expressed as a perturbative series
in $\alpha_S(\muR)$ reads
\begin{eqnarray}
  \lefteqn{F_{\gamma^* \gamma \pi}(Q^2)}
        \nonumber \\ 
      &=& 
       F_{\gamma^* \gamma \pi}^{(0)}(Q^2)
        + \frac{\alpha_S(\muR)}{4 \pi}  \, 
       F_{\gamma^* \gamma \pi}^{(1)}(Q^2)
    \nonumber \\ && 
        + \, \frac{\alpha_S^2(\muR)}{(4 \pi)^2} 
          \left[  \left( -\frac{2}{3} n_f \right) \, 
       F_{\gamma^* \gamma \pi}^{(2,n_f)}(Q^2,\muR)
              + \cdots \right] 
           \nonumber \\ & &
         + \cdots 
             \, ,
\label{eq:Finres}
\end{eqnarray}
where
\begin{subequations}
\begin{eqnarray}
     F_{\gamma^* \gamma \pi}^{(0)}(Q^2)   &=&
    T_H^{(0)}(x,Q^2) \otimes \Phi^{LO}(x,Q^2) \, ,
\label{eq:finres0}
        \\[0.2cm] 
     F_{\gamma^* \gamma \pi}^{(1)}(Q^2)   &=&
    T_H^{(1)}(x,Q^2,Q^2) \otimes \Phi^{LO}(x,Q^2) 
        \nonumber \\ & &
      + 
    T_H^{(0)}(x,Q^2) \otimes \Phi^{NLO}(x,Q^2) \, ,
           \nonumber \\ & &
\label{eq:finres1}
        \\[0.2cm]
     F_{\gamma^* \gamma \pi}^{(2,n_f)}(Q^2,\muR) &=&
         T_H^{(2, n_f)}(x,Q^2,Q^2,\muR) \, \otimes
                          \Phi^{LO}(x,Q^2)
        \nonumber \\ & &
               + 
    T_H^{(0)}(x,Q^2) \otimes \Phi^{NLO}(x,Q^2) 
          \ln \frac{\muR}{Q^2} 
            \, .
           \nonumber \\ & &
\label{eq:finres2nf}
\end{eqnarray}
\label{eq:finres}
\end{subequations}
As it is seen, we are left with one expansion parameter:
$\alpha_S(\muR)$. 
The scale $\muR$ actually represents
the renormalization scale of the complete
perturbatively calculable part of the pion
transition form factor \req{eq:Fpi2q2}, i.e., 
of 
\begin{equation}
T_H(y, Q^2,\mu_0^2)=
 T_H(y, Q^2,Q^2) \otimes \phi_V(y,s, Q^2, \mu_0^2)
       \, .
\label{eq:THQ2mu0}
\end{equation}
Although, the physical pion transition form factor
$F_{\gamma^* \gamma \pi}(Q^2)$
does not depend on the choice of the
renormalization scale $\muR$,
when calculating to any finite order
a residual dependence on the $\muR$ scale
remains.
 
\section{Numerical predictions}

\subsection{
Fixing the renormalization scale according to 
the BLM procedure}
\label{s:BLM}

The dependence of finite order predictions
on the renormalization scale introduces a
theoretical uncertainty in their interpretation
(see \cite{MNP99} for a detailed discussion),
which is especially evident in
calculations to lowest order in order $\alpha_S$.
It would be advantageous to optimize the 
scale choice according to some sensible criteria. 
The BLM procedure \cite{BLM83,BrLu95,BrJ98} offers such
criteria. 
The essence of the BLM procedure is that 
all vacuum-polarization effects
(gluon vacuum polarization contributions, analogous to QED,
as well as, quark vacuum polarization and vertex corrections)
from the QCD $\beta$ function are incorporated into the
running coupling constant rather than into the coefficients 
of the perturbative expansion. In practice, this amounts to
computing quark-loop insertions in the diagrams of
that order (since $\beta_0=11-2/3 n_f$,$\cdots$)
and setting the scale by demanding that 
the coefficients of the perturbative expansion are
$n_f$ independent.

Hence, according to the BLM scale setting prescription,
the renormalization scale for the pion transition form factor
entering at the NLO  is determined
from the NNLO $n_f$-proportional terms,
and is fixed by the requirement
\begin{equation}
 F_{\gamma^* \gamma \pi}^{(2,n_f)}(Q^2,\muR)=0
      \, .
\label{eq:BLMcon}
\end{equation}
Note that 
in the present calculation
the effective nature of $\muR$
has been implicitly assumed, i.e., 
$\muR$ has been treated as independent of the 
momentum fractions $x$ throughout the paper;
otherwise the factorization of singularities
would take a cumbersome form (if menagable at all). 
Apart from that we would be faced with the problem of a
clear separation of the short- and long-
distance effects \cite{BaRS00}.
\begin{widetext}
Therefore, the only consistent way to asses the BLM scale
is to solve Eq. \req{eq:BLMcon},
resulting in some mean value BLM scale
\begin{subequations}
\begin{equation}
 \muR=\mu_{BLM}^2 = a_{BLM}(Q^2) \; Q^2 
        \, ,
\label{eq:muBLM}
\end{equation}
where
\begin{equation}
    a_{BLM}(Q^2)  
     = \exp \left( -\frac{A^{(2,n_f)}(x) \otimes \Phi_{LO}(x,Q^2)}{
            A^{(1)}(x) \otimes \Phi_{LO}(x,Q^2) +
            A^{(0)}(x) \otimes \Phi_{NLO}(x,Q^2)}
                         \right) 
               \, .
\label{eq:aBLM}
\end{equation}
\label{eq:BLMsc}
\end{subequations}
As it is seen, the scale $\mu_{BLM}^2$ 
depends on the specific form of the distribution amplitude.

In Section \ref{s:formalism} the nonperturbative
input, i.e., the distribution amplitude 
$\phi(x,\mu_0^2)= \Phi(x,\mu_0)/N_{\phi}$
determined at the scale $\mu_0^2$,
has been presented in the form of an expansion over Gegenbauer
polynomials \req{eq:phiexp}. 
The evolution to the scale $\muF$ has been
described by Eqs. (\ref{eq:evDA}-\ref{eq:phiNLOexp}).
Retaining only the first three terms in the general
expansion of the pion DA given in Eq. \req{eq:phiexp} 
\begin{equation}
\phi(x,\mu_0^2) = 6 x (1-x)
   \left[1 + B_2 C_2^{3/2}(2 x - 1) + B_4 C_4^{3/2}(2 x -1)
   \right]
         \, ,
\label{eq:phiB24}
\end{equation}
the LO and NLO contributions to 
$Q^2 F_{\gamma^* \gamma \pi}(Q^2)$ 
%obtained from \req{eq:Finres} and given by
%\begin{eqnarray}
%  \lefteqn{Q^2 F_{\gamma^* \gamma \pi}(Q^2)} 
%    \nonumber \\ &=& 
%       Q^2 F_{\gamma^* \gamma \pi}^{(0)}(Q^2)
%        + \frac{\alpha_S(\muR)}{4 \pi}  \, 
%       Q^2 F_{\gamma^* \gamma \pi}^{(1)}(Q^2)
%        + \frac{\alpha_S^2(\muR)}{(4 \pi)^2} 
%          \left[  \left( -\frac{2}{3} n_f \right) \, 
%       Q^2 F_{\gamma^* \gamma \pi}^{(2,n_f)}(Q^2,\muR)
%              + \cdots \right] + \cdots 
%             \, ,
%           \nonumber \\ & &
%\label{eq:Q2Finres}
%\end{eqnarray}
take the form
\begin{eqnarray}
  Q^2 F_{\gamma^* \gamma \pi}^{(0)}(Q^2) &=& 
      2 C_{\pi} f_{\pi}
     \left[ 3 \left( 1 + B_2^{LO}(Q^2) + B_4^{LO}(Q^2) \right) 
         \right] 
\label{eq:numres0B24} \\ 
  Q^2 F_{\gamma^* \gamma \pi}^{(1)}(Q^2) &=& 
      2 C_{\pi} f_{\pi}
               \left[
       \left(-20 + \frac{295}{18} B_2^{LO}(Q^2)  
                 + \frac{10487}{225} B_4^{LO}(Q^2) \right)
                    + 3
       \sum_{k=2}^{\infty}{}' B_k^{NLO}(Q^2)
               \right]
              \, ,
        \nonumber \\
\label{eq:numres1B24}
\end{eqnarray}
while the $n_f$ proportional NNLO contribution
amounts to
\begin{eqnarray}
  Q^2 F_{\gamma^* \gamma \pi}^{(2,n_f)}(Q^2) &=& 
      2 C_{\pi} f_{\pi}
               \left[
       \left( -43.47 + 78.47 B_2^{LO}(Q^2)+ 197.165 B_4^{LO}(Q^2)
               \right)
               \right.
        \nonumber \\ & & 
               \left.
       \quad + \left(-20 + \frac{295}{18} B_2^{LO}(Q^2)  
                 + \frac{10487}{225} B_4^{LO}(Q^2) \right)
            \ln \frac{\muR}{Q^2} 
                \right.
          \nonumber \\ & & 
                \left.
         \quad            + \left( 3
               \sum_{k=2}^{\infty}{}' B_k^{NLO}(Q^2) \right)
            \ln \frac{\muR}{Q^2} 
                 \right]
               \, .
\label{eq:numres2nfB24}
\end{eqnarray}

Therefore, on the basis of Eqs. \req{eq:BLMcon} and
\req{eq:numres2nfB24} one finds 
that the BLM scale for the pion transition form factor
is given by \req{eq:muBLM}
with 
\begin{equation}
    a_{BLM}(Q^2)  
     = \exp \left( 
  -\frac{-43.47 + 78.47 B_2^{LO}(Q^2)+ 197.165 B_4^{LO}(Q^2)}{
       -20 + \displaystyle\frac{295}{18} B_2^{LO}(Q^2)  
                 + \frac{10487}{225} B_4^{LO}(Q^2) 
       + 3 \sum_{k=2}^{\infty}{}' B_k^{NLO}(Q^2)} 
                         \right) 
               \, .
\label{eq:aBLMB24}
\end{equation}
\label{eq:BLMsc24}

\end{widetext}

Expressions \req{eq:Finres} and 
(\ref{eq:numres0B24}-\ref{eq:numres2nfB24}),
representing the complete NLO prediction for the
pion transition form factor,
together with the expressions \req{eq:muBLM}
and \req{eq:aBLMB24}, specifying the corresponding
BLM scale, are valid for arbitrary
distribution amplitude (with the evolutional effects included)
and represent the main results of this paper.

\subsection{Numerical predictions
in the $\overline{MS}$ and $\alpha_V$ schemes}

Based on the general expressions derived in a preceding
subsection, we now proceed to
obtain numerical predictions for the pion transition form
factor using two specific distribution amplitudes:
the asymptotic DA and the CZ distribution amplitude.

There is increasing theoretical evidence coming from
different calculations
\cite{soft,Rad98,Rad95etc,DArec}
that the low energy pion distribution
amplitude does not differ much from its 
asymptotic form
$\phi_{as}(x) = 6 x (1-x)$
(which represents the solution of the evolution
equation \req{eq:eveq} for $\muF\to \infty$).
The distribution $\phi(x,\mu_0^2)=\phi_{as}(x)$
is characterized by the fact that 
at the LO it has no evolution,
while the NLO evolutional effects are tiny \cite{MNP99},
and for the purpose of this calculation these effects
can safely be neglected.

The expression for the pion transition form factor 
$Q^2 F_{\gamma^* \gamma \pi}(Q^2)$,
based on the Eqs. \req{eq:Finres} and 
(\ref{eq:numres0B24}-\ref{eq:numres2nfB24}),
and  corresponding to the asymptotic distribution
then reads
\begin{eqnarray}
  \lefteqn{Q^2 F_{\gamma^* \gamma \pi}(Q^2)} 
    \nonumber \\  &=& 2 C_{\pi} f_{\pi}
     \left\{ 3  
    +\frac{\alpha_S(\muR)}{4 \pi}
       (-20) 
         \right. \nonumber \\ & & \left.
    +\frac{\alpha_S^2(\muR)}{(4 \pi)^2}
       \left[ \left( -\frac{2}{3} n_f \right)
       \left( -43.47 - 20 \ln \frac{\muR}{Q^2} \right)
         + \cdots \right] 
        \right. \nonumber \\ & & \left.
       + \cdots \right\}
            \, .
\label{eq:numresAS}
\end{eqnarray}
The $n_f$-proportional NNLO contribution
determines the value of the BLM scale 
\begin{equation}
 \muR = \left(\mu_{BLM}^2\right)^{as}
   \approx 0.114 Q^2
   \approx \frac{Q^2}{9}
        \, .
\label{eq:BLMas}
\end{equation}
One notes that this scale is considerably softer
than the total momentum transfer $Q^2$, 
which is consistent with partitioning of $Q^2$
among the pion constituents.
It should be pointed out, however, 
that in the $\overline{MS}$ scheme
the BLM scale does not reflect the mean gluon momenta.
Based on \req{eq:numresAS}, the NLO prediction amounts to
\begin{equation}
  Q^2 F_{\gamma^* \gamma \pi}(Q^2)= 
      0.185 
    +\frac{\alpha_S(\muR)}{\pi} (-0.309)
    + \cdots
        \, .
\label{eq:numResAS}
\end{equation}

\begin{figure}
  \centerline{\epsfig{file=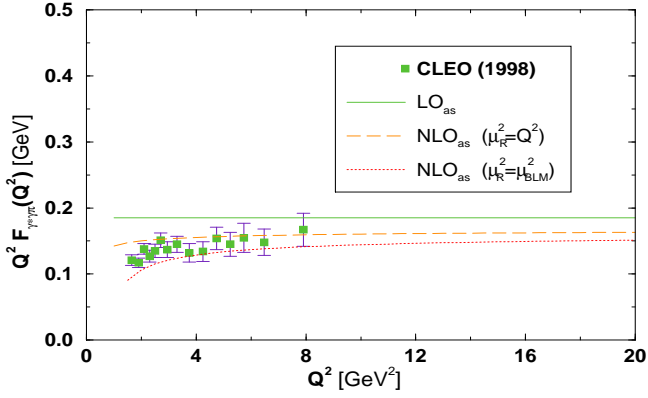,height=5.7cm,width=8.5cm,silent=}}
%  \centerline{\epsfig{file=slike/ascz.eps,height=8cm,width=11cm,silent=}}
 \caption{The LO and NLO predictions for the pion transition
   form factor (scaled with $Q^2$)
   $Q^2 F_{\gamma^* \gamma \pi}(Q^2)$, 
   obtained using the asymptotic DA. 
   The NLO predictions are obtained using
   the BLM scale
   $\muR=\mu_{BLM}^2 \approx Q^2/9$ \protect\req{eq:BLMas} and
   the commonly used choice $\muR=Q^2$.
   The experimental data are taken from \protect\cite{Gr98ea}.}
 \label{f:numAS}
\end{figure}

This prediction obtained with the 
asymptotic distribution and the $\overline{MS}$
definition of the strong coupling renormalized
at the BLM scale given by \req{eq:BLMas} is displayed 
in Fig. \ref{f:numAS},
along with the CLEO experimental data.
For comparison also included in Fig. \ref{f:numAS}
is the NLO prediction obtained by employing the 
widely used choice $\muR=Q^2$.
The usual one-loop formula
for the QCD running coupling constant \req{eq:alphaS}
has been used with $\Lambda=\Lambda_{\overline{MS}}=0.2$ GeV$^2$.

As it is seen from Fig. \ref{f:numAS}, the NLO
results for the pion transition form factor
display the following features.
First, inclusion of the NLO contributions
decreases the LO prediction.
Second, predictions based on the asymptotic distribution 
are in reasonably good agreement with currently available experimental data.

In comparison with the choice $\muR=Q^2$,
the BLM scale choice increases the absolute
value of the ratio of the NLO to LO prediction
by $\approx 11-6\%$ for the values of $Q^2$ between $6$ and $20$ GeV$^2$.
As an extension of the BLM scale-fixing prescription 
to all orders in perturbation theory, 
in \cite{GoK98} 
all the $(\beta_0 \alpha_s)^n$ contributions 
to the pion transition form factor were resummed 
under the assumption of "Naive Non-Abelianization" (NNA). 
Our results cannot be directly compared with \cite{GoK98},
but there are indications that 
the usual BLM scale-setting overestimates 
the size of higher-order 
contributions associated with the one-loop running
coupling \cite{resum}.

\begin{figure}
  \centerline{\epsfig{file=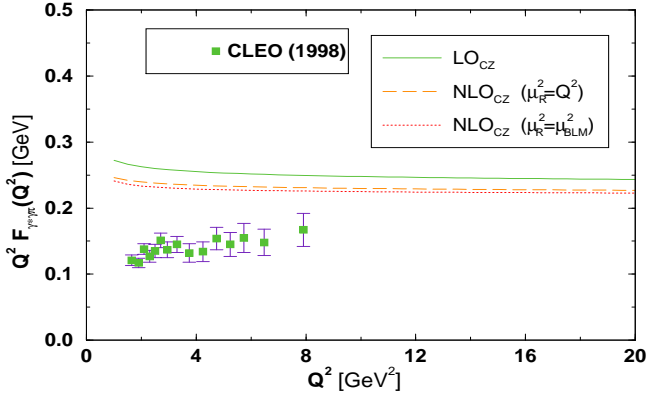,height=5.7cm,width=8.5cm,silent=}}
%  \centerline{\epsfig{file=slike/ascz.eps,height=8cm,width=11cm,silent=}}
 \caption{The LO and NLO predictions for the pion transition
   form factor (scaled with $Q^2$)
   $Q^2 F_{\gamma^* \gamma \pi}(Q^2)$, 
   obtained using the CZ distribution amplitude. 
   The NLO predictions are obtained using
   the BLM scale determined from
   \protect\req{eq:aBLMB24}, 
   and the commonly used choice $\muR=Q^2$.
   The experimental data are taken from \protect\cite{Gr98ea}.}
 \label{f:numCZ}
\end{figure}

The complete NLO prediction for the pion transition form
factor corresponding to 
the end-point concentrated $CZ$ distribution amplitude \cite{ChZ84},
given by Eqs. (\ref{eq:numres0B24}-\ref{eq:numres2nfB24})
with the coefficients $B_2=2/3$ and $B_4=0$,
is shown in Fig. \ref{f:numCZ}.
Owing to the fact that the LO and NLO evolutional corrections
to the CZ distribution are considerable \cite{MNP99},
they have been taken into account.
The BLM scale for the CZ distribution is higher than
the BLM scale for the asymptotic DA,
and it varies from $Q^2/1.84$ to  $Q^2/2.37$
for $Q^2$ between $2$ and $20$ GeV$^2$.

As it is evident from Fig. \ref{f:numCZ},
the complete NLO prediction for the pion transition
form factor derived from the CZ distribution
exceeds the experimental data significantly.
This result can be considered as a serious failure
of the CZ distribution amplitude.
Comparing Figs. \ref{f:numAS} and \ref{f:numCZ},
one observes that the difference between our results
for the asymptotic and CZ distribution amplitudes
is sufficiently large for an unambiguous experimental
discrimination between the two possibilities.
Therefore, one expects that the pion distribution
amplitude is closer to the asymptotic form than
to the strongly end-point concentrated  DA's like
CZ \cite{Rad95etc}.


The size of higher-order QCD corrections
represents the missing ingredient in assessing
the validity of the perturbative prediction and 
the convergence of the expansion.
One hopes that  the BLM prescription
offers a systematic way to choose
the renormalization scale and minimize
higher-order contributions.
In order to check this for the case of the pion
transition form factor, one would have to evaluate
the complete NNLO contribution, which is a
very demanding task.
Another sensible indicator of the applicability
of the perturbative calculation is the
size of the expansion parameter $\alpha_S(\muR)$.
The rather low BLM scale given in \req{eq:BLMas},
and consequently the large $\alpha_S(\mu_{BLM}^2)$,
questions the applicability 
of the perturbative prediction at experimentally
accessible momentum transfers.
Namely, the NLO predictions obtained in this paper
assuming the asymptotic DA and the BLM scale \req{eq:BLMas}
satisfy the requirement $\alpha_S(\muR)<0.5$
for $Q^2 \geq 6$ GeV$^2$.
It should be pointed out, however, that
there is an intrinsic disadvantage
in using the $\overline{MS}$ running coupling
(given by \req{eq:alphaS}) as an expansion parameter,
since it has a simple pole at $\muR=\Lambda^2$. 
This does not reflect the nonperturbative
behavior 
of $\alpha_S(\muR)$ for small $\muR$, and a number of proposals have
been suggested for the form of the coupling constant in
this regime \cite{alphaSmod,alphaSmodn},
but its implementation demands caution \cite{MNP99}.
For a recent application of \cite{alphaSmodn}
to the calculation of the pion transition form
factor, see \cite{SteSch00}.

\begin{figure}
  \centerline{\epsfig{file=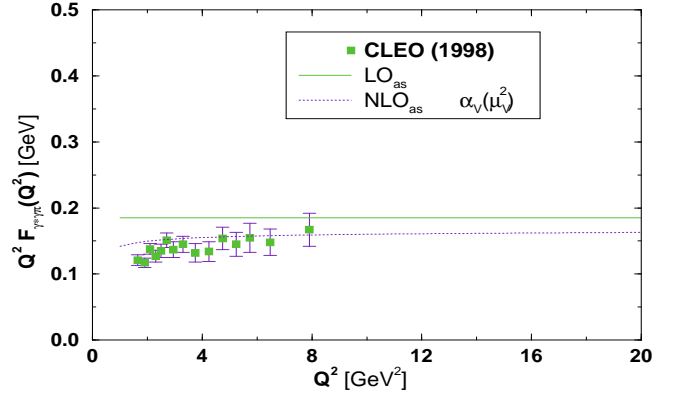,height=5.7cm,width=8.5cm,silent=}}
%  \centerline{\epsfig{file=slike/asV1.eps,height=8cm,width=11cm,silent=}}
 \caption{The LO and NLO predictions for the pion transition
   form factor (scaled with $Q^2$)
   $Q^2 F_{\gamma^* \gamma \pi}(Q^2)$, 
   obtained corresponding to the asymptotic DA in the $\alpha_V$ scheme 
   with $\mu_V^2\approx Q^2/1.7$ \protect\req{eq:Vas}.
   The experimental data are taken from \protect\cite{Gr98ea}.}
 \label{f:numV}
\end{figure}
So far nothing has been said concerning the
renormalization scheme dependence of the predictions.
It is known that the renormalization scheme dependence can be 
avoided by considering relations between physical observables, 
which must be independent of the choice of the scheme and scale
to any fixed order of perturbation theory.
This requirement can be expressed in the form of 
"commensurate scale relations" (CSR), 
in which the BLM scale-setting method is used to fix the 
renormalization scale \cite{BrLu95}. 
In practice, a CSR for two observables is obtained 
by relating their respective perturbative predictions calculated in, 
for example, the $\overline{MS}$ scheme, and then by algebraically 
eliminating $\alpha_{\overline{MS}}$.
The choice of the BLM scale ensures
that the resulting CSR is independent of the choice 
of the intermediate renormalization scheme.
Following this approach, in \cite{BrJ98}
the exclusive hadronic amplitudes
were calculated in the
$\alpha_V$ scheme, 
in which the effective coupling $\alpha_V(\mu^2)$ is defined
from the heavy-quark potential $V(\mu^2)$.
The $\alpha_V$ scheme is a natural, physically based scheme, 
which by definition automatically incorporates vacuum polarization 
effects %due to the fermion pairs 
into the coupling. 
The $\mu_V^2$ scale which then appears in 
the $\alpha_V$ coupling reflects the mean virtuality of the 
exchanged gluons.

If use is made of the scale-fixed relation 
between  the couplings 
$\alpha_{\overline{MS}}$ 
and $\alpha_V$ \cite{BrJ98}
\begin{equation}
\alpha_{\overline{MS}}(e^{-5/3} \mu_V^2)
=
\alpha_V(\mu_V^2) \left( 1 + \frac{\alpha_V(\mu_V^2)}{4 \pi}
          \, \frac{8 C_A}{3} + \cdots \right) \, ,
\label{eq:defalphaV}
\end{equation}
then,
to the order we are calculating, 
the prediction for the pion transition form
factor 
takes the form
\begin{equation}
  F_{\gamma^* \gamma \pi}(Q^2) = 
       F_{\gamma^* \gamma \pi}^{(0)}(Q^2)
        + \frac{\alpha_V(\mu_V^2)}{4 \pi}  \, 
       F_{\gamma^* \gamma \pi}^{(1)}(Q^2)
               + \cdots 
             \, ,
\label{eq:FinresV}
\end{equation}
with the ``V'' scale being given by
\begin{equation}
         \mu_V^2 = e^{5/3} \, \mu_{BLM}^2
              \, .
\label{eq:Vsc}
\end{equation}
Now, on the basis of \req{eq:BLMas} and \req{eq:Vsc}
one finds that
the V scale, corresponding to the asymptotic distribution
and reflecting the mean (NLO) gluon momentum,
is 
\begin{equation}
         \left(\mu_V^2 \right)^{as} \approx 0.6025\, Q^2 
               \approx \frac{Q^2}{1.7} 
              \, . 
\label{eq:Vas}
\end{equation}
Just for comparison, it is worth mentioning here
that the corresponding $\mu_V^2$ scale for the
pion electromagnetic form factor amounts to
$Q^2/20$ \cite{BrJ98,MNP99}.

Furthermore, since $\alpha_V$ is an effective running coupling defined
from the physical observable it must be finite at low momenta,
and the appropriate parameterization of the low-energy region 
should in principle be included.
Nevertheless, in the energy region we are interested
in, the usual one-loop conventional solution of the
renormalization group equation for the QCD coupling 
\req{eq:alphaS} can be employed.
The numerical NLO prediction for the pion transition
form factor obtained from Eqs. (\ref{eq:FinresV}-\ref{eq:Vas})
and calculated with $\Lambda_V=0.16$ GeV$^2$
is depicted in Fig. \ref{f:numV}.
As can be seen, it is in good agreement with experimental data.
The LO QCD correction, i.e., the NLO contribution,
lowers the LO prediction for $\approx 16\%$ 
for $Q^2 \approx 6$ GeV$^2$, i.e., for $\alpha_V(\mu_V^2) \approx 0.3$.

\section{Summary and conclusions}
\label{s:concl}

In this paper we have determined 
the NLO Brodsky-Lepage-Mackenzie (BLM) scale
and obtained the complete NLO prediction
for the pion transition form factor.

To determine the NLO BLM scale, 
a consistent and detailed calculation 
of both the hard-scattering and the 
perturbatively calculable part of 
the pion distribution amplitude has been performed
up to $n_f$-proportional NNLO contributions.
The calculation has been carried out in the Feynman gauge.
To control both the UV and collinear divergences
the dimensional regularization method has been employed.
By combining, according to Eq. \req{eq:Fpiur1},
and matching the results for the hard-scattering amplitude with
the corresponding results obtained for the distribution
amplitude, a proper cancellation of collinear singularities
has been established
and the $\gamma_5$ ambiguity problem (related to the
use of dimensional regularization) has been resolved
using the naive-$\gamma_5$ as well as 
the 't Hooft-Veltman (HV) schemes.
As a result, the complete leading-twist NLO prediction
for the pion transition form factor 
%has been obtained and is
has been obtained as given by  Eqs. \req{eq:Finres} and  
(\ref{eq:numres0B24}-\ref{eq:numres2nfB24}),
and Eqs. \req{eq:muBLM} and \req{eq:aBLMB24}, 
specifying the corresponding BLM scale.
Derived in the $\overline{MS}$ scheme 
(for which suitable compact form \req{eq:renalphaS}
has been adopted), these expressions
are valid for arbitrary form of the distribution amplitude
(with the evolutional effects taken into account),
and represent the main result of the present paper.

It has been demonstrated that the prediction for the 
leading-twist perturbative QCD prediction
for the pion transition form factor is 
independent of the factorization scale $\muF$ 
at every order in the strong coupling constant
$\alpha_S$ provided both the hard-scattering amplitude and the
pion distribution amplitude are treated consistently 
regarding their $\muF$ dependence.
The factorization scale  
disappears from the final prediction at every order in $\alpha_S$
without introducing any theoretical uncertainty.
Consequently, for practical purposes the
simplest and commonly used choice $\muF=Q^2$
is justified at the intermediate steps of the calculation.

Based on the general expressions
\req{eq:Finres} and (\ref{eq:numres0B24}-\ref{eq:numres2nfB24}),
the NLO predictions for the pion transition form factor 
have been obtained using the asymptotic and the CZ distribution
amplitudes, with the renormalization scales
being given by the respective BLM scales determined from
Eqs. \req{eq:muBLM} and \req{eq:aBLMB24}.
These predictions are shown in Figs. \ref{f:numAS} and
\ref{f:numCZ}, respectively.
By comparing these figures, one observes that, while,
on one hand, the prediction derived from the asymptotic 
distribution is in good agreement with the presently
available experimental data, on the other hand, the prediction
obtained assuming the CZ distribution exceeds the data
significantly, clearly demonstrating the inadequacy of the CZ
distribution.
This is in accordance with the conclusions reached in
\cite{soft,Rad98,Rad95etc,DArec},
according to which the distribution amplitude is closer 
to the asymptotic form than to the end-point concentrated
distribution of the CZ type. 
The renormalization scale $\muR$  fixed
according to the BLM scale setting prescription
within the $\overline{MS}$ scheme
and corresponding to the asymptotic pion distribution amplitude,
turns out to be
$\mu_{R,BLM}^2 \approx Q^2/9$.
Thus, in the region of $Q^2 < 8$ GeV$^2$,
in which the experimental data exist,
$\mu_{BLM}^2 < 1$ GeV$^2$.
Consequently, the prediction obtained with $\muR=\mu_{BLM}^2$
cannot, in this region, be considered reliable.

In addition to the results calculated with the
asymptotic distribution and $\overline{MS}$ renormalization
scheme, the numerical prediction assuming the same distribution
but in the  $\alpha_V$ scheme, 
with the renormalization scale 
$\muR=\mu_V^2 = e^{5/3} \mu_{BLM}^2 \approx Q^2/2$,
has also been obtained.
It is displayed in Fig. \ref{f:numV} and
as seen, is in good agreement with experimental data.
Due to the fact that the scale $\mu_V^2$ reflects
the mean gluon momentum in the NLO diagrams,
it is to be expected that the higher-order QCD
corrections are minimized, so that the leading order
QCD term gives a good approximation to the complete sum.

\acknowledgments

  One of us (B.M.) acknowledges the support 
  by the Alexander von Humboldt Foundation and the hospitality
  of the theory groups at the Institut f\"{u}r Physik,
  Universit\"{a}t Mainz \& Institut f\"{u}r Theoretische Physik,
  Universit\"{a}t W\"{u}rzburg.
  This work was supported by the Ministry of Science and Technology
  of the Republic of Croatia under Contract No. 00980102.

\appendix

\section{$\mathbf{\gamma_5}$ problem}
\label{app:gamma5}

\subsection{General remarks}

When using dimensional regularization, one runs
into trouble with quantities that have the well-defined
properties only in $D=4$ space-time dimensions, that is, with the
Levi-Civita tensor
$\eps_{\mu \nu \lambda \kappa}$, which is a genuine
4 dimensional object,
and consequently with the pseudoscalar $\gamma_5$ Dirac matrix.
The generalization of the $\gamma_5$ matrix 
in $D$ dimensions represents a problem,
since it is not possible to simultaneously retain
its anticommuting and trace properties.
In practice, the ambiguity arises
when evaluating a trace
containing  a  $\gamma_5$
and pairs of 
contracted $\gamma$ matrices
and/or pairs of Dirac slashed loop momenta
\footnote{
The presence of a pair of Dirac slashed loop momenta
leads in fact  to the  appearance of 
a pair of contracted $\gamma$ matrices,
since the loop integration 
\begin{displaymath}
   \int \frac{d^D l}{(2 \pi)^D} \, 
     \frac{ l^{\kappa} l^{\tau} }{(Denominator)} =
    g^{\kappa \tau} I_g + \cdots
\end{displaymath}
transforms
$l^{\kappa} l^{\tau} \gamma_{\kappa} \cdots \gamma_{\tau}$ into
$\gamma_{\kappa} \cdots \gamma^{\kappa} I_g + \cdots$. 
Apart from the contracting $\gamma$ matrices and pairs of 
Dirac slashed loop momenta, the rest of the trace elements
could be treated as 4 dimensional, so their (anti)commutation
with $\gamma_5$ does not make a difference. 
}.

To deal with a $\gamma_5$ matrix,
several possible schemes have been proposed
in the literature. 
Following the previous calculation 
of the pion transition form factor 
\cite{Bra83}, we have compared two of
them in the present calculation.

In the so-called naive-$\gamma_5$ scheme \cite{ChF79},
the anticommutation
property of $\gamma_5$ 
\begin{equation}
   \left\{ \gamma_{\mu}, \gamma_5 \right\} = 0
\label{eq:ng5sh}
\end{equation}
is retained, while 
the cyclicity of the trace 
is abandoned, so that, for example,
\begin{subequations}
\begin{eqnarray}
  \text{Tr} \left[ \gamma_5
          \not \! a  \gamma_{\mu}
      \not \! b  \not \! c \not \! d \gamma^{\mu} \right] &=&
        (D-6) \,
  \text{Tr} \left[ \gamma_5
       \not \! a  \not \! b  \not \! c  \not \! d   \right] 
           \, ,  \quad   \label{eq:Trexamp1} \\
  \text{Tr} \left[ \gamma^{\mu} \gamma_5
          \not \! a \gamma_{\mu}
      \not \! b \not \! c \not \! d  \right] &=&
        (2-D) \,
  \text{Tr} \left[ \gamma_5
        \not \! a  \not \! b  \not \! c  \not \! d  \right] 
           \, . \quad \label{eq:Trexamp2}
\end{eqnarray}
The traces obtained by cyclic permutation of
the matrices 
$\text{$\gamma_5$, 
$ \not \! a$,   
$\gamma_{\mu}$, $\not \! b$,  $ \not \! c$, $\not \! d$, 
$ \gamma^{\mu}$}$
can be  divided into two classes, depending on 
the location of $\gamma_5$ with respect to the contracted $\gamma$
matrices: 
%and, consequently, considering the results in
%naive-$\gamma_5$ scheme:
those in which 
$\gamma_5$ is outside the contracted pair 
as in \req{eq:Trexamp1}, and those
where $\gamma$s are contracted through $\gamma_5$ as
in \req{eq:Trexamp2}.
As is seen, the result \req{eq:Trexamp1}
and the result \req{eq:Trexamp2},
in which  the anticommuting property \req{eq:ng5sh}
of $\gamma_5$ had to be used before the contraction
of $\gamma$ matrices can be performed, differ by $D-4$. 
Consequently, if the
trace is multiplied by a pole in $D-4$,
there appears a finite ambiguity in the result.
\label{eq:Trexamp}
\end{subequations}
 
An alternative scheme has been proposed
in the original paper on the dimensional regularization
by 't Hooft and Veltman \cite{tHV72}, and further
systematized by Breitenlohner and Maison \cite{BM77}.
In this scheme, to which we refer as HV scheme,
the anticommutativity of
$\gamma_5$ is abandoned and replaced by
\begin{eqnarray}
   \left\{ \gamma_{\mu}, \gamma_5 \right\} &=& 0 
           \text{ for }  \mu=0, \cdots, 4
       \nonumber \\
   \left[ \gamma_{\mu}, \gamma_5 \right] &=& 0 
           \text{ for }  \mu=4, \cdots, D
         \, .
\label{eq:BMsh1}
\end{eqnarray}
For calculational purposes, it proves useful 
to introduce the following notation \cite{BM77}:
\begin{subequations}
\begin{eqnarray}
g_{\mu \nu}&=&\hat{\hat{g}}_{\mu \nu} + \hat{g}_{\mu \nu} \, ,
              \\
\gamma_{\mu}&=&\hat{\hat{\gamma}}_{\mu} + \hat{\gamma}_{\mu} \, ,
              \\
l_{\mu} &=&\hat{\hat{l}}_{\mu} + \hat{l}_{\mu} \, ,
\end{eqnarray}
\end{subequations}
where
\begin{eqnarray}
\hat{\hat{g}}_{\mu}^{\mu}=4 \, , &  \quad&
\hat{g}_{\mu}^{\mu}=D-4 \, ,
\end{eqnarray}
and
\begin{eqnarray}
g_{\mu \kappa} \hat{g}^{\kappa}_{\nu}= \hat{g}_{\mu \nu} \, ,
                & \quad &
\hat{g}_{\mu \nu} \gamma^{\nu}(l^{\nu}) =
\hat{\gamma}_{\mu}(\hat{l}_{\mu}) \, .
\end{eqnarray}
The relation \req{eq:BMsh1} can then be written as
\begin{equation}
 \gamma_{\mu} \gamma_{5} = - \gamma_5 \hat{\hat{\gamma}}_{\mu}
       + \gamma_5 \hat{\gamma}_{\mu}
            \, .
\label{eq:BMsh2}
\end{equation}
This prescription for $\gamma_5$ violates the Ward identities
and introduces ``spurious'' anomalies which violate chiral
symmetry.
% and hence gauge invariance 
To restore the Ward identities, 
finite counterterms should be added order by order 
in perturbation theory \cite{Bo80etc}.
In this scheme, 
the cyclicity of the trace is retained and
the traces given in \req{eq:Trexamp}
become 
\begin{eqnarray}
  \text{Tr} \left[ \gamma_5
          \not \! a  \gamma_{\mu}
      \not \! b  \not \! c \not \! d \gamma^{\mu} \right] &=&
  \text{Tr} \left[ \gamma^{\mu} \gamma_5
          \not \! a \gamma_{\mu}
      \not \! b \not \! c \not \! d  \right] 
         \nonumber \\ &=&
        (D-6) \,
  \text{Tr} \left[ \gamma_5
       \not \! a  \not \! b  \not \! c  \not \! d   \right] 
       \, . \quad
\label{eq:TrexampBM}
\end{eqnarray}
As is seen,  the result \req{eq:TrexampBM}, obtained
in the HV scheme, corresponds to the result \req{eq:Trexamp1},
obtained using the naive-$\gamma_5$ scheme
\footnote{
As stated in \protect\cite{tHV72}, should we allow
to anticommute $\gamma_5$ before continuation
to $D$ dimensions and after that use the HV scheme defined by
\protect\req{eq:BMsh1}, 
different results would emerge and ambiguity would reappear.
We would again obtain two classes of results:
the result \protect\req{eq:TrexampBM} for 
$\gamma_5$ outside the contracting pair of $\gamma$ matrices, 
and the result \req{eq:Trexamp2}
for anticommuting the matrix $\gamma_5$ 
in between the contracting $\gamma$s.
In this sense, the HV scheme would also lead
to ambiguous results.}.

If a trace contains an even number of $\gamma_5$
matrices, then the property $\gamma_5^2=1$ 
can be used to eliminate $\gamma_5$'s from the trace,
and the Ward identities are preserved if the
naive-$\gamma_5$ scheme \req{eq:ng5sh} 
is used \cite{ChF79} 
(the cyclicity of the trace is restored and
the corresponding results are unambiguous).
On the other hand, in the HV scheme 
the ``spurious'' anomalies can occur
owing to the 
non-anticommuting property of $\gamma_5$
\footnote{
Again, should we allow to anticommute 
$\gamma_5$ before continuing to $D$ dimensions
and then to use the HV scheme, different classes
of results could be obtained, 
depending on the position of $\gamma_5$s.
Hence, in this scheme, the ambiguity would still be present.}.

As for the traces containing an odd number of
$\gamma_5$ matrices, we are left with the
above mentioned ambiguities in the results.

\subsection{The $\gamma_5$ ambiguity in the 
$\gamma^* \gamma \to q \overline{q}$ amplitude}

There are two approaches one can take in order
to resolve the $\gamma_5$ ambiguity problem in
practical calculations (such as the calculation
of the hard-scattering amplitude of the
pion transition form factor). 
For each diagram, 
one can determine the way of manipulating the $\gamma_5$ matrices
so that the Ward identities are preserved
\cite{Bra83,AgC81}.
Alternatively, one can perform the calculation
using the HV scheme, and then introduce
an additional renormalization constant
which eliminates the ``spurious'' anomalies
introduced by this prescription.
In \cite{Bra83}, the counterterm at the NLO order
was calculated for
the special case of hard-scattering amplitude
already convoluted with the asymptotic
distribution amplitude. 

We have calculated the LO, NLO, and $n_f$-proportional
NNLO terms to the hard-scattering amplitude
using both the naive
and the HV prescription for $\gamma_5$. 
By combining these results with the results for the
distribution amplitude obtained in the same order,
ambiguity in the naive-$\gamma_5$ scheme
has been resolved and the HV renormalization constant
determined.

Although our end NLO result for the hard-scattering amplitude
agrees with the result
given in \cite{Bra83} (determined up to $O(\eps^0)$),
the same is not true for the contributions of the 
individual one-loop diagrams of Fig. \ref{f:Tnlo}
when calculated using the naive prescription for $\gamma_5$.

Namely, owing to the fact that the relative position of $\gamma_5$
with respect to
the Dirac slashed loop momenta was ignored, 
classes of terms appearing in the contribution
of individual diagrams were omitted in \cite{Bra83}.

For example, the trace corresponding to the diagram 
$A23$ of Fig. \ref{f:Tnlo} 
leads to three
different results depending on the position of $\gamma_5$: 
\begin{widetext}
\begin{subequations}
\begin{eqnarray}
\lefteqn{
  \text{Tr} \left[ \gamma_5 \not \! P
      \gamma^{\mu} \left( x \not \! P - \not \! q \right)
   \gamma_{\lambda} \left( x \not \! P - \not \! q - \not  l \right)
   \gamma^{\nu} \left( -(1-x) \not \! P - \not  l \right)
   \gamma^{\lambda} \right] 
}
\label{eq:TrA23a}
            \\ & \neq &
  \text{Tr} \left[ \gamma^{\lambda} \gamma_5 \not \! P
      \gamma^{\mu} \left( x \not \! P - \not \! q \right)
   \gamma_{\lambda} \left( x \not \! P - \not \! q - \not  l \right)
   \gamma^{\nu} \left( -(1-x) \not \! P - \not  l \right)
   \right] 
\label{eq:TrA23b}
            \\ & \neq &
  \text{Tr} \left[ \left( -(1-x) \not \! P - \not  l \right)
      \gamma^{\lambda} \gamma_5 \not \! P
      \gamma^{\mu} \left( x \not \! P - \not \! q \right)
   \gamma_{\lambda} \left( x \not \! P - \not \! q - \not  l \right)
   \gamma^{\nu} 
   \right] 
        \, .
\label{eq:TrA23c}
\end{eqnarray}
\end{subequations}
\end{widetext}
In Ref. \cite{Bra83}, however, only the results for
traces \req{eq:TrA23a} and \req{eq:TrA23b} were given.
\label{eq:TrA23}
To conveniently describe the $\gamma_5$ ambiguity
present in the contributions of $A23$ and other diagrams,
we have introduced 
the parameters $\delta$ and $\delta'$.
Thus, $\delta=0$ corresponds to the
situation where $\gamma_5$ lies outside the contracting
$\gamma$s and pairs of $\not l$s, as in \req{eq:TrA23a}.
Corresponding to the case when $\gamma$'s are contracted
through the $\gamma_5$ matrix
is $\delta=1$.
For $\gamma_5$ not being placed between the pairs of  $\not  l$s,
as in \req{eq:TrA23b},
$\delta'=0$, while for
$\gamma_5$ placed between the pairs $\not  l$s,
as in \req{eq:TrA23c},
$\delta'=1$.
The contributions obtained using the HV scheme
correspond to $\delta=0$.
The contributions 
defined in \req{eq:TAij}
of the individual NLO diagrams 
of Fig. \ref{f:Tnlo}
can then be parameterized by
\begin{widetext}
\begin{eqnarray}
\widetilde{T}_{A11(22)} &=&
  \left[ \,   -\frac{1}{2}(\left[ \, \eta_{UV} + 2 \delta \, \right]- 1 )
              -\frac{1}{2}(\left[ \, \eta_{IR} - 2 \delta \, \right]+ 1 )
                \, \right]   + O(\eps) \, \label{eq:tTA11Oe0}\\[0.2cm]
\widetilde{T}_{A33} &=&
   \bigg[ \, -\left[ \, \eta_{UV} + 2 \delta \, \right] +\ln (1-u) - 1  
                 \, \bigg]     + O(\eps)\, \label{eq:tTA33Oe0}\\[0.2cm]
\widetilde{T}_{A23} &=&
    \bigg[ \, \left[ \, \eta_{UV} +2 \delta (1 + \delta') \, \right] + 2 \eta_{IR}+\ln (1-u) -4
            \, \bigg] + O(\eps)   \,  \label{eq:tTA23Oe0} \\[0.2cm]
\widetilde{T}_{A13} &=&
   \left[ \, \left[ \, \eta_{UV} + 2 \delta (1 + \delta') \, \right] +
     2 \eta_{IR}\left(1 + \frac{1}{u} \ln (1-u)\right) 
        \nonumber \right. \\ & & \left. 
     +  \frac{1}{u} \ln^2 (1-u)
     - \frac{3-u}{u} \ln (1-u)-4  
             \, \right] + O(\eps)  \,  \label{eq:tTA13Oe0} \\[0.2cm]
\widetilde{T}_{A12} &=&
   \left[ \, -2  \frac{1-u}{u} \ln (1-u)
    \left[ \, \eta_{IR} + 2 \delta (1 +  \delta') \, \right]  
            \nonumber \right. \\ & & \left. 
     -\frac{1-u}{u} \ln^2 (1-u) 
    + 10 \frac{1-u}{u} \ln (1-u)
              \, \right]+ O(\eps) \,  ,
\label{eq:tTA12Oe0}
\end{eqnarray}
\end{widetext}
where to facilitate the comparison with the 
results of Ref. \cite{Bra83}, we have introduced
$\eta_{UV}= 1/\hat{\eps}+ \ln \mu^2/Q^2$,
$\eta_{IR}= -1/\hat{\eps}- \ln \mu^2/Q^2$, and
$1/\hat{\eps} =1/\eps-\gamma + \ln (4 \pi)$.
In \cite{Bra83} the results corresponding to
$\delta'=1$ in  \req{eq:tTA23Oe0} and \req{eq:tTA13Oe0},
as well as to $\delta'=0$ in \req{eq:tTA12Oe0}, were omitted.

%Note that the parameters $\delta$ and $\delta`$ 
%can generally differ from diagram to diagram.
%i.e., that, as mentioned before, the treatment of
%$\gamma_5$ matrix differs.
It was argued in \cite{Bra83} that since
the quark propagator and the photon vertex corrections 
were related by the Ward identity of QED,
they should be calculated as if they were not part of
the trace with $\gamma_5$. 
This then determines the choice $\delta=0$ in
diagrams $A11$, $A22$, $A33$, $A23$, and $A13$.
The remaining ambiguity associated with the 
IR (collinear) pole in diagram $A12$ was resolved by 
repeating the calculation in the 
``equal-mass regularization'', in which the
quark and the gluon were given the same small mass $m$. 
It was found that the correct choice for diagram $A12$ 
corresponded to the result obtained by
contracting the $\gamma$s through $\gamma_5$,
and for $\gamma_5$ placed between the $\not l$s 
(although the latter requirement was not stressed in \cite{Bra83}), 
that is, for
$\delta=1$ and $\delta'=1$.

We have confirmed these choices by our calculation.
In presenting our results in Sec. \ref{s:T},
we have already adopted the $\delta=0$ choice (or, equivalently, 
the HV results) for diagrams
$A11$, $A22$, $A33$, $A23$, and $A13$, as well as
for the diagrams obtained from these by inserting
the quark vacuum polarization loop.
The ambiguity present in diagram $A12$
and corresponding vacuum polarization diagram,
is parameterized only by $\delta$, 
with $\delta`$ taken as $1$.

A remark concerning the presentation
of the $\gamma^* \gamma \rightarrow \overline{q} q$
amplitude in the form given by \req{eq:Gmunu}
is in order.

The multiplication of the amplitude by the factor
$1/(i e^2) \, N \, \eps_{\mu \nu \alpha' \beta'} P^{\alpha'}q^{\beta'}$
and the contraction of the Levi-Civita tensors 
before the loop integrals are evaluated 
simplifies the calculation 
(since then at most three-point Feynman integrals appear).
Since the generalization of the Levi-Civita tensors
to $D$ dimensions is not unique,
caution should be exercised when contracting these tensors.
When using the HV scheme, the relation \cite{BM77}
\begin{equation}
\eps^{\mu_1 \mu_2 \mu_3 \mu_4} \, \eps_{\nu_1 \nu_2 \nu_3 \nu_4}
= - \sum_{\pi \in S_4} \text{sign} \pi \prod_{i=1}^4
     \hat{\hat{g}}^{\mu_i}_{\nu_{\pi(i)}}
\label{eq:epskontr}
\end{equation}
should be employed, in which case $N=2/Q^4$. 
Note that the loop integrals with $\hat{\hat{l}}$ terms
in the numerator of the integrand will be encountered.
Only the one-loop integrals containing $\hat{\hat{l}}^2$
terms are different from the corresponding counterparts
with $l$ in place of $\hat{\hat{l}}$, while
the presence of the $\hat{\hat{l}}^{\mu_1}\cdots\hat{\hat{l}}^{\mu_n}$
terms does not alter the usual results. 
When using the naive-$\gamma_5$ scheme, the 
$\hat{\hat{g}}^{\mu_i}_{\nu_{\pi(i)}}$ from \req{eq:epskontr}
can be replaced by $g^{\mu_i}_{\nu_{\pi(i)}}$,
and then $N=  2/((D-2)(D-3))\; 2/Q^4$.
We have checked our results 
by evaluating the contributions with and without the
Levi-Civita contraction. The results agree.

\section{Feynman rules for the perturbatively calculable part of the DA}
\label{app:DA}

In this section we list the Feynman rules 
for the $\tilde{\phi}(u,t)$ operator \req{eq:phiOqq}
rederived following \cite{Ka85etc}.

In Section \ref{s:formalism} the 
distribution amplitude $\tilde{\phi}(u,t)$
for a state composed
of a free quark and antiquark 
has been introduced \req{eq:phiOqq}
%%%%%
\begin{eqnarray*}
  \tilde{\phi} (u, t)
& = &  \int \frac{dz^-}{2 \pi} e^{i(u-(1-u))z^- /2}
   \nonumber \\ & & \times
    \left< 0 \left| 
  \overline{\Psi}(-z) \, \frac{\gamma^+ \gamma_5}{2\sqrt{2}}
           \, \Omega \, \Psi(z) \right| q \overline{q}; t \right>
          \, \frac{1}{\sqrt{N_c}}
              \, . \qquad
%\label{eq:phiOqq}
\end{eqnarray*}
with $z^+=z_{\perp}=0$.
The quark and antiquark
carry the
momenta $t P$ and $(1-t) P$, respectively,
and
the frame in which
$P^+=P^0+P^3=1$, $P^-=P^0-P^3=0$, $P_{\perp}=0$ 
has been chosen.
The path-ordered factor $\Omega$ \req{eq:Omega}
\begin{displaymath}
 \Omega  =   \text{exp} \left\{ i g \int_{-1}^{1} ds A^+(z s)z^-/2 \right\}
      \, ,
%\label{eq:Omega}
\end{displaymath}
makes $\tilde{\phi}(u,t)$ gauge invariant,
and it can be expanded in perturbation
theory as
%%%%%
\begin{equation}
 \Omega  =  \sum_{n=0}^{\infty}
     \frac{(i g)^n}{n!} \int_{-1}^1 \prod_k^n
      d s_k \,  A^+(s_k z) z^-/2 
            \, .
\label{eq:expOmega}
\end{equation}
The path-ordering is immaterial since $A^+$ 
fields commute. In the light-cone gauge
($A^+=0$), this operator is unity,
but generally (for example, in
the Feynman gauge we are using) introduces
extra diagrams. The $n$-th order
term in the \req{eq:expOmega} series
will correspond to diagrams with 
$n$ gluon lines attached to the 
operator vertex. 
By inserting the term
\begin{equation}
   \theta(1-s_k) \theta(1+s_k) = i \int
     \frac{d r}{2 \pi}  e^{i r s_k} \frac{1}{r}
        \left( e^{-i r} - e^{i r} \right)
\label{eq:thetatheta}
\end{equation}
the limits of the integration 
in \req{eq:expOmega} are changed,
and the function $\tilde{\phi} (u, t)$
takes the form
\begin{widetext}
\begin{eqnarray}
  \tilde{\phi} (u, t) &=&  \int \frac{dz^-}{2 \pi} e^{i(u-(1-u))z^- /2}
             \nonumber \\ & & \times
   \sum_{n=0}^{\infty} \frac{(i g)^n}{n!} 
    \, i \int
     \frac{d r}{2 \pi}  \frac{1}{r}
        \left( e^{-i r} - e^{i r} \right)
      \int_{-\infty}^{\infty}
      ds_1 \cdots ds_n e^{\imath r(s_1+ \cdots+s_n)}
             \nonumber \\ & & \times
    \left< 0 \left| 
  \overline{\Psi}(-z) \, \frac{\gamma^+ \gamma_5}{2\sqrt{2}}
           \, A^+(s_1 z) \cdots A^+(s_n z) \, 
            \Psi(z) \right| q \overline{q} \right>
          \left( \frac{z^-}{2} \right)^n \,
          \, \frac{1}{\sqrt{N_c}}
              \, .
\label{eq:phiOqqFr}
\end{eqnarray}
\end{widetext}
The fields in \req{eq:phiOqqFr}, 
along with the standard quark-gluon interaction
insertions, are
contracted in all possible ways, yielding
different Feynman diagrams contributions.
Here we list the Feynman rules
derived from \req{eq:phiOqqFr}. 

The operator
$ \displaystyle \left< 0 \left| 
 \overline{\Psi}(-z) \, \frac{\gamma^+ \gamma_5}{2\sqrt{2}}
  \Psi(z) \right. \right.$
from Eq. \req{eq:phiOqqFr} will be represented by
the crossed circle $\otimes$ in Feynman diagrams.
The $\otimes$ vertex 
has a quark line entering
and leaving the $\otimes$ vertex, and an arbitrary
number of attached gluon lines 
($A^+$ field can be contracted only with $A^-$,
so the gluon line does not re-enter the vertex).
 
The general form of the $\otimes$ vertex 
with $n$ gluons attached to it
is given by
\begin{subequations}
\begin{equation}
  \frac{\gamma^+ \gamma_5}{2 \sqrt{2}}
  \int \frac{d z^-}{2 \pi} e^{i S z^-/2}
  \prod_{j=1}^{n} (i g) \frac{i}{q_j^+} 
           \left( 1 - e^{i z^- q_j^+} \right)
            \, ,
\label{eq:Overtex}
\end{equation}
where
\begin{equation}
  S=u-(1-u)- k_1^+ - k_2^+ - \sum_{j=1}^{n} q_j^+
         \, ,
\label{eq:S}
\end{equation}
\label{eq:OvertexS}
\end{subequations}
and
$q_1, \cdots, q_n$ are 4-momenta of the gluons entering the
circle, 
while $k_1$ and $k_2$ are the 4-momenta of the quarks 
entering and leaving the circle, respectively.
The form given in \cite{Ka85etc} is slightly
different and incomplete regarding the 
gluon 4-momenta sign convention.
In the special case when there are
no gluons attached to the $\otimes$ vertex,
and the notation $k_1=k$ and $k_2=k-P$
is used, the expression \req{eq:OvertexS}
takes the form
\begin{equation}
  \frac{\gamma^+ \gamma_5}{2 \sqrt{2}}
   \, \delta(u-k^+)
       \, .
\label{eq:Fr0}
\end{equation}

The gluon propagator for the gluon attached to the
$\otimes$ vertex
(stemming from the $A^+ A_{\nu}$ contraction)
takes the form
\begin{equation}
       -i \frac{g^+_{\nu}}{q^2 + i \eta} \delta_{a b}
                 \, ,
\end{equation}
so after the $g^+_{\nu} \gamma^{\nu}$ contraction, the gluon-quark vertex 
for a gluon attached to the $\otimes$ vertex becomes 
\begin{equation}
        -i g \gamma^{+} \frac{\lambda_a}{2}
              \, .
\end{equation}

We will not specify here 
the usual Feynman rules already used in the calculation of the
hard-scattering amplitude

Let us just note that
similarly to the calculation of the hard-scattering amplitude,
the correct spin and parity state of the
$q \overline{q}$ state has been projected by
multiplying the diagrams by 
\begin{equation}
      \frac{\gamma_5 \gamma^{-}}{2 \sqrt{2}}
         =
      \frac{\gamma_5 \not \! P}{\sqrt{2}}
\label{eq:gamma5Tr}
\end{equation}
and taking the trace.
The color singlet nature of the $q \overline{q}$ state 
is taken into account by including the factor
\begin{equation}
       \sum_{\alpha=1}^{3} \frac{\delta_{\alpha \beta}}{\sqrt{N_c}}
\label{eq:colorTr}
\end{equation}
and consequently the trace over color indices must be taken.
The result should be multiplied by an extra factor $1/\sqrt{N_c}$ 
(see Eq. \req{eq:phiOqqFr}), 
which takes into account that
$\tilde{\phi}(u,t)$ is normalized to give the LO result
$\delta(u-t)$ (i.e., normalized to $\openone$).
Finally, note that we are investigating
the meson flavor nonsinglet distribution amplitude.

\section{On the coupling constant renormalization}
\label{app:alphaS}

In this section we briefly resume
the relevant ingredients of
the coupling constant renormalization,
which are frequently obscured in practical applications
found in the literature, and
we also introduce our specific representation \req{eq:renalphaS}
for the $\overline{MS}$ renormalization used in this
calculation.

\subsection{Coupling constant renormalization}

It is well known that in $D=4$ dimensions, the QCD coupling
strength $g$ is a dimensionless quantity
$[g]=M^0$ ($M$ denotes the mass unit), while in
$D=4-2 \eps$ dimensions $[g]=M^{\eps}$.
Obviously, 
the dimension of the ``bare''coupling constant $\alpha_S$, 
related to ``bare'' $g$ by
$g^2= 4 \pi \alpha_S$, 
corresponds to $[\alpha_S]=M^{2 \eps}$.
The renormalized coupling,
i.e., the running coupling $\alpha_S(\mu^2)$,
is, naturally, a dimensionless
quantity and the coupling constant renormalization
introduces the additional scale $\mu^2$, whose presence
balances the dimensions in
the renormalization equation
$\alpha_S \; \mu^{-2 \eps} = \alpha_S(\mu^2) \, Z_{\alpha}$.
The scale introduced by the renormalization of the
coupling constant is often referred to as renormalization
(or coupling constant) scale and denoted by $\muR$,
while the renormalization of the coupling constant
in the MS scheme (the simplest renormalization scheme) reads 
\begin{equation}
   \alpha_S  = \mu_R^{2 \eps} \;
                \alpha_S(\muR)
        \left( 1 - \frac{\alpha_S(\muR)}{4 \pi} \beta_0
                \frac{1}{\eps} + O(\alpha_S^2) \right)
        \, ,
\label{eq:alren1}
\end{equation}
where $\beta_0=11-2/3 \, n_f$.

In practical calculations, the additional scale $\mu^2$
is often introduced before the coupling constant
renormalization (the presence of $\mu^2$ in
Feynman integrals is quite standard),
which corresponds to introducing the
dimensionless ($[\alpha_S]=M^0$)
``bare'' coupling constant $\alpha_S$ related
to $g$ by 
$g^2= 4 \pi \alpha_S \, \mu^{2 \eps}$,
and consequently 
$\alpha_S \sim \mu^{-2 \eps}$.
The ``bare'' and renormalized coupling constants
are then related by
$\alpha_S  = \alpha_S(\mu^2) \, Z_{\alpha}$,
and, if we choose to renormalize the coupling at
the renormalization scale $\muR$ different from
the scale $\mu^2$ introduced by regularization, 
the renormalization in MS scheme reads
\begin{equation}
   \alpha_S  =  \left(\frac{\muR}{\mu^2}\right)^\eps
                 \alpha_S(\muR)
        \left( 1 - \frac{\alpha_S(\muR)}{4 \pi} \beta_0
                \frac{1}{\eps} + O(\alpha_S^2) \right)
               \, .
\label{eq:alren2}
\end{equation}
We adopt the latter definition of the ``bare'' 
coupling constant as dimensionless quantity,
and neglect the
$O(\alpha_S^3)$ terms in further considerations.

From \req{eq:alren2} one trivially obtains
the scale changing relation
\begin{subequations}
\begin{eqnarray}
 \lefteqn{\alpha_S(\mu^2)}\nonumber \\ 
      & = & \left( \frac{\muR}{\mu^2} \right)^\eps 
            \alpha_S(\muR)
     \left[ 1 + \frac{\alpha_S(\muR)}{4 \pi} \beta_0
           \frac{1}{\eps} 
         \left( \left(\frac{\muR}{\mu^2} \right)^\eps
                -1 \right) \right]
                \nonumber \\ & &
\label{eq:chsc}  \\ &=&
           \alpha_S(\muR) 
     \left( 1 + \frac{\alpha_S(\muR)}{4 \pi} \beta_0
              \ln \frac{\muR}{\mu^2} \right)
                 + O(\eps)
             \, ,
\label{eq:chscLO}
\end{eqnarray}
\label{eq:ChSc}
\end{subequations}
the $\beta$ function
\begin{eqnarray}
\beta(\alpha_S(\mu^2),\eps) &=& \mu^2 \frac{\partial }{\partial \mu^2}
        \alpha_S(\mu^2) 
        \nonumber \\ &=&
   - \eps \, \alpha_S(\mu^2) 
   - \frac{\alpha_S^2(\mu^2)}{4 \pi} \beta_0 
          \, ,
\label{eq:beta}
\end{eqnarray}
and, consequently, the running coupling
\begin{equation}
  \alpha_S(\mu^2)=\frac{4 \pi}{\beta_0 
   \left( \left(\frac{\mu^2}{\Lambda^2}\right)^\eps -1 \right) \frac{1}{\eps} }
                 =\frac{4 \pi}{\beta_0 \ln \frac{\mu^2}{\Lambda^2}}
                       + O(\eps)
                  \, .
\label{eq:alphaS}
\end{equation}
It is safe to ignore the $O(\eps)$ terms in \req{eq:alphaS},
since the expression for the running coupling 
is usually introduced
after all singularities have been removed 
(by renormalization and/or factorization).
Generally, one cannot neglect the $O(\eps)$ terms in
\req{eq:alren2} or \req{eq:ChSc} 
(neither in order $\alpha_S$ nor in $\alpha_S^2$), 
and the use of the compact
forms \req{eq:alren2} and \req{eq:chsc} 
is preferred until all singularities are removed%
\footnote{
One of the motivations for this short summary
on the coupling constant renormalization 
was the appearance of different forms
of $\alpha_S$ renormalization equations 
in the literature.
The quite often used form %\cite{MNP99,?}
\begin{displaymath}
    \alpha_S = 
           \alpha_S(\muR) 
     \left[ 1 - \frac{\alpha_S(\muR)}{4 \pi} \beta_0
              \left(\frac{1}{\eps}-\ln \frac{\muR}{\mu^2}\right)
             \right]
%\label{eq:renalSMSLO}
\end{displaymath}
represents an ``effective'' expression,
which can be strictly used only for calculations 
in which there are no singularities apart from those
that get renormalized by this
coupling constant renormalization.
The curious looking 
$\alpha_S$ renormalization procedure
given in, for example, \cite{Neer},
presumably represents an attempt to %naively 
generalize the above given ``effective'' form
to all orders in $\eps$.
In the presence of additional UV or IR singularities,
the terms containing both scales $\muR$ and $\mu^2$
remain after the $\alpha_S$ renormalization,
which is clearly inconsistent.
The final finite results are correct since these curious
looking terms are moved to 
renormalization and/or factorization constants.
}.  

\subsection{Renormalization schemes}

Next we turn to the choice of the renormalization
scheme and the representations of that choice.

One can introduce different renormalization
schemes by modifying the 
renormalization constant $Z_{\alpha}$ to
$\bar{Z}_{\alpha}$ as a function of
\begin{equation}
         \qquad f(\eps) = 1 + \eps \, f^{(1)} + \eps^2 f^{(2)} + \cdots
            \, ,
\label{eq:f}
\end{equation}
where $f(\eps)$ defines the renormalization scheme choice,
while
$\alpha_S  = \alpha_S(\mu^2) \, \bar{Z}_{\alpha}$. 
Equation \req{eq:alren2} is then generalized to
\begin{equation}
   \alpha_S  =  \left( \frac{\muR}{\mu^2} \right)^\eps \alpha_S(\muR)
        \left( 1 - \frac{\alpha_S(\muR)}{4 \pi} \beta_0
                \, f(\eps) \, \frac{1}{\eps} \right)
             \, ,
\label{eq:renalSf}
\end{equation}
and the choice $f(\eps) = 1 \equiv f_{MS}(\eps)$ corresponds
to the MS scheme.  
Equations (\ref{eq:ChSc}-\ref{eq:alphaS}) get modified by
$\beta_0 \rightarrow \beta_0 \, f(\eps)$.

The $\overline{MS}$ scheme is 
defined by
\begin{equation}
 f_{\overline{MS}}(\eps) = 1 + \eps(-\gamma+\ln (4 \pi)) 
                             + O(\eps^2)
                  \, ,
\label{eq:fMS}
\end{equation}
and, to the order we are calculating, 
the $O(\eps^2)$ terms are arbitrary.
Different definitions can be found in the literature:
the original definition of 
the $\overline{MS}$ scheme \cite{MSb}
$f_{\overline{MS}}(\eps) =1 + \eps(-\gamma+\ln (4 \pi))$
or the choices
$f_{\overline{MS}}(\eps) =\exp \left(\eps(-\gamma+\ln 4 \pi)\right)$ 
and 
$f_{\overline{MS}}(\eps) =(4 \pi)^\eps/\Gamma(1-\eps)$ 
(for example, \cite{Neer,Cat98} and \cite{KuS94}, respectively),
which mimic the $\eps$ dependence of the 
$\gamma^n$, $\ln^n 4 \pi$, $(\pi^2)^n$ proportional terms
introduced by dimensional regularization.
Although they are all
valid choices leading to the same ($\overline{MS}$) result
(after renormalization and factorization of singularities),
they unnecessarily complicate the calculation,
since
the intermediate results should be expanded over
$\eps$ and since they do not contain the 
($\Gamma$) functions  originally introduced by dimensional regularization.
A more appropriate choice would be
the one that contains combinations
of $\Gamma$s that naturally emerge in the calculation.
For this calculation, in which
both UV and IR singularities were regularized
by dimensional regularization,
the appropriate choice is (see \req{eq:M12} and \req{eq:trik})
\begin{equation}
 f_{\overline{MS}}(\eps) = \eps \; \Gamma(\eps)\Gamma(1-\eps)
              \frac{\Gamma(1-\eps)}{\Gamma(1-2 \eps)}
                     (4 \pi)^\eps
                         = \eps \; \Gamma_{UV}^{(0)}(\eps)
                         \, ,
\label{eq:fMSour}
\end{equation}
while, for example, for the calculation in which 
only UV singularities
were regularized by dimensional regularization,
the choice $f_{\overline{MS}}(\eps)=\eps \Gamma(\eps)$
would be appropriate.

Alternatively, we can represent the dependence
of the coupling constant renormalization 
on the renormalization scheme by
$\alpha_S \, f(\eps)  = \alpha_S(\mu^2) \, Z_{\alpha}$
(see, for example, \cite{Cat98}).
The generalization of \req{eq:alren2} is then given by
\begin{equation}
   \alpha_S  =  \left( \frac{\muR}{\mu^2} \right)^\eps 
                   (f(\eps))^{-1} \, \alpha_S(\muR)
        \left( 1 - \frac{\alpha_S(\muR)}{4 \pi} \beta_0
                \frac{1}{\eps} \right)
            \, ,
\label{eq:renalSf1}
\end{equation}
and it represents a representation alternative
to \req{eq:renalSf}.
Equations (\ref{eq:ChSc}-\ref{eq:alphaS}) are  valid
for the \req{eq:renalSf1} representation.

Since $\beta_0$ (and $\beta_1$) does not dependent on the
choice of the renormalization scheme, 
the renormalization scheme and scale
dependence can be described,
to the order we are calculating, 
only by one parameter, (for example, the scale).
The renormalization scheme and the renormalization scale 
are treated on the same footing
in representation \req{eq:renalSf1}
and their equivalence is explicit. 
For example, by substituting
$\muR  = \tilde{\mu}_R^2\; (f(\eps))^{\frac{1}{\eps}}$,
$\Lambda^2 = \Lambda_{MS}^2 \; (f(\eps))^{\frac{1}{\eps}}$
into \req{eq:renalSf1}, the renormalization
equation \req{eq:alren2} in the MS scheme is obtained.

The another advantage of the \req{eq:renalSf1} representation
is that  
after the coupling constant renormalization is performed,
the dimensional parameter $\eps$ remains as the only
artifact of dimensional regularization.
Consequently,
when using Eq. \req{eq:renalSf}, the renormalization
and factorization constants 
$Z_{{\cal M}, ren}, Z_{{\cal M}, col}$ contain
$\gamma^n$, $\ln^n 4 \pi$, $(\pi^2)^n$ proportional terms
apart from the $\frac{1}{\eps^n}$ poles.
In contrast,
when using Eq. \req{eq:renalSf1} 
$Z_{{\cal M}, ren}, Z_{{\cal M}, col}$ contain
only simple $\frac{1}{\eps^n}$ poles.

In this work the representation
\req{eq:renalSf1} along with 
the  definition \req{eq:fMSour}
of the $\overline{MS}$ scheme is used.

\end{document}